\documentstyle[12pt]{l-aa}
\input{psfig}

\def\eqno(#1){\label{#1}}
\def\eqref(#1){(\ref{#1})}

\def\hmove{\hspace{-.35cm}}

\def\disp{\displaystyle}

\def\pp{\parshape 2 0truecm 15truecm 2truecm 13truecm}
\def\apjref#1;#2;#3;#4; {\par\pp#1, {\it #2}, {\bf #3}, #4. \par}
\def\bookref#1;#2;#3; {\par\pp#1, {\it #2}, {\rm #3}. \par}
\def\prepref#1;#2; {\par\pp#1, {\it #2}. \par}
\def\ie{{\it i.e.}}
\def\eg{{\it e.g.}}

\def\ltsima{$\; \buildrel < \over \sim \;$}
\def\simlt{\lower.5ex\hbox{\ltsima}}
\def\gtsima{$\; \buildrel > \over \sim \;$}
\def\simgt{\lower.5ex\hbox{\gtsima}}
\def\tiy{y}
\def\ka{\kappa}
\def\d{{\rm d}}
\def\i{{\rm i}}

\def\gm{\gamma}
\def\la{\lambda}

\def\xib{\overline{\xi}}
\def\Mb{\overline{m}}

\def\Nb{\overline{N}}
\def\nub{\overline{\nu}}
\def\hb{\overline{h}}
\def\bb{\overline{b}}
\def\mg{\big <}
\def\md{\big >}
\def\sym{{\rm cyc}}
\def\mP{{\cal P}}

\def\mN{{\cal N}}
\def\Mpc{{\rm Mpc}}
\def\kpc{{\rm kpc}}
\def\Nc{N_{{\rm c}}}
\def\Mc{m_{{\rm c}}}
\def\Mcu{m_{{\rm c}1}}
\def\Mcd{m_{{\rm c}2}}
\def\Mv{m_{{\rm v}}}
\def\Nv{N_{{\rm v}}}
\def\Mci{m_{{\rm c}i}}
\def\Mvi{m_{{\rm v}i}}
\def\Mv{m_{{\rm v}}}
\def\P{P}
\def\Q{Q}

\def\vx{{\rm \bf x}}
\def\vr{{\rm \bf r}}

\def\ob{{\rm halo}}
\def\tnu{\tilde{\nu}}
\def\tka{\tilde{\kappa}}
\def\zetax{\zeta}
\def\tzeta{\tilde{\zeta}}
\def\tzetax{\tilde{\zeta}}
\def\bzeta{\overline{\zeta}}
\def\VP{P}

\def\bp{\overline{p}}

\def\xia{\overline{\xi}}

\begin{document}

\thesaurus{2(12.04.1; 12.12.1; 12.04.1; 11.03.1)} 
\title{Halo correlations in nonlinear cosmic density fields}
\author{F. Bernardeau, R. Schaeffer}    
\institute{Service de Physique Th\'eorique,
C.E. de Saclay, F-91191 Gif-sur-Yvette Cedex France}

\offprints{F. Bernardeau}
\maketitle

\markboth{Halo correlations in nonlinear cosmic density
field}{F. Bernardeau \& R. Schaeffer}

\begin{abstract}
The question we address in this paper is the determination of the
correlation properties of the dark matter halos appearing in cosmic
density fields once they underwent a strongly nonlinear evolution
induced by gravitational dynamics. A series of previous works have
given indications of what non-Gaussian features are induced by
nonlinear evolution in term of the high-order correlation functions.
Assuming such patterns for the matter field, that is that the
high-order correlation functions  behave as products of two-body
correlation functions, we derive the correlation properties of the
halos, that are assumed to represent the correlation properties of
galaxies or clusters.

 The hierarchical pattern originally induced by gravity is shown to be
conserved for the halos.  The strength of their correlations at any
order varies, however,  but is found to depend only on their internal
properties, namely on the parameter $x\propto m/r^{3-\gm}$ where $m$
is the mass of the halo, $r$ its size and $\gm$ is the power law index
of the two-body correlation function. This internal parameter is seen
to be close to the depth of the internal potential well of virialized
objects.  We were able to derive the explicit form of the generating
function of the moments of the halo counts probability distribution
function. In particular we show explicitely that, generically,
$S_P(x)\to P^{P-2}$ in the rare halo limit. 

Various illustrations of  our general results are presented. As a
function of the properties of the underlying matter field, we
construct the count probabilities for halos and in particular discuss
the halo  void probability. We evaluate the dependence  of the halo
mass function  on the environment: within clusters, hierarchical
clustering implies the higher masses are favored.  These properties
solely arise from what is a natural bias (\ie, naturally induced by
gravity) between the observed objects and the unseen matter field, and
how it manifests itself depending on which selection effects are
imposed. 


\end{abstract}

\section{Introduction}

One of the major pending cosmological problem for the formation of galaxies and
clusters, their distribution and evolution, is the relation of the latter
with the initial conditions prevailing in the early universe. To this sake,
most of the observational constraints are relative to the luminous part of the
universe. On the other hand, theoretical model of primordial fluctuations
are for the matter distribution. Whether light traces matter, so that the
former can be used to constrain the latter, has been a long debate over
the past decade. It was clear that 
from the constraint on the small-scale peculiar 
velocity dispersion of galaxies (Davis \& Peebles 1983), matter cannot
trace light (Kaiser 1984, Davis et al. 1985, Schaeffer \& Silk 1985)
if the density of the universe is to be equal to the critical density.

That galaxies might be biased with respect to the matter field 
has therefore been quite popular in the past years especially 
in the frame  of the Cold Dark Matter cosmological model. Standard
CDM model provides indeed enough power at small scale for hierarchical
clustering to occur but it does not
not produce enough power at large scales, when correctly
normalized at small scales, to explain the large scale galaxy counts
(Saunders et al. 1991, Davis et al. 1992). This problem has been confirmed
by the measurement of the very large scale cosmological
fluctuations by Smoot et al. (1992). 

It is therefore now customary
 to work within the assumption of a low-density  universe (with a possible
cosmological constant), that preserves
the hierarchical clustering scenario, and with the idea that
light might trace matter. It is then crucial
to understand the possible existence of biases at smaller scales and
their possible variation with scale
to establish reliable constraints on the shape of the power initial
spectrum.

 Bardeen et al. (1986) proposed a 
mechanism for the galaxies to
be more correlated that the matter. It relies of the correlation
properties of the peaks in the initial Gaussian density field. 
This approach was further extended by Mo, Jing \& White (1997).
The idea is
that galaxies form at the location of the density peaks
and thus are biased from the beginning compared to the whole matter field.
However such a description is far from being complete since the subsequent
complex dynamical 
evolution of the density field is ignored. At scales up to
8 $h^{-1}\Mpc$ the density field is completely nonlinear so that the
statistical properties of the peaks appearing in the initial Gaussian
fluctuation may have been completely transformed.
We present below arguments that take into account the nonlinear
evolution of the matter field to show that the halos in such an evolved
density field indeed are distributed differently than the matter. But
{\it definitely not} in the way predicted by gaussian fluctuations.
We present the complete
correlation properties that are to be expected and
the expression of the biases that appear at various levels in  
the nonlinear regime. 

The small-scale power law behavior of the galaxy correlation
function is specific to hierarchical
clustering scenarios and arises from the nonlinear instabilities in an
expanding universe. The value of the small-scale power law index is
likely to come from the shape of the primordial power spectrum
(Davis \& Peebles 1977). Some authors, however, (Saslaw \& Hamilton 1984) 
propose
an explanation based on thermodynamical arguments to explain
the emergence of such a power law behavior. In any case it is
thought to be due of a relaxation process in the nonlinear
regime. Generalizing a relation of  Peebles (1980), Hamilton
et al. (1991) propose (see also Valageas \& Schaeffer 1997
-hereafter VS97-)
a universal transformation
to get the evolved non-linear two-point correlation function
from the initial spectrum, based
on empirical observations in numerical simulations.
The strength of the matter two-body correlation function is obviously
an important information to derive the initial matter power
spectrum, but the differences between linear regime and
nonlinear regime is unlikely to be reducible solely to a transformation
of the fluctuation spectrum.
Indeed the observed three-point correlation function of the galaxies,
for instance,
also takes a specific form, as product of two two-body correlation functions
(Groth \& Peebles 1977)
 and can provide alternative checks
for the scenario of structure formation. These features 
cannot be predicted by analytical calculations using the simple linear 
approximation  with initial Gaussian fluctuations. This failure
is not due to inapproriate initial conditions but to the fact that 
the linear approximation is inadequate. Perturbative 
calculations introducing higher order of the overdensity field
have demonstrated that the gravity can induce a full hierarchy
of correlations starting with Gaussian initial conditions (Fry 1984b, 
Bernardeau 1992, 1994). 

The scaling due to hierarchical clustering
can be expressed through the behavior of the
mean $p$-body connected correlation functions of the matter field
within a volume $v$, $\xib_p(v)$,
as a function of the two-body one (see Balian \& Schaeffer 1989
-hereafter BaS89- for 
a description of this statistical tools). This relation can be written,
\begin{equation}
\xib_p(v)=S_p\left[\xib_2(v)\right]^{p-1},\eqno(1.1)
\end{equation}
where the coefficient $S_p$ are independent of the scale. 
When the fluctuations are small ($\xib_2(v)\to 0$), 
 one can derive the full series of the coefficients
$S_p$.
Unfortunately such results in the quasi-Gaussian regime are irrelevant for
the fully nonlinear regime where numerous shell crossings and
relaxation processes have to be taken into account.
Explanations for the observed
structures  describing the dynamics of pressure-less
particles in 
gravitational interaction that do not assume the existence of coherent
velocity flows are to be invoked. This hierarchy of equations (The Born, Bogolubov,
Green, Kirkwood, Yvon -BBGKY- hierarchy) concerns the $p$-body density
functions (in the full phase space) 
and has been established by Peebles (1980) for matter in an
expanding universe. It cannot be solved in general,
although there exist some interesting attempts
 (Hamilton 1988b, Balian \& Schaeffer 1988, Hamilton et al. 1991) 
but admit the so-called self-similar solutions (Davis \& Peebles 1977).
The latter solutions contain a precious  indication since 
it can be shown that, in the limit
of large fluctuations, a similar relationship as in \eqref(1.1) is expected
to occur
between the mean correlation functions. Models of matter fluctuations
in which such a behavior is seen are called hierarchical models.
That such solutions are relevant for the evolved density field
was also not obvious and had been the subject of many papers.
After the success of early models (Schaeffer 1984,
1985),
 a complete description of the
properties that are to be expected for the counts in cell
was given by BaS89
(see \S 2.2.1 of this paper for a very short review).
These predictions have now been widely checked, for points
 in numerical simulations
(see Valageas, Lacey \& Schaeffer 1999) and in catalogues
(see Benoist et al. 1998).
We take now for granted that the hierarchy \eqref(1.1) holds

For the analysis of galaxy catalogues, however, it is more or 
less assumed that 
the galaxy field traces the underlying density field, or at least that
the properties of the matter should be preserved in the galaxy field.
The question we then have to address is
to confidently relate properties concerning matter 
distributions to observational statistical quantities such as the two
and three-point galaxy or cluster correlation functions, their 
cross-correlations or the galaxy and cluster void probability functions...
In the Gaussian approximation, the knowledge of the $p$-body density
functions reduces to the behavior of the two-body function. The bias,
measuring the ratio of the strength of the two-point correlation function
of the galaxies to the one of the matter is then the only relevant
parameter, as stressed previously,
to quantify the departure between mass fluctuations and would-be
light fluctuations. It appears to be a crucial parameter to constrain
the cosmological models. However, the actual departure between galaxy
distribution and matter distribution cannot be reduced to this sole parameter.
More generally we have to search how the $p$-body correlation functions
are modified for the observed galaxy distribution.

Previous work (Bernardeau \& Schaeffer 1991 -hereafter BeS91-, VS97,
Valageas \& Schaeffer 1998) 
suggests a natural way to identify the 
visible objects among the matter fluctuations. 
It is based on the assumption that 
dissipative processes are epiphenomena that do not modify the distribution
of matter, but just mark, with luminous stars, the dense matter 
concentrations. 
At scales smaller than the correlation length
it appears (BaS89) that most of the volume of the universe is (nearly) empty 
whereas a small fraction of its volume contains almost all the matter.
The latter fraction of the universe is identified with the observed luminous
objects. It happens that there is a unique parameter that governs the mass 
distribution function of the objects,
\begin{equation}
x=\disp{m\over \overline{\rho}\ v\ \xib},
\eqno(1.2)
\end{equation}
where $m$ is the mass of the object, $ \overline{\rho}  $ the mean
mass density of the  
universe, $v$ the volume of the objects and $\xib$ the mean value of the 
matter two-point correlation function within a cell of volume $v$.
As $\xib$ is seen to behave as $r^{-\gm}$ where $r$ is the size of the
object and $\gm$, the power-law index of the two-body correlation function
is close to 1.8, $x$ scales as $m/r^{3-\gm}$ and is then nearly proportional
to the internal velocity dispersion of these objects, (which by definition
are supposed to be dense enough to be virialized).

The main purpose of this paper is then to show how the galaxy and cluster 
correlations are related to the fluctuations of the matter density in the
deeply nonlinear regime. 
	In practice, we divide the universe into cells and 
calculate the correlation function of those cells containing 
a given amount of matter (above a rather small minimum). In 
the following we may refer to these dense cells as ``halos''.
All the calculations have been done
for hierarchical models.
More specifically we introduced the tree-hierarchical models
for which a complete description of the biases
can be derived.
These models form a large part of the hierarchical 
models where the many-body correlation functions behave as a power of scale,
and can be parameterized as products of the two-body correlation functions.
They are described in Sect. \ref{Nonlinbe}. 

A calculation of the bias parameter was performed (Schaeffer 1985, 1987) 
using specific tree-hierarchical based models.
In a previous paper Bernardeau \& Schaeffer (1992, hereafter BeS92) 
performed the calculation of the two- and three-point correlation functions
of the dense spots in the matter field using a much more general form of
tree-hierarchical models. These calculations were recently extended by
Munshi et al. (1998b) up to the 6-point functions exploiting the same
techniques and models. Once again
the scaling parameter $x$ was seen to play a central role. The bias, 
the parameter $Q_3$ measuring the strength of the three-point correlation 
function for halos,
 were both shown to depend only on their internal scaling parameter $x$.
In Sect. \ref{GenFunc} 
of this paper we present the generalization of these results to the
correlations at any order. The complete $p$-body correlations of condensed
objects are thus derived from the matter $p$-body correlation functions.

Sect. \ref{Models} is devoted to the discussion of the various possible models, still within the tree form for the matter correlations, that can
be realistically (or less realistically) used to describe the matter field.
The range of the prediction for the halo correlations depending on which model is adopted is discussed in great detail.
 
In Sect. \ref{ObsCons} we discuss various possible
 applications of these results, that may be checked either in galaxy catalogues or in numerical simulations. 
We consider the first few moments of the counts of objects, 
 the void probability functions. Special care is devoted to the determination 
of the  mass multiplicity
function of halos 
 in over-dense areas (rich clusters). 

In the last section we strike
the balance of what have been obtained and present some possible
developments. The mathematical aspects of the derivations are given in the
appendices.

Preliminary account of this work, that contains all the
major results presented here, has been given by Bernardeau
(PhD Thesis, Paris 1992). 
After near final completion of this paper, we learned about
the overlapping work by Munshi et al. (1998b).

\section{The nonlinear behavior}
\label{Nonlinbe}
\subsection{The shape of the matter correlation functions}

We assume that the correlation functions in the matter field follow
a hierarchical pattern as in Eq. \eqref(1.1). More 
precisely we assume that there are solutions of the equations of motion 
for the dynamical evolution
of the matter distribution, for which all many-body correlation functions
exhibit a scaling law,
\begin{eqnarray}
&&\hmove \Omega=1,\ \ \ \disp{
\xi_p(\lambda \vr_1,\dots,\lambda \vr_p,\mu t)=\left[{\mu^{2/3(3-\gm)}
\over \lambda^{\gm}}\right]^{p-1}}\nonumber\\
&&\hmove \ \ \disp{\times\xi_p(\vr_1,\dots,\vr_p,t)},
\eqno(2.1)
\end{eqnarray}
where $\xi_p$ is the connected $p$-point correlation function as a function
of the $p$ comoving coordinates $\vr_1,\dots,\vr_p$ and time $t$
(Davis \& Peebles 1977). The scaling
laws shown in this equation are valid for an Einstein-de Sitter universe
and power-law initial conditions. For the latter, in
case of $\Omega\ll 1$, an other similar scaling is reached in which
the coefficient $2/3$ appearing in equation \eqref(2.1) has to be changed in 1,
\begin{eqnarray}
&&\hmove \Omega\ll 1,\ \ \ 
\disp{\xi_p(\lambda \vr_1,\dots,\lambda \vr_p,\mu t)=\left[{\mu^{3-\gm}
\over \lambda^{\gm}}\right]^{p-1}}\nonumber\\
&&\hmove \ \ \times \disp{\xi_p(\vr_1,\dots,\vr_p,t)},
\eqno(2.2)
\end{eqnarray}
which only changes the time dependence of these functions
 (Balian \& Schaeffer 1988). The latter
have argued that, provided
 either $\Omega=1$ or $\Omega\ll1$ so that the above solutions exist,
they may be reached for more general initial conditions.
Numerical simulations show
(Davis et al. 1985, Bouchet et al. 1991,
Colombi et al. 1994, 1995, 1996, Munshi et al. 1998a, 
Valageas et al. 1999) that \eqref(2.1) is indeed relevant at late
times, at least for $\Omega=1$ and $p=2,3$. The matter
correlation functions cannot be obtained from observation, but the galaxy
correlations for $p=2,3$ also are seen to obey Eq. (\ref{2.1}) or Eq. 
(\ref{2.2}) (for the scale 
dependence). It is thus fair to assume that either Eq. (\ref{2.1}) or
Eq. (\ref{2.2}) holds also for the higher $p$-body correlation functions, 
and can be used as a model to describe the matter fluctuations in the 
nonlinear regime.

Once the two-point correlation function
is assumed to obey \eqref(2.1) or \eqref(2.2), a further simplification
can be introduced by assuming
 that $\xi_p$ can be written 
as a product of $p-1$ two-body correlation functions,
\begin{eqnarray}
&&\hmove \xi_p(\vr_1,\dots,\vr_p)=\disp{
\sum_{{\rm trees}\ (\alpha)} Q_p^{(\alpha)}(\vr_1,\dots,\vr_p)}\nonumber\\
&&\hmove \times \disp{\sum_{{\rm labels}\ t_{\alpha}}\ 
\prod_{{\rm links}} \xi_2(\vr_i,\vr_j)},
\eqno(2.3)
\end{eqnarray}
where $(\alpha)$ is a particular tree topology (Fig. 1) connecting the $p$
points without making any loop,
$t_{\alpha}$ is a particular labeling of the $p$ points 
by the coordinates $\vr_1, \dots, \vr_p$,
the sum corresponding
to all different possible labelings for
the given topology $(\alpha)$
and the last product is made over the
$p-1$ links between the $p$ points with two-body correlation functions.
 We note for later use that, for each  topology $(\alpha)$,
we can list  the number of vertices having a given number, $q$, of
outgoing lines, $q=1, \dots, \infty$ .
 The weight $Q_p^{(\alpha)}$ is independent of time 
and is a homogeneous
function of the positions, 
\begin{equation}
Q_p^{(\alpha)}(\lambda \vr_1,\dots,\lambda \vr_p)=
Q_p^{(\alpha)}(\vr_1,\dots,\vr_p), \eqno(2.4)
\end{equation}
 associated with the order of the correlation
and the topology involved.

\begin{figure}
\centerline{
\psfig{figure=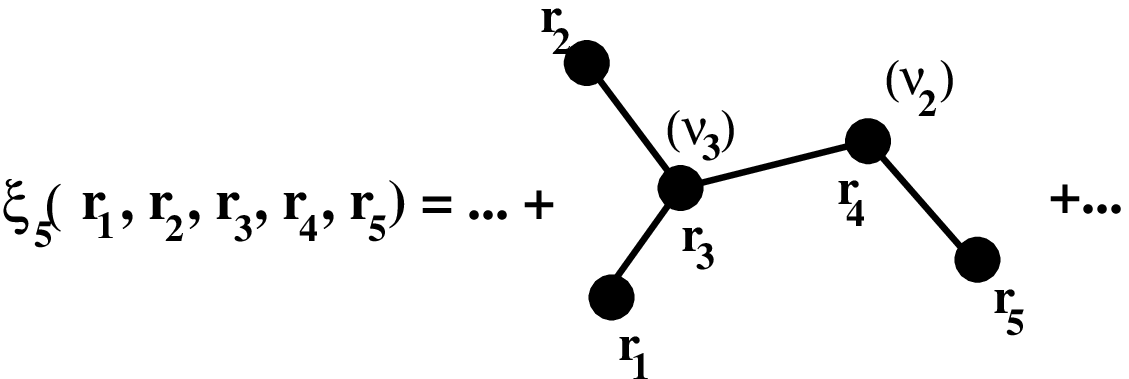,width=7cm}}
\caption{An example of contribution to the 5-point correlation function.
The term represented in the figure reads 
$
\xi(\vr_1,\vr_2)\allowbreak
\nu_3(\vr_1\!-\!\vr_2,\vr_3\!-\!\vr_2,\vr_4\!-\!\vr_2)\allowbreak
\xi(\vr_2,\vr_3)\allowbreak
\xi(\vr_4,\vr_2)\allowbreak
\nu_2(\vr_2\!-\!\vr_4,\vr_5\!-\!\vr_2)\allowbreak
\xi(\vr_2,\vr_5),$
with the convention $\nu_1 = 1$. It has 3 vertices with 1 outgoing line,
each being
weighted by $\nu_1$, 1 vertex with 2 lines weighted by $\nu_2$ and 
1 vertex with 3 lines weighted by $\nu_3$. Each link is weighted 
by a factor $\xi$.  These weights depend on the labels $\vr_1 \dots \vr_5$
attached to the vertices.
In case of the minimal tree-hierarchical model the vertices $\nu_2$, 
$\nu_3,\dots$
are independent of the positions of the points; otherwise they are homogeneous
functions of the lines to which they are connected.
}
\end{figure}

Several attempts have been made using the form \eqref(2.3). The functions
$Q_p^{(\alpha)}$ are generally assumed to be independent of the 
coordinates $\vr_1,\dots,\vr_p$, and sometimes
also, for simplicity, to be
 independent of $\alpha$.
For instance Schaeffer (1984, 1985) uses 
$Q_p=\left(2^{p-1}/ p\right)^{\nu},\,\nu\!=\!1\pm 1$,
constant or moderately increasing with $p$,
justified on observational grounds, so as 
to produce enough power towards the small masses (power-law increase)
and a slow enough
fall-off of the induced multiplicity function, 
which must not be faster than exponential.
This implies a very strong growth, as $p!$, of the $p$-body correlation function amplitude.
 Fry (1984a) proposed $Q_p\!=\!{1/2}\,
\allowbreak\left( {4Q/p}\right)^{p-2} {p/(p-1)}$,
decreasing roughly like
 $1/p!$
as a solution of the BBGKY hierarchy. 
This
implies correlation functions that grow at most as a power of $p$.
The associated  mass function, for counts far from the mean,
then  differs by several orders of magnitude from the previous one. 
This result, however, was obtained by means of approximations
that were later seen (Hamilton 1988b) to be unjustified,
the accumulated error on $Q_p$ being precisely of the order of $p!$.
The latter found an approximate solution where $Q_p^{(\alpha)}$
vanishes for all tree topologies, except for the particular line-topology,
for which $Q_p^{(\alpha)}=Q^{p-2}$, indeed of moderate variation as a 
function of $p$ since $Q  \ge 1/2$, in this case,  is of order unity.

There are, however, strong observational constraints on these coefficients
due to the fact that they determine the shape of the mass function
of objects (Schaeffer 1984, 1985, BaS89, BeS91, VS97). For instance the solution
proposed by Hamilton (1988b) is  seen to produce still  too sharp a cut-off at 
high masses (although exponential, as desired) as well as at low masses,
where  the power-law behavior is lacking
 and is  not realistic enough. 
For this reason, BaS89 made a study valid for  all models that satisfy
a well-established set of known constraints arising either from
mathematical consistency (such as the positivity of probabilities) 
or from the requirement that the mass function should look like the
Schechter (1976) function.

\subsubsection{The minimal tree-hierarchical model}

The model proposed by Schaeffer as well as the form
proposed by Hamilton are both a particular case of a more general form,
\begin{equation}
Q_{p,{\rm minimal}}^{(\alpha)}=\prod_{{\rm vertices\ of\ }(\alpha)}\nu_q,
\eqno(2.5)
\end{equation}
where $\nu_q$ is a constant weight associated to a vertex of the tree topology
with $q$ outgoing lines. The value of
$Q_p^{(\alpha)}$ is then obtained by the product on all the
weights associated with the topology $(\alpha)$. The form \eqref(2.5), assumed
to hold for the matter correlations, has been used (BeS92) to calculate
the two- and three-body correlation functions for galaxies or clusters.
We call it here the {\it minimal tree-hierarchical model}. 
This form is supported by numerical evidences for at least the
bi-spectrum which is found to be very weakly
dependent on the geometry in the nonlinear regime (see Scoccimarro et
al. 1998 and references therein). This is also assumed for the
construction of the so-called Hyperextended Perturbation Theory model
(Scoccimarro \& Frieman 1999).

\subsubsection{The general tree-hierarchical model}

But the aim of this article
is to find the relation between the distribution of the matter and the
luminous (dense) objects, using the most general possible form for the 
many-body correlation functions compatible with \eqref(2.3). The most general 
form that is relevant to the kind of calculations we develop in
this paper is of the form
\begin{equation}
Q_p^{(\alpha)}=\prod_{{\rm vertices}} 
\nu_q(\vr_{s_1}\!-\!\vr_i,\dots,\vr_{s_q}\!-\!\vr_i),
\eqno(2.6)
\end{equation}
the product being over all vertices associated to the tree-topology 
$(\alpha)$; the position of the $i^{th}$ vertex is at $\vr_i$
and $\vr_{s_1},\dots,\vr_{s_q}$ are the coordinates of the end-points of its
$q$ outgoing lines. The function 
$\nu_q(\vr_{s_1}\!-\!\vr_i,\dots,\vr_{s_q}\!-\!\vr_i)$
is supposed to be an homogeneous function of the positions. In general
it is then supposed that the weight associated to the vertices may depend
not only on the number of outgoing lines as in the minimal models, but
also on the geometry of these lines such as their relative angles, or
the ratios of their lengths. This very general form of the correlations
will be referred to as  the {\it tree hierarchical model}.

If there is no definitive proof that the many-body correlation functions
take the form \eqref(2.6) in the strongly nonlinear regime,
this kind of structure is however obtained in the quasi-linear regime
(Fry 1984b, Bernardeau 1992) where the correlation functions follow
a tree like pattern. 
In the following we will mention the properties that are specifically due
to the hypothesis \eqref(2.6) once the form \eqref(2.1) has been assumed.

\subsection{ Matter distribution in cells}

To calculate the number density of the dense and under-dense cells as
a function 
of the mass (\ie, the number of matter particles) it has been shown 
(BaS89) that it is sufficient to know the average of the
many-body correlations within a cell, as is exemplified by \eqref(1.1).
We recall here these findings, in a language that is somewhat
different from the one used in these earlier publications, but
that will be similar to the one relevant to the purpose of this paper.

Let us divide the universe in cells of given scale, or volume $v$. The
counts in cells, $p(m) \d m$, that is the probability of finding a given
amount of matter in a randomly chosen cell, can be obtained (White
1979, BaS89)
from a generating function that depends only on the averages,
\begin{equation}
\xib_p=\int_v{\d^3\vr_1\over v}\dots{\d^3\vr_p\over v}\ 
\xi_p(\vr_1,\dots,\vr_p).
\eqno(2.7)
\end{equation}
This implies that the counts are defined by the knowledge of a set of numbers
$S_p$, $p=3,\dots,\infty$ (to which we add $S_1=S_2=1$ for convenience),
\begin{equation}
S_p={\xib_p\over \xib_2^{p-1}}
\eqno(2.8)
\end{equation}
which are independent of cell size and of time, provided \eqref(2.1) holds. The
counts in cells are then given by (BaS89),
\begin{eqnarray}
&&\hmove p(m)\d m=\disp{ {\d m\over m_{\rm
c}}\int_{-\i \infty}^{+\i \infty} {\d y\over 2\pi\i}\ 
e^{xy-\varphi(y)/\xib}\ },\eqno(2.9)\\
&&\hmove {\rm with}\ \ \  
x={m\over m_{\rm c}},\ \ \ m_{\rm c}=\overline {\rho}\ v\ \xib,\nonumber
\end{eqnarray}
where $\varphi(y)$  reads,
\begin{equation}
\varphi(y)= - \disp{\sum_{p=1}^{\infty}(-1)^p S_p {y^p  \over p!}}.
\eqno(2.10)
\end{equation}
We assume in this paper the coefficients $S_p$ are given.

Within the minimal tree-hierarchical model \eqref(2.5), 
each of the coefficients
$S_p$ is related to the $p-1$ first coefficients $\nu_q$ 
(Appendix \ref{Aappend}). 
As a result $\nu_2$ can then be calculated from $S_3$, ...
$\nu_{p-1}$ from $S_p$. In case of the minimal tree-hierarchical model, 
the matter correlation
functions are thus entirely determined by the values of the coefficient $S_p$,
and thus by the shape of the mass distribution.
For the general tree hierarchical models, however, the count-in-cells are seen 
\eqref(A.19) to depend
on a suitably defined average over the scale-dependent vertices $\nu_q$
\begin{equation}
\nu_q=\disp{\int_v{\d^3\vr\over v}\,\nu_q(\vr-\vr_1,\dots,\vr-\vr_q)
\prod_{i=1}^{q}{\xi(\vr,\vr_i)\over\xib}\,{\d^3\vr_i\over v}}.
\eqno(2.10.1)
\end{equation}
So, in case only the count-in-cells are considered, the tree
hierarchical model and the minimal tree hierarchical model lead to
similar, and in practice,  
indistinguishable results. However, a priori, 
the underlying correlations of the general tree
hierarchical models might be
quite different and more information than the count-in-cells is needed to 
determine all other probabilities.
Nevertheless, for the count-in-cells, both models can be described in
terms of constant coefficients $\nu_q$. 
The relation between $S_p$ and $\nu_q$  can be explicited through the
generating function of the 
vertices $\nu_q$,
\begin{equation}
\zeta(\tau)=\sum_{q=0}^{\infty} (-1)^q \nu_q {\tau^q\over q!},
\eqno(2.11)
\end{equation}
from which $\varphi(y)$ is easily obtained (BeS92) by summing (see \eg,
Jannink \& des Cloiseaux 1987) the so-called ``tree-graph series'':
\begin{equation}
\varphi(y)=y \zeta(\tau)-{1\over2} y\tau{\d\zeta\over\d\tau}(\tau),
\eqno(2.12)
\end{equation}
where $\tau$ is given as a function of $y$ by the equation,
\begin{equation}
\tau=-y\,\zeta'(\tau).
\eqno(2.13)
\end{equation}
The knowledge of the function $\varphi(y)$ or $\zeta(\tau)$ determines
the multiplicity of cells as a function of their matter content. 
A useful relation, that is readily derived from \eqref(2.12) and 
\eqref(2.13), is
\begin{equation}
\disp{{\partial \varphi(y) \over \partial y }=\zeta(\tau)\ .}
\eqno(2.13.1) 
\end{equation}

\subsection{Model-independent properties of the probability distribution}
\label{ModIndProp}

The general
properties of the cell distribution $p(m)$, Gaussian or non-Gaussian regime,
behavior for small or large masses in the nonlinear regime can be 
established, provided $\varphi$ (and hence $\zeta$) satisfies some very general
constraints (BaS89). 

These properties, called ``model independent'', concern
any hierarchical model (and not only the tree models considered here).
 First, $\varphi(y)$ 
should be 
 a power-law at large $y$, or $\tau$ respectively:
\begin{equation}
\varphi(y)\propto a y^{1-\omega},\ \ 0\le\omega\le1,\ \ a\approx 1  \ .
\eqno(2.14)
\end{equation}
This implies for  $\zeta(\tau)$, at large $\tau$ 

\begin{equation}
\zeta(\tau)\propto c\tau^{-\ka},\ \ 
\ka={2\omega\over 1-\omega}\ge 0,\ \ 
c\approx 1. 
\eqno(2.15)
\end{equation}
Secondly, the series \eqref(2.10) must have a very small radius of convergence,
equal to $1/x_*\sim 0.1$. This may be due to a rapid divergence of
$\zeta(\tau)$ but more likely because of a singularity brought by the
implicit equation 
\eqref(2.13)  at $y = y_s=-1/x_*$, which must thus be quite close to the origin. This calls also for a small value of the associated value $\tau_s$.
The latter situation is the one encountered
in the quasi-Gaussian regime (Bernardeau 1992).

Finally, the choice \eqref(2.9) for defining $x$ and $m_{\rm c}$ implies, for
small $y$ and $\tau$, that the following  expansions are valid,
\begin{equation}
\varphi(y) \sim y - y^2/2 \ ,
\eqno(2.15.1) 
\end{equation}
\begin{equation}
\zeta(\tau) \sim 1 - \tau \ .
\eqno(2.15.2)
\end{equation}

For the dense cells, the multiplicity function 
in the non-linear regime, $\xib\gg1$, takes the scaling form (BaS89),
\begin{equation}
p(m)=\disp{-{1\over m_{\rm c} \xib}\int{d y \over 2\pi\i} 
\varphi(y)\,e^{xy}},
\end{equation}
that can be written,
\begin{equation}
p(m)=\disp{{1\over m_{\rm c} \xib}\ h\left(m\over m_{\rm c}\right)},
\eqno(2.16)
\end{equation}
provided 
\begin{equation}
\disp{m \ll m_{\rm v} \ , \
m_{\rm v}=m_{\rm c} \left(a\over\xib\right)^{1/(1-\omega)} \ }.
\eqno(2.17)
\end{equation}
 The latter constraint ensures that 
the matter content of the cells is large enough.
We will refer to these cells as ``halos''.

The function
$h(x)$ shows that the dependences on size, density and mass only 
enter through 
the ratio $x = m/m_{\rm c}$ (see Eqs. [\ref{1.1}] and [\ref{2.9}]). 
It is related to $\varphi(y)$ by 
\begin{equation}
h(x)=\disp{-\int_{-\i \infty}^{+\i\infty}
{\d y \over 2\pi\i} \  e^{xy}\varphi(y) \ }. 
\eqno(2.18)
\end{equation}
It is defined only for $x$ strictly positive and behaves as
\begin{equation}
h(x)\propto a(1-\omega)/\Gamma(\omega) \ x^{\omega-2}\ \ {\rm for}\
\ 0<x_v\ll x\ll1,
\end{equation}
with
\begin{equation}
x_v = {m_{\rm v}\over m_{\rm c}} = 
\left({a\over\xib}\right)^{1/(1-\omega)}
\end{equation}
and
\begin{equation}
h(x)\propto \disp{x^{(\omega_s-1)} e^{-x/x_*}}\ \ {\rm for}\ \ x\gg 1 \ ,
\eqno(2.19)
\end{equation}
with, generically, for tree-hierarchical models $\omega_s=-3/2$ .
The mass multiplicity function thus indeed is quite similar to the usual
Schechter (1976) parameterization (exponential times power-law) of the galaxy
and cluster luminosity functions.

These predictions have been successfully checked against direct 
galaxy counts (Alimi et al. 1990, Maurogordato et al. 1992, Benoist et
al. 1998). Numerical simulations also stongly back this description of the scaling properties induced by hierarchical clustering
  (Bouchet et al. 1991,
Colombi et al. 1994, 1995). Some departures from the scaling law however
were found by 
Colombi et al.  1996, whereas Munshi et al. 1998a find consistency with the numerical data, see discussion of this point by
Valageas et al. 1999.

A quite convenient form of $\zeta(\tau)$, that bears all the desired feature, asymptotic form at small $x$ and exponential with large $x_*$, is (BeS92)
\begin{equation}
\zeta(\tau)=(1+{\tau \over \ka })^{-\ka}\ , 
\eqno(2.20) 
\end{equation}
which leads to 
\begin{eqnarray}
\omega&=&{\kappa\over\kappa+2},\\
x_*&=&{1\over \kappa}{(\kappa+2)^{\kappa+2}\over(\kappa+1)^{\kappa+1}}.
\end{eqnarray}
The choice of $\ka=1.3$ insures a good agreement with the counts
in cell statistics obtained in CDM
numerical simulations (Bouchet et al. 1991b), whereas power-law
initial conditions (Colombi et al. 1996, Munshi et al. 1998b, Valageas
et al. 1999) lead to  
$\ka \sim 1.6, 1.3, 0.9 $, for $\Omega = 1 $ and, respectively, 
 a spectral index $ n = 0, -1, -2$.

These forms are very close to 
the form proposed by Bernardeau (1992) derived from the
equations of the dynamics in the quasi-linear approximation. The latter finds
\eqref(2.20) with  $\ka = 1.5$ is relevant already in  the early
non-linear regime (when smoothing effects are neglected that is for $n=-3$).  

\subsection{Sum rules}
\label{sumrules}

The normalization properties for $h(x)$ are worth being recalled.
Indeed we have,
\begin{equation}
\int_{0}^{\infty} x\,h(x)\d x=1\ ,
\eqno(2.21)     
\end{equation}
reminiscent  of,
\begin{equation}
\int_0^{\infty} m\,p(m)\d m =\rho v\ . 
\eqno(2.22)   
\end{equation}
Both relations are exact sum rules, showing that whenever $p(m)$ in
the previous integral is replaced by its approximate form, \eqref(2.16),
which is valid
for the distribution of dense cells, there is no error in computing the total
mass, which is thus shown to essentially be contained in these denser
cells.
The normalization property,
\begin{equation}
\int_0^{\infty}p(m)\d m=1\ , 
\eqno(2.22.1)   
\end{equation}
which means the cells occupy all the volume on the universe,
however has no simple analogue for $h(x)$.
Indeed we have,
\begin{eqnarray} 
&&\hmove \disp{\int_{x_{min} \gg x_v}^{\infty}h(x)\d x}\nonumber\\
&&\hmove \ \ \ \ =\disp{{a\over \Gamma(\omega)}\ \ x_{min}^{\omega - 1}}\disp{
\ll {a\over \Gamma(\omega)}\ x_v^{\omega - 1} 
= {\xib\over\Gamma(\omega)}}\ ,
\eqno(2.22.2)
\end{eqnarray}
that implies that the volume fraction occupied by the dense cells
described by $h(x)$ is, 
\begin{equation}
\int_{m_{min} \gg m_{\rm v}}^{\infty} p(m)\d m \ll {1\over \Gamma(\omega)}\ . 
\eqno(2.22.3)   
\end{equation}
which is small as compared to unity. This shows, indeed, that $h(x)$
describes the denser cells, that contain all the matter but,
occupying a negligible fraction of the volume, may be expected to be
surrounded by regions with a negligible fraction of the matter. 
 This justifies the referrence to these cells as
``halos''.
It
will be the correlations of these dense cells
that we aim to
calculate in the following. 
Clearly, would we use the full
expression \eqref(2.9) for the probability, which holds in all cases, even
for the nearly empty regions with $x \sim x_v$, and include the latter
in the sum, the 
normalization would of course be unity, Eq. \eqref(2.22.1).  

For completeness, we recall also that the normalization of $m_{\rm c}$ 
(Eq. \ref{2.9}) that is used to define $x$ has been chosen so as to yield
\begin{equation}
\int_{0}^{\infty} x^2\ h(x)\d x  =1   \ .
\eqno(2.22.4)     
\end{equation}
And more generally we have,
\begin{equation}
\int_{0}^{\infty} x^p\ h(x)\d x  =S_p   \ .
\eqno(2.22.4.1)
\end{equation}

Important constraints can be obtained from the sum rule
\eqref(2.22.4.1). Indeed,  
writing that,
\begin{equation}
\int_{0}^{\infty} (x-a)^2x^p\ h(x)\d x  \ge 0   \ ,
\eqno(2.22.4.2)
\end{equation}
that is 
\begin{equation}
S_{p+2} - 2 a S_{p+1}+ a^2 S_p \ge 0
\eqno(2.22.4.3)
\end{equation}
whatever $a$. For the value $ a = S_{p+1}/S_p$
that minimises the above expression, we get the constraint,
\begin{equation}
S_{p+2}S_p - S_{p+1}^2 \ge 0 .
\eqno(2.22.4.4)
\end{equation}
The same constraints hold if the counts are discrete, 
still in the limit where $\xib$ is large. One simply has to use the
exact  expression \eqref(A.2) of $p(N)$ for the probability  
and the weight $N(N-1) ... (N-p) p(N)$ instead of $x^ph(x)$ so as to
take advantage of the properties (BaS89) of the factorial moments. 

Such constraints were first derived in a slightly different form
by Peebles (1980) for $S_3$ and  Fry (1986) for all the terms. They imply 
for instance,        
\begin{equation}
S_3 \ge 1,
\eqno(2.22.4.5)
\end{equation}
and also 
that the many-body coefficients $S_p$ can {\it never vanish}, as is
directly obvious from \eqref(2.22.4.1) since $h(x)$ is  positive.  

There are also several mathematical forms for $h(x)$ that will be useful.
Integrating \eqref(2.18) by parts, we get,
\begin{eqnarray} x\,h(x) 
&=&  \disp{\int_{-\i \infty}^{+\i\infty}
{\d y \over 2\pi\i} \  e^{xy}{\partial   \varphi(y) \over \partial y}}\nonumber\\
&=&  \disp{\int_{-\i \infty}^{+\i\infty}
{\d y \over 2\pi\i} \  e^{xy} \zeta(\tau)}\ .
\eqno(2.22.5)
\end{eqnarray}
Using $\tau$ as a variable, a second integration by parts gives,
\begin{equation}
x^2h(x)= - \int
{\d \tau \over 2\pi\i} \  e^{xy} {\partial \zeta(\tau) \over \partial \tau } \ ,
\eqno(2.22.6)
\end{equation}
where however the integration contour of $\tau$ has to be suitably defined
according to the function $\tau(y)$ defined by \eqref(2.13).

\section{Generating functions for conditional cell counts}
\label{GenFunc}

We still consider the Universe divided in small cells. The central
problem we solve in this paper is the calculation of the correlations, at
any order, among cells that contain a specific amount of matter.
 
We consider a large volume $V$ that contains $\P$ cells, 
labeled by $i$ = $1 \dots \P$,
of volume $v_i$, respectively.
Our aim is to derive the joint mass distribution functions
$p(m_1,\dots,m_{\P})\d m_1\dots\d m_{\P}$, probability that the $\P$
cells contain 
respectively the masses $m_1\dots m_{\P}$. We will then assume that
such a joint 
distribution function is the joint density distribution functions of the
halos of the density field.

\subsection{Biased two-body correlation function}

We assign a position $\vr_i$ to each of the cells, to which we associate
respectively the typical masses   
$\Mci$ and $\Mvi$ defined as previously. We then restrict our calculation
to the case $\Mvi\ll m_i$ which corresponds to the condensed objects.
The case of the joint mass distribution function for two cells has already
been addressed in a previous paper (BeS92). We have
\begin{eqnarray}
&&\hmove \disp{p(m_1,\vr_1;m_2,\vr_2)\d m_1\d m_2=p(m_1)p(m_2)\d m_1\d m_2}
\nonumber\\
&&\hmove \ \ \disp{+p^{(1)}(m_1)\xi_2(\vr_1,\vr_2)p^{(1)}(m_2)\d m_1 \d m_2}.
\eqno(2.23)
\end{eqnarray}
When $m \gg \Mv$, so that the scaling \eqref(2.16) applies,
the function $p^{(1)}(m)$ is related to a new series 
of parameter $S_p^{(1)}$, that instead of being averages over a unique
cell \eqref(2.8) are now defined as mixed averages within two (small) 
cells of volume $v_1$ and $v_2$, respectively, under the constraint that
the center $ \vr_1$ and  $\vr_2 $ of these two (distant)  cells remain
fixed. In general, for tree models, the $S_p^{(1)}$ parameters
depend only on the average of the $p+1$-point correlation
function when $p$ points are in one cell and the other in the second.
A similar situation is encountered in the quasi-linear regime
(Bernardeau 1996).
We thus have generically, 
\begin{eqnarray}
&&\hmove S_p^{(1)}=\disp{{1\over \xib^{p-1}}{1\over
\xi_2(\vr_1,\vr_2)}}\nonumber\\ 
&&\hmove \ \ \times\ \disp{\int_{v_1} {\d^3\vr'_1\over
v_1}\dots{\d^3\vr'_p\over v_1} 
\xi_{p+1}(\vr'_1,\dots,\vr'_p,\vr_2)}.
\eqno(2.24)
\end{eqnarray}
In the case of the hierarchical models, the $S_p^{(1)}$
 parameters are independent
of both the cell volume and $\vert \vr_1-\vr_2\vert$ (as soon as the
finite cell size effect are neglected, \ie, $v_1^{1/3}$ (resp. $v_2^{1/3}$)   
$\ll\vert \vr_1-\vr_2\vert$) and are specific
numbers related to the matter correlation functions. 
The function $p^{(1)}(m)$ is then related to 
$S_p^{(1)}$ exactly by the same way $p(m)$ is related to $S_p$.
We can as well defined the generating function of this new series
\begin{equation}
\varphi^{(1)}(y)=-\sum_{p=1}^{\infty} (-1)^p S_p^{(1)} {y^p\over p!},
\eqno(2.25)
\end{equation}
which permits the calculation of $p^{(1)}(m)$,
\begin{equation}
p^{(1)}(m)=\disp{-{1\over m_{\rm c} \xib}\int
{\d y \over 2\pi\i} \varphi^{(1)}(y)\,e^{xy}}.
\end{equation}
In case of the minimal tree-hierarchical model we have (BeS92 and Appendix 
\ref{Bappend}),
\begin{equation}
\varphi^{(1)}(y)=-y{\d \zeta\over \d \tau}(\tau),
\eqno(2.26)
\end{equation}
where $\zeta$ is given by \eqref(2.11) and
$\tau$ is solution of the equation \eqref(2.13), so that we have the simple
relation $\varphi^{(1)}(y)=\tau(y)$. 

 For the general tree-hierarchical models the situation is slightly
more complicated and the $S_p^{(1)}$ coefficients  depend on 
vertices $\nu_q$. With outgoing lines starting at $ \vr'$,
one ending at $\vr_2$ (far away) outside, whereas
$q-1$ ones end within the cell centered at $\vr_1$,
at $\vr'_1 \dots \vr'_{q-1}$ that
are averaged over the cell volume, in a procedure similar to
\eqref(2.24), we get,
\begin{eqnarray}
&&\hmove \disp{\nu_q^{(1)}=\int_{v_1}  
\nu_q(\vr'_1-\vr',\dots,\vr'_{q-1}-\vr',\vr_2-\vr')}
\nonumber\\
&&\hmove \ \ \disp{{\xi_2(\vr'_1-\vr')\over\xib}\dots
{\xi_2(\vr'_{q-1}-\vr')\over\xib}{\d^3\vr'_1\over v_1}
\dots{\d^3\vr'_{q-1}\over v_1}{\d^3\vr'\over v_1}}.
\eqno(2.27)
\end{eqnarray}   
We can  define the generating function for these vertex weights
($\nu_1^{(1)}=1$),
\begin{equation}
\zeta^{(1)}(\tau)=\sum_{q=1}^{\infty}(-1)^q  \nu_q^{(1)}{\tau^{q-1}\over (q-1)!}.
\eqno(2.28)
\end{equation}
The function $\varphi^{(1)}(y)$ is then related to the function 
$\zeta^{(1)}(\tau)$ by the relationship,
\begin{equation}
\varphi^{(1)}(y)=-y{\zeta^{(1)}}(\tau).
\eqno(2.29)
\end{equation} 
This equation defines $\varphi^{(1)}(y)$ in terms of
the function $\tau(y)$  given by \eqref(2.13) which depends only on the
statistics within one cell and is thus 
considered here
to be known. The problem is then reduced to the construction of
$\zeta^{(1)}(\tau)$, that is to calculate the averages \eqref(2.27). 
For the minimal tree models where $\nu_q$ is constant, this is straightforward
and the quantities $\nu^{(1)}_q$ just identify to the mere vertices
 $\nu_q$, so we have $\zeta^{(1)}(\tau) = \d\zeta(\tau)/\d\tau$.

The relation \eqref(2.23) together with the function $\varphi^{(1)}(y)$
gives the correlation function between two cells,
\begin{equation}
\xi_2(m_1,\vr_1;m_2,\vr_2)={p^{(1)}(m_1)\over p(m_1)}\xi_2(\vr_1,\vr_2)
{p^{(1)}(m_2)\over p(m_2)}
\eqno(2.30)
\end{equation}
which has the same spatial dependence than the matter with a bias factor
$b(m)$  given by,
\begin{equation}
b(m)={p^{(1)}(m)\over p(m)}.
\eqno(2.31)
\end{equation}
This function describes the departure between the matter correlation
and the halo correlation for the two-point function. It turns
out to be a function of $x=m/m_{\rm c} \propto m/r^{3-\gm}$ only,
\begin{equation}
b(x)=\int_{-\i\infty}^{+\i\infty}
\d y\ \varphi^{(1)}(y)\ e^{xy}\big/\int_{-\i\infty}^{+\i\infty}
\d y\ \varphi(y)\ e^{xy},
\eqno(2.32)  
\end{equation}
so that \eqref(2.23) reads,
\begin{eqnarray}
&&\hmove p(m_1,\vr_1;m_2,\vr_2)\d m_1\d m_2=p(m_1)p(m_2)\d m_1\d m_2\nonumber\\
&&\hmove \ \ \ \left[1 + b(x_1) b(x_2)\xi_2(\vr_1,\vr_2)\right].
\eqno(2.33)
\end{eqnarray}
where $x_1$ (resp. $x_2$ ) is the internal $x$ parameter of cell 1 (resp. 2).

Two {\it model independent} properties (in the sense they do not
depend on the  function $\zeta^{(1)}(\tau)$ to be constructed in a
specific model) result thus from hierarchical clustering in the
non-linear regime: a {\it factorization property}, the total bias
being a product of two factors that refer to the internal properties
of each cell, and a {\it scaling property}, these factors depending
only on the internal scaling parameter $x$ of each cell.

The observational consequences of these results have been widely discussed
in BeS92. These predictions have been checked against observations.
 Benoist et al. (1996)  show
that the amplitude of the galaxy
correlation function is definitely dependent on their 
luminosity (the existence of a luminosity dependence 
of  the galaxy correlations showing their distribution is biased
with respect to the matter  
was by no means obvious: 
see Hamilton (1988a) and Valls-Gabaud et al. (1989) for an early discussion on 
this point).
With the uncertainty related to the 
transformation of our $x$ parameter into luminosity (see BeS91, VS98), the
observed bias follows very closely the predicted trend. 
Numerical simulations also (Munshi et al. 1999) exhibit the predicted
scaling of $b(x)$ with $x$ as well as the factorization property
\eqref(2.33). 

\subsection{The bias for many-body correlations}

The above considerations can be generalized to calculate the
correlations of an arbitrary number of cells, again under the
condition that 
their distance is much larger than their size.
In general, for a given number of cells the connected part of the joint
mass distribution function follows the formation rules of the tree formalism
(the demonstrations are given in detail in the appendices). Indeed the
result is,
\begin{eqnarray}
&&\hmove \xi_{\P}(m_1,\vr_1;\dots;m_{\P},\vr_{\P})=\nonumber\\
&&\hmove \disp{\sum_{{\rm trees} (\alpha)}
Q_{\P}^{(\alpha)}(m_1,\vr_1;\dots;m_{\P},\vr_{\P})
\prod_{{\rm links}}\xi_2(\vr_i,\vr_j),}
\eqno(2.33.1)
\end{eqnarray}
with the same structure as the one encountered in  equation \eqref(2.3).
The normalization parameters $Q^{(\alpha)}$ are homogeneous
functions of the geometry
of the positions of the cells $\vr_1,\dots,\vr_{\P}$, and functions of
the masses $m_1,\dots,m_{\P}$ attributed to each of the cells. For
each particular tree connecting the $\P$ cells, $Q_{\P}^{(\alpha)}$ is
obtained by a product over the set of vertices (whose order we label
by $\Q$) associated to the tree $(\alpha)$,
\begin{eqnarray}
&&\hmove 
Q^{(\alpha)}(m_1,\vr_1;\dots;m_{\P},\vr_{\P})=\nonumber\\
&&\hmove \ \ \ \prod
\disp{p^{(\Q)}(m_i,\vr_{s_1}\!-\!\vr_i,\dots,\vr_{s_{\Q}}\!-\!\vr_i)\over 
p(m_i)}.
\eqno(2.34a)
\end{eqnarray}
This form of $Q$ defines ``effective'' vertices $\nu$,
\begin{equation}
\nu_{\Q}(\vr_1\!-\!\vr,\dots,\vr_{\Q}\!-\!\vr) =
\disp{
p^{(\Q)}(m,\vr_1\!-\!\vr,\dots,\vr_{\Q}\!-\!\vr)\over p(m)}.
\eqno(2.34b)
\end{equation}

The vertex associated with the cell labeled $(i)$ depends on the mass $m_i$
and possibly on the geometry of the $\Q$ outgoing lines that appear
in the tree representation of each particular term. 

The functions $p^{(\Q)}(m)$ are related to the matter correlation function
properties. More specifically these functions are generated by an $S_p^{(\Q)}$
series, defined by,
\begin{eqnarray}
&&\hmove S_p^{(\Q)}(\vr_{1}-\vr,\dots,\vr_{\Q}-\vr)=
{1\over \xib^{p-1}}
{1\over \xi_2(\vr_{1},\vr)\dots\xi_2(\vr_{\Q},\vr)}\nonumber\\
&&\hmove \ \ \
\times\int_v\xi_{p+\Q}(\vr'_1,\dots,\vr'_p,\vr_{1},\dots,\vr_{\Q})  
{\d^3 \vr'_1\over v}\dots{\d^3 \vr'_p\over v}.
\eqno(2.35)
\end{eqnarray}
In case of the minimal tree model, the parameters $S_p^{(\Q)}$ are pure
numbers that do not depend on the size of the cells ($v_i$), neither on the
geometries of the $\Q$ outgoing lines. For the general
hierarchical models a dependence with the geometry of these lines 
 is expected since the matter correlation function are
supposed to present such a dependence. The derivation of the value
associated to the dressed vertices $p^{(\Q)}(m)$ is based on the same
principle as for $p(m)$ and $p^{(1)}(m)$,
\begin{equation}
p^{(\Q)}(m)\d m=
-\ {\d m\over \xib\Mc}\int_{-\i \infty}^{+\i \infty}
{\d y \over 2\pi\i} \varphi^{(\Q)}(y)\ e^{xy},
\eqno(2.36)
\end{equation}
with
\begin{equation}
\varphi^{(\Q)}(y)=-\sum_{p=1}^{\infty} \ (-1)^p S^{(\Q)}_p\ {y^p\over p!}.
\eqno(2.37)
\end{equation}

As can be seen by comparing \eqref(2.6) and \eqref(2.34a), 
the dressed $Q^{(\alpha)}$  can then be written as
 a product of vertices given by
\begin{equation}\nu_{\Q}(x)=
\int_{-\i\infty}^{+\i\infty}
\d y\ \varphi^{(\Q)}(y) e^{xy}\big/\int_{-\i\infty}^{+\i\infty}
\d y\ \varphi(y) e^{xy}.
\eqno(2.38)
\end{equation}
 We have obviously $\nu_0(x)=1$ and $\nu_1(x)=b(x)$.The correlation
functions of the condensed objects are thus based on a tree 
structure similar to the one of the matter.
As the distribution of the matter field is given by the
``microscopic'' vertices $\nu_q$, that depend on position unless we
use the minimal tree hierarchical model, the distribution of the cells
is given by these new ``dressed'' vertices $\nu_{\Q}$. These vertices
depend on the internal cell properties through the scaling parameter
$x$ of the cell. 
They do not depend on position because all the calculations are done
in the limit where the cell size is much smaller than the distance
between cells.

In the following the relationship between $\nu_{\Q}(x)$ and the matter
correlation functions will be explicited. 
We naturally expect that these results obtained for cells randomly 
distributed in the Universe are also valid for the counts of
astrophysical objects. A precise calculation of the halo correlations 
would require a more complicated calculation, but the general properties
we find here are expected to be valid. Some 
departures cannot naturally be excluded although they are expected to be small
since the densest spots of the present universe contrast extremely well
with the matter field. 

\subsection{General properties of the galaxy and cluster
correlation functions}

Before calculating the vertices $\nu_{\Q}(x)$, we can already derive
a  set of general properties of the dressed correlation functions,
that are independent of the specific model used for the matter
correlation functions. They depend only on the assumption of an
underlying tree 
structure of the latter.
These results
are the core of what has to be retained from these calculations. 

First of all in the strongly nonlinear regime the
correlation functions of the halos have the same
scale, and time, dependence as the matter:
\begin{eqnarray}
&&\hmove {\rm for\ }\Omega=1,\ \ 
\xi_P^{\ob}(\lambda \vr_1,\dots,\lambda \vr_P,\mu t)=\nonumber\\
&&\hmove \ \ \ \ \disp{\left[{\mu^{2/3(3-\gm)}
\over \lambda^{\gm}}\right]^{P-1} \xi_P^{\ob}(\vr_1,\dots,\vr_P,t)},
\eqno(2.39a)
\end{eqnarray}
and so that the hierarchical properties assumed to the matter field
should also be present in the galaxy field, or in the cluster field.
As a result the ratios,
\begin{equation}
S_P^{\ob}={\xib_P^{\ob}\over \left(\xib_2^{\ob}\right)^{P-1}},
\eqno(2.39b)
\end{equation}
are also scale independent.
The fact that indeed the observations confirm the existence of such property is
an indication of the validity of the hierarchical hypothesis at the
matter distribution level. It is thus a confirmation of the validity of the
hypothesis \eqref(2.6). 

The index $\gm$ measuring the slope of the two-point galaxy correlation 
function in the nonlinear regime has also to remain identical to the one
of the matter field, although the normalizations are not the same
(see BeS92 for a complete discussion). As a result the pair velocity 
correlation between galaxies is expected to have the distance dependence 
expected in the hypothesis of no-bias. This is what is indeed observed
as stressed by Peebles (1987), 
and is by no means in contradiction the existence
of a non-trivial natural bias.
However it does not mean that the large scale (quasilinear, see
Bernardeau 1996) 
and small scale (nonlinear, considered here) biases should
be the same. A short examination of the equation \eqref(2.27) proves that
the parameters involved for the values of $b(x)$ are completely determined
by the nonlinear regime in case of the small-scale bias, whereas they
are given by a mixture of the small scale and large scale correlation
function behaviors for the bias at large scale. A quantitative, if not
qualitative, change is expected to occur when one goes from one regime
to the other, which may lead to a variation of $b$. 
This is not in contradiction with the previous remark on the index
$\gm$ 
since in the former case, we suppose to be only in the strongly nonlinear 
regime.

The normalization parameters (the $S_{\P}$ in the equation [\ref{2.39b}])
that are measurable for the galaxy or cluster fields
have no reason to be identical to the ones of the matter field.
As a result it would be
a complete nonsense to try to discriminate
theories with the use, for instance, of the skewness of the matter field
obtained in numerical simulation with
the one of the galaxy field. 

The strength of the correlation functions, so the biases at any level,
depend only on the internal parameter, $x$, of the objects. This is
a property of great interest for observational checks. Indeed objects of
different natures can have the same $x$ parameter. For instance
the brightest galaxies and the rich clusters should have the same
biases, the ``common'' galaxies should have biases comparable to ones
of the groups (BeS92). Such features can easily be checked in the present
or coming catalogues.

The last property concerns the $\nu_{\Q}(x)$ parameters that govern the
correlation strength. It has just been stressed that they are a function
of $x$ only. We also point out that they describe as well
the auto-correlation functions as the cross-correlation functions between
objects of different kind or between objects and the matter field.
What does it mean in practice? Consider for instance three fields of any
kind, galaxy, cluster or matter. Then one may want to know what is the
three-point density function of these fields, \ie, what is the joint expected
densities of these three fields.
The three-point density function 
$n^{(3)}(\vr_1,\vr_2,\vr_3)$ can be written 
\begin{eqnarray}
&&\hmove n^{(3)}(\vr_1,\vr_2,\vr_3)=n_1 n_2 n_3
+n_1 b_1 \xi_2(\vr_1,\vr_2) b_2 n_2 n_3+\sym.\nonumber\\
&&\hmove \ \ \ 
+n_1 b_1 \xi_2(\vr_1,\vr_2) \nu_{2,2} n_2 \xi_2(\vr_2,\vr_3) b_3 n_3+\sym.
\eqno(2.40)
\end{eqnarray}
where $n_1$ is the density of the objects of kind 1, $n_2$ of kind 2 and
$n_3$ of kind 3, $b_1$, $b_2$ and $b_3$ are their respective bias parameters.
The coefficients $\nu_{2,i}$ are the parameters describing the three-point 
correlation function of the objects of kind $i$. The possible geometric
dependence 
In Fig. (2) we give
a diagrammatic representation of this relation.

\begin{figure}
\centerline{
\psfig{figure=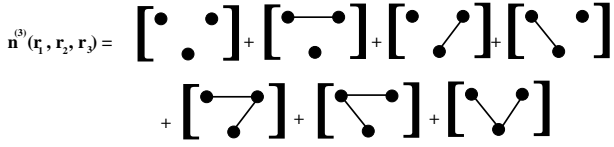,width=8cm}}
\caption{Diagrammatic representation of the three-point halo density function.
Each filled point represents a halo at the position $\vr_1$, $\vr_2$ or
$\vr_3$. Each line equals  the matter two-point correlation function, the
points with one line have a factor $n_i\,b_i$ depending to which point $i$ it
is attached, and the points with two lines have the factor $n_i\,\nu_{2,i}$.
The latter is independent of the positions of the points in case of the
minimal tree-hierarchical model, it is an homogeneous function of the positions
in the general case.}
\end{figure}

These kinds of relations based on tree constructions can be done
at any order where the correlation properties of any kind of object is 
fully determined by the series of the vertices $\nu_{\Q}$. The knowledge
of the vertices $\nu_{\Q}$ then give very diverse results concerning
the auto-correlation properties as well as the cross-correlation properties
of any kind. This may be of great interest since one can imagine observational
tests to check such a property. The simplest is to check that the
auto-correlation function of galaxies, $\xi_{gg}$, the
cross-correlation between galaxies and clusters, $\xi_{gc}$, and the
auto-correlation function of clusters, $\xi_{cc}$, verify the relation,
\begin{equation}
\xi_{cc}\ \xi_{gg}=\xi_{gc}^2.
\eqno(2.41a)
\end{equation}
One can also construct some relations involving the three-point correlation
functions between galaxies and clusters that should be verified in 
catalogues,
\begin{equation}
\xi_{ccg}={2 \over 3}
{\xi_{gg}\over \xi_{gc}}\xi_{ccc}+{1\over3}{\xi_{cc}\over \xi_{gg}}
\xi_{ggg}.
\eqno(2.41b)
\end{equation}
The verification of these relations is a check of the hypothesis \eqref(2.1)
as well as of the hypothesis \eqref(2.6). It would also be the indication that
the clustering properties of the astrophysical objects have 
a pure gravitational origin.

The relation \eqref(2.40), in case the objects are of the same kind, 
reduces to,
\begin{eqnarray}
&&\hmove n^{(3)}(\vr_1,\vr_2,\vr_3)=n^3\left[
1+\xi_2^{\ob}(\vr_1,\vr_2)+\sym.\right.\nonumber\\
&&\hmove \ \ \ \left.+Q_3\ \xi_2^{\ob}(\vr_1,\vr_2)\ \xi_2^{\ob}(\vr_2,\vr_3)
+\sym.\right],
\eqno(2.42)
\end{eqnarray}
where $\xi_2^{\ob}=b^2\xi_2$ is the measurable two-point correlation 
function and $Q_3\equiv \nu_2/b^2$. The latter is the parameter describing
the three-point correlation function and can be measured directly. Such
a remark can be made at any order so that we are led to define
effective vertex weights,
\begin{equation}
\tnu_{\Q}={\nu_{\Q}\over b^{\Q}},
\eqno(2.43)
\end{equation}
(note that $Q_3=\tnu_2$). These parameters are the only ones that can be
measured directly in catalogues and the only quantity that cannot be measured
directly is $b$.

\subsection{The halo correlation functions}
\subsubsection{The general expression of the vertices}

As it has been discussed previously the correlation functions of the
astrophysical objects are completely determined by the parameters
$\nu_{\Q}(x)$ (Eq. [\ref{2.38}]). The latter are determined by the parameters
$S_p^{(\Q)}$ (Eq. [\ref{2.35}]) that depend on the vertices of the
matter correlation 
functions. They depend more precisely on averages over the elementary vertices
 $\nu_q$, located at $\vr$ in $V$,
with $q$ outgoing lines, $\Q$ of which end up in $V$ far from $\vr$ and $q-\Q$
of which end within a volume $v$ centered at $\vr$  
\begin{eqnarray}
&&\hmove \nu_{q,\Q}=
\int_V{\d^3\vr\over V}{\d^3\vr_1\over V}\dots{\d^3\vr_{\Q}\over V}
\int_v{\d^3\vr_{\Q+1}\over v}\dots{\d^3\vr_q\over v}\nonumber\\
&&\hmove \ \ \ \times
\nu_q(\vr_1-\vr,\dots,\vr_{\Q}\!-\!\vr,\vr_{\Q+1}\!-\!\vr,\dots,\vr_q\!-\!\vr)
\nonumber\\
&&\hmove \ \ \ \times{\xi(\vr_1,\vr)\over\xib(V)} \dots
{\xi(\vr_{\Q},\vr)\over\xib(V)}{\xi(\vr_{\Q+1},\vr)\over\xib(v)}\dots
{\xi(\vr_{q},\vr)\over\xib(v)},
\eqno(2.44)
\end{eqnarray}
where $q$ represents the total number of lines connected to the vertex
and $\Q$ is the number of long lines (corresponding to the volume $V$).
In the tree models $\nu_{q,\Q}$ is a parameter independent of the size
of the volumes $v$ and $V$. In case of the minimal tree model, one obtains 
$\nu_{q,\Q}=\nu_{q}$ but we prefer to work without such a restrictive
hypothesis. These parameters define a function of two variables,
\begin{equation}
\zeta(\tau,\theta)=\sum_{q,\Q}\ \nu_{q,\Q}
{(-1)^{q-\Q}\tau^{q-\Q}\over (q-\Q)!}{(-1)^{\Q}\theta^{\Q}\over \Q!}
\eqno(2.45)
\end{equation}
In case of the minimal tree model the function $\zeta(\tau,\theta)$ is simply
a function of $\tau+\theta$ and can be directly deduced from the
knowledge of the count-in-cells statistics, that is the knowledge of
the function $\zeta$,
\begin{equation}
\zeta^{\rm minimal}(\tau,\theta)=\zeta(\tau+\theta) \ . 
\eqno(2.45.1)
\end{equation}
In general, all the informations that are required for the
calculations of the many-body correlation function bias
we want to do are contained in the function 
$\zeta(\tau,\theta)$.

On the other hand the halo correlations are entirely driven by the parameters
$\nu_{\Q}(x)$ that, as for the matter field, define a generating function
\begin{equation}
\zetax(x,\theta)=\sum_{\Q=0}^{\infty}\ (-1)^{\Q}\nu_{\Q}(x)
{\theta^{\Q}\over\Q!}.  
\eqno(2.46)
\end{equation}
As a result, the halos of the density field of scaling parameter $x$
have correlations given by the function $\zetax(x,\theta)$
with the meaning of the equations (\ref{2.3}-\ref{2.6}). 
In the appendices, we derive
the expression of 
$\zetax(x,\theta)$ from the expression of $\zeta(\tau,\theta)$.
The result can be expressed in terms of an auxiliary
function, $h(x,\theta)$, through,
\begin{equation}
\zetax(x,\theta)= {h(x,\theta) \over h(x)}\ .
\eqno(2.47a)
\end{equation}
The function $h(x,\theta)$ generalizes the function $h(x)$ 
with an extra dependence with the parameter $\theta$ 
that describes the environment. We have,
\begin{eqnarray} 
h(x,\theta) &=& -  \int_{-\i\infty}^{+\i\infty}
{\d y \over 2\pi\i} \varphi(y,\theta)\ e^{xy} \ ,\eqno(2.47b)\\
\varphi(y,\theta)&=&y\zeta(\tau,\theta)-{1\over 2}y\tau{\partial \zeta\over 
\partial\tau}(\tau,\theta)\ ,\eqno(2.47c)\\
\tau&=&-y{\partial \zeta\over \partial\tau}(\tau,\theta) \ .
\eqno(2.47d)
\end{eqnarray}
We note here that, since $\zeta(\tau,\theta=0) = \zeta(\tau)$, we have
$\varphi(y,\theta=0) = \varphi(y)$ and
$h(x,\theta=0)=h(x)$ (this justifies the notation we
adopted).

This is the central result of this paper. It is established in detail
in the appendix \ref{Bappend}. 
It is valid  only provided (\ref{2.17}) is satisfied so as the scaling
in $x$ holds, as we generically assume here for the objects we call halos.

The 
expression for the vertex generating function in the general case,
when the lower masses are included, and even  in case 
the density field is represented by discrete points, 
is given in App. \ref{Bappend}. It amounts in general 
to use $\xib \exp(-{\varphi(y,\theta)/\xib})$  instead of
$-\varphi(y,\theta)$ in Eq.  (\ref{2.47b}), the definition of $\zeta$, 
Eq.  (\ref{2.47a}), being changed accordingly. In this case there is
no scaling in $x$ for the underdense cells, and the $x$-dependence is
to be replaced by the dependence on mass $m$. 

The equations (\ref{2.47a}-\ref{2.47d}) explicit the relationship,
\begin{equation}
\varphi(y,\theta)=\sum_{\Q}(-1)^{\Q}\varphi^{(\Q)}(y){\theta^{\Q}\over \Q!},
\eqno(2.47.4)
\end{equation}
between $\varphi^{(\Q)}(y)$  and the vertices $\nu_{q,\Q}$ that 
are directly derived from the matter correlations
as a relation between  $\varphi^{(\Q)}(y)$ and $\zeta(\tau,\theta)$.
From $\varphi(y,\theta)$, it is then possible to deduce 
$\zetax(x,\theta)$. The latter may be directly related to $\zeta(\tau, \theta)$
by means of Eqs. \eqref(2.47.2) or \eqref(2.47.3).
The vertex generating function $\zetax(x,\theta)$
plays the same role for describing the statistics of the cells labeled by 
their parameter $x$ as the generating function 
$\zeta(\tau)$ which gives the statistics of the matter field (\ref{2.9}, 
\ref{2.12}).
The expansion of $\varphi(y,\theta)$ in powers of $\theta$ defines the dressed
$S_\P(x)$ coefficients, whose equations can also be obtained by directly
expanding (\ref{2.47a}-\ref{2.47d}) .
All properties of the halo correlations can be derived from this equation.
In particular it  is easy to see that,
\begin{equation}
\varphi^{(1)}(y)=y{\partial \zeta\over\partial \theta}
(\tau,\theta)\vert_{\theta=0},
\end{equation}
which is the generating function of the trees 
with one {\it long} external line.

These relations can be written in several equivalent forms.
From,
\begin{equation}
{\partial \varphi(y, \theta) \over \partial y } = \zeta(\tau, \theta)\ , 
\eqno(2.47.1) 
\end{equation}
which is analogous to \eqref(2.13.1), Eq. \eqref(2.47b) can also be written
\begin{equation} 
xh(x,\theta) =  \int_{-\i\infty}^{+\i\infty}
{\d y \over 2\pi\i} \ \zeta(\tau , \theta)\ e^{xy}\ ,
\eqno(2.47.2)
\end{equation}
as well as, after a second integration by parts,
\begin{equation} 
x^2h(x,\theta)= -  \int
{\d \tau \over 2\pi\i} \ {\partial \zeta(\tau,\theta) \over \partial \tau }
\ e^{xy} \ .
\eqno(2.47.3)
\end{equation}
As in \eqref(2.22.6), 
the integration is to be made over a contour that is defined by the
relation \eqref(2.47d) which determines the function $y(\tau, \theta)$.

\subsubsection{The normalization properties}

Considering the function $\zetax(x,\theta)$ defined in equation 
\eqref(2.46) and given
in (\ref{2.47a}-\ref{2.47d}) it is straightforward to demonstrate that,
\begin{equation}\int_0^{\infty}
x\ \zetax(x,\theta)\ h(x)\ \d x=\zeta(\tau,\theta) \vert_{\tau=0} ,
\eqno(2.48)
\end{equation}
which means that whatever $\Q$ we have,
\begin{equation}\int_0^{\infty}
x\,\nu_\Q(x)\,h(x)\,\d x={\nu_q}_{\vert_{q=\Q}} \ .
\eqno(2.49)
\end{equation}
These normalization properties (Eqs. [\ref{2.48}, \ref{2.49}]) 
are quite important and deserve 
attention. It means that at a given scale {\it the mean correlations of the
dense cells are the ones of the matter field}, once the mean is
calculated in a proper way: each object has to be weighted be its 
own internal scaling parameter $x$ which is equivalent to have the probability weighted by the mass. It is thus clearly seen that the dense cells represent the matter content of the universe : when properly averaged, the dense 
cell correlations 
are identical to the correlations of the matter points, at all orders.
It implies that the halos are not biased as a whole compared to the
matter field. Nevertheless, a population of particular object is always
biased
because there is a segregation in the correlation properties of the
halos \ie, $\nu_{\Q}(x)$ varies with $x$. It is clear from \eqref(2.49) 
that an $x$-independent vertex $\nu_\Q$ implies $\nu_\Q(x)=
{\nu_q}_{\vert_{q=\Q}}$.
The fact that the clusters are
more correlated that the galaxies (Bahcall 1979)
or a luminosity-dependence of the galaxy correlation function (Benoist et al. 1998)
 is an indication that such a segregation
indeed exists in the Universe and implies that galaxies are not
expected to be correlated as the matter field.

Another sum rule involves also the underdense cells (App. \ref{Bappend}), 
and in the continuous limit  
(where the $x$-dependence of the vertices is to be replaced by their
$m$-dependence) reads 
\begin{equation}\int_0^{\infty}
\ \zetax(m,\theta)\ p(m)  \ \d m= 1 ,
\eqno(2.49.1)
\end{equation}
which means that whatever $\Q$ we have,
\begin{equation}\int_0^{\infty}
\,\nu_\Q(m)\,p(m)\,\d m= 0 .
\eqno(2.49.2)
\end{equation}
So, {\it the underdense cells must be necessarily antibiased} ($\nu_{\Q} < 0$),
as a generalisation of the antibiaising property found (BeS92) for $b(m)$
($\nu_1(m)$ here) that was seen to change sign for $m \sim \Mv$.

Similar relations can be written for a distribution of points,
and are given in App. \ref{Bappend}.

\subsubsection{The validity limits}

The calculations that are presented in the previous section are valid under
three conditions:

\begin{itemize}
\item The spatial size  of the halo has to be smaller that the correlation
length so that the mean value of the two-point matter correlation function
in the volume occupied by the object is  larger than unity.
\item The space dependence of the correlation functions is valid
when the size of the halos is smaller than their (mean) relative distances.
More precisely the calculations are made to leading order
in the limit of vanishing
values of $\xib(V)/\xib(v_i)$ where $\xib(v_i)$ is the mean value of the
matter correlation function within the object $i$ and $\xib(V)$
is its mean value at a scale at which we want to derive the halo correlations.
For instance if galaxies are assumed to have a size of $100h^{-1}\kpc$
including their dark halo, at the $1h^{-1}\Mpc$ scale the corrective terms
corresponding to finite cell
size effects for their correlation functions
are expected to be of the order of 1\%.
\item The scale dependence in $x$ is valid when the internal mass
exceeds a mass threshold $\Mv(v_i)$ depending on the size of the object.
At the $100h^{-1}\kpc$ scale we have
\begin{equation}
\Mv(R)=2\ 10^7\ M_{\odot}\ {\Omega\over b^2} \left({R\over 100 h^{-1}\kpc}
\right)^{3+{\gm\omega/(1-\omega)}},
\eqno(2.50)
\end{equation} 
so that galaxies are far above that limit.
\end{itemize}

\subsection{The rare halo correlation functions}
\label{RaHalo}

The only parameters that can be directly measured in the catalogues are
the parameters $\tnu_{\Q}(x)$ as defined in equation 
\eqref(2.43). Their generating function,
\begin{equation}\tzetax(x,\theta)=1-\theta+\tnu_2{\theta^2\over2}
-\tnu_3\ {\theta^3\over 3!}+\dots
\eqno(2.51)
\end{equation}
is related to $\zetax(x,\theta)$ by
\begin{equation}\tzetax(x,\theta)=\zetax(x,\theta/b(x)).
\eqno(2.52)
\end{equation}
The main analytical result that can be derived from the equation
(\ref{2.47a}-\ref{2.47d}) 
concerns the large $x$ behavior. With the sole assumption that 
$\zeta(\tau,\theta)$ is a regular function that does not contain any
spurious singularity for the useful\footnote{so that  
$\vert y_s\vert$  corresponds to a singularity of implicit
equation \eqref(2.47d) for $\tau$ rather than to a singularity of $\zeta$
 and hence  to  a finite value of $\tau_s$ around
which $\zeta$ can be expanded, as is required by the model-independent
condition, Sect. \ref{ModIndProp}. The required condition are not fulfilled, 
for instance, by the model of Hamilton where $\tau_s$ is infinite.}
values of $\tau$, we obtain
a general result that is valid for any tree-hierarchical model (Appendix 
\ref{Dappend}),
\begin{eqnarray}
h(x)&\propto& x^{-5/2}e^{-x/x_*},\ \ \ x_*\approx 5\sim10\nonumber\\
b(x)&\propto& x\nonumber\\
\tnu_{\Q}(x)&\to& 1\ \ \ {\rm whatever}\ \ \Q\ \ {\rm and}\ \ S_P(x)\to P^{P-2}
\nonumber\\
\zetax(x,\theta)&\sim& e^{-\theta b(x)}\nonumber\\
\tzetax(x,\theta)&\sim& e^{-\theta},
\label{hp}
\eqno(2.53)
\end{eqnarray}
when $x\gg1$.
Note that this limit concerns fully nonlinear objects but is valid
for any correlation scale (linear or nonlinear).
This result has been presented in an 
early work (Bernardeau, PhD thesis, Paris 1992), 
and is suggested by the results obtained by Munshi et al. (1998b).
It is relevant a priori for describing the correlation of 
very bright galaxies.  Indeed, the observations called immediately at the 
time these concepts were discovered for such a modeling
of the distribution of bright galaxies:
the $\nu = 0$ model of  Schaeffer (1984, 1985)  
for the galaxy many-body correlations is equivalent to $\tnu_Q(x) =  1$
in the language of the present paper.

\subsubsection{Counting overdense cells}

We are now all set to give, as one of the possible applications of the formulae established previously, the statistics 
of the number of cells of size $v$
within volume $V$ that contain a mass larger than $m$,
(which is labeled by a scaling parameter larger than $x$).
For simplicity we call these cells  ``full'' cells .
The probability $P(>m, N)$ to have $N$ of such full cells can be written
\begin{equation}
P(>m, N) =  {1 \over 2\pi\i} \int { \d \lambda \over  \lambda^{N+1}}
e^{\chi_V(\la)} \ ,
\eqno(2.53.1)
\end{equation}
with (App. \ref{Bappend}),
\begin{eqnarray}
\chi_V(\la)&=&-\Phi(>x,Y)/\xib(V) \eqno(2.53.2)\\
\Phi(>x,Y)&=&Y\zeta(>x,\theta)-{1\over2}Y\,\theta
{\partial \zeta\over \partial \theta}(>x,\theta) \eqno(2.53.3)\\
\theta&=&-Y{\partial \zeta\over \partial \theta}(>x,\theta);\eqno(2.53.4)\\
Y&=&(1-\la)\,N_c \ , \ 
N_{\rm c}=n(>m)\,V\,\xib(V);
\eqno(2.53.5)
\end{eqnarray}
where 
\begin{equation}
n(>m) ={1\over v} p(>m) = { \rho \over m_{\rm c} } h(>x)
\eqno(2.53.6)
\end{equation}
 is the average number density of full cells.
This is the analogue of  \eqref(2.9)  for discrete counts.
For $N$ large enough we have
\begin{equation}
P(>m, N) = {1 \over N_c \xib(V)} H(>x,X) , X={N \over N_{\rm c}}
\eqno(2.53.7)
\end{equation}
with
\begin{equation}
H(>x,X) = -   \int_{-\i\infty}^{+\i\infty}
{\d Y \over 2\pi\i} \ \Phi(>x,Y)  \ e^{XY}\ ,
\eqno(2.53.8)
\end{equation}
corresponding exactly to \eqref(2.9).
The moments of $X$ then are by construction
\begin{equation}
\int \d X X^{\P} \ H(>x,X) = S_{\P}(>x).
\eqno(2.53.9)
\end{equation}
The positivity of $H(>x,X)$ then implies the positivity 
of the $S_{\P}$ coefficients as well as the constraints 
\eqref(2.22.4.4) for the latter.

Identical formulae can be written for halos  containing a mass
between $m$ and $m+\Delta m$, using $\zeta(x,\theta)$ in place of
$\zeta(>x,\theta)$ and $n(m)\Delta m$ in place of $n(>m)$,
with the same constraints 
\eqref(2.22.4.4) on $S_{\P}(x)$.

\section{Models for the matter correlations}
\label{Models}

\subsection{Specific models for $\zeta(\tau,\theta)$}

As can be seen in the equation 
\eqref(2.43) the form of the halo correlation functions
is entirely determined by $\zeta(\tau,\theta)$. This function is expected
to obey to various rules. For the general case, two scales are involved.
When both are in the nonlinear regime, one expects that,
\begin{equation}
\nu_{q,0}=\nu_{q,q}\ \ {\rm so\ that}\ \ \zeta(\tau,0)=\zeta(0,\theta)_{\vert\theta=\tau}.
\eqno(2.54)
\end{equation}

The numerical results presented in the following part have then
been partly obtained with,
\begin{equation}
\zeta=\left[1+{\tau+\theta\over \ka}\right]^{-\ka}
\ \ \ {\rm with}\ \ \ \ka = 1.3,
\eqno(2.55)
\end{equation}
relevant (BeS92) to the non-linear regime with CDM initial conditions.

However, we do not exclude that the larger scale may reach the linear 
regime. In such a case the relation \eqref(2.54) is not expected to be valid,
but the parameters $\nu_{q,q}$ are expected to take the values found
in the quasi-Gaussian regime from perturbative calculations (Bernardeau 1992,
1994) that is for instance 
$\zeta(0,\theta)\approx[1+2\theta/3]^{-3/2}$ independently
of $\Omega$ for $n=-3$.

We also introduce an other explicit model that does not lead to 
mathematical inconsistencies. This one is based on a property of
``scale factorizability'' which means that we assume that
$\nu_{q,\Q}=\nu_{q-\Q}\ \nu_{\Q}$. This can be understood qualitatively since 
we have  two scales $v$ and $V$  different enough, so that the averages 
are in some sense independent. In such a case the function 
$\zeta(\tau,\theta)$ is factorized and 
\begin{equation}
\zeta^{\rm fact.}(\tau,\theta)=\zeta(\tau)\zeta(\theta). 
\eqno(2.55.1)
\end{equation} 
We can then adopt the form,
\begin{equation}
\zeta^{\rm fact.}(\tau,\theta)=\left[1+{\tau\over \ka}\right]^{-\ka}
\left[1+{\theta\over \ka}\right]^{-\ka}.
\eqno(2.56)
\end{equation}
In both cases the function $\zeta(\tau,\theta=0)$ is known to agree with
all the constraints on $h(x)$ (BeS92). The form \eqref(2.56) is in agreement
with equation \eqref(2.54).

The numerical results presented in the following have been obtained with
these two forms.

We will also present ``model-independent'' predictions, obtained by
the mere assumption,
\begin{equation}
\zeta(\tau) \propto \tau^{-\ka} 
\end{equation}
for large  $\tau$, with,
\begin{equation} 0 \le \ka \le \infty \ . 
\end{equation} 
This leaves out the case where $\zeta(\tau)$ is a polynomial in $\tau$,
that will be discussed below as a generalization of the Hamilton (1988b) 
model.

\subsection{The minimal tree hierarchical model}

The minimal tree model is a priori an attractive model,
for which the geometrical dependence of the matter correlation
functions are well defined and entirely determined by the mass
function. It is thus natural to try to calculate the cell 
correlations in such a model.

The function $\varphi(y)$ for the dressed vertices in the 
minimal tree-hierarchical  model, given by using,
\begin{equation}
t=\tau+\theta,
\eqno(E.1) 
\end{equation}
as a new function depending on $y$ and $\theta$ through,
\begin{equation}
t=\theta-y\zeta'(t),
\label{tdey}
\end{equation} 
can be written,
\begin{equation}
\varphi(y,\theta)=y\zeta(t)-{1\over 2}y(t-\theta)\zeta'(t).
\end{equation}
Successive derivatives of $\varphi(y,\theta)$
with respect to $\theta$ will give
\begin{eqnarray}
\varphi^{(1)}(y)&=&t(y,\theta)_{\vert\theta=0}\label{phi1min}\\
\varphi^{(\Q)}(y)&=&
{\d^{\Q-1}\over\d\theta\ ^{\Q-1}}t(y,\theta)_{\vert\theta=0}.
\label{phiQmin}
\end{eqnarray}
As an application of these results we have,
\begin{eqnarray}
\varphi^{(1)}(y)&=&\tau(y);\\
\varphi^{(2)}(y)&=&-{y\zeta''\over1+y\zeta''};\\
\varphi^{(3)}(y)&=&-{y\zeta'''\over(1+y\zeta'')^3};\\
\varphi^{(4)}(y)&=&{3(y\zeta''')^2\over(1+y\zeta'')^5}-
{y\zeta^{(4)}\over(1+y\zeta'')^4};\\
\varphi^{(5)}(y)&=&-{15(y\zeta''')^3\over(1+y\zeta'')^7}+10
{y^2\zeta^{(4)}\zeta'''\over(1+y\zeta'')^6}-\nonumber\\
&&\ \ -{y\zeta^{(5)}\over(1+y\zeta'')^5},
\end{eqnarray}
where the derivatives are with respect to $t$,
now obviously related to $y$ by (\ref{tdey}) with $\theta=0$,
\begin{equation} 
t=-y\zeta'(t).
\eqno(E.3) 
\end{equation}

The results for $\varphi^{(1)}$ and $\varphi^{(2)}$
have been given in BeS92 and the three others 
by Munshi et al. (1998b)
from an order by order expansion of the  tree summation.
The general expression (\ref{phiQmin})
obtained here allow a simple and direct calculation of 
these functions to an arbitrary order\footnote{Note 
that, as for the general case given by 
Eqs. (\ref{2.47a}-\ref{2.47d}), these 
expressions are valid only provided (\ref{2.17}) holds, as we assume
throughout for what we call halos. 
In general (see App. \ref{Bappend}), 
$t(y,\theta)$ in Eqs. (\ref{phi1min}-\ref{phiQmin})
is to be replaced by  $t(y,\theta) \exp[-{\varphi(y,\theta)/\xib}]$.}.

It is also possible to express $h(x,\theta)$ directly as a function of
$\zeta(t)$ 
\begin{equation} 
x^2h(x,\theta) =  \   \int
{\d t \over 2\pi\i} \  \zeta'(t) \ e^{x {\theta-t \over \zeta'(t)}} \ .
\eqno(E.2)
\end{equation}
This equation can be written,
\begin{equation} 
xh(x,\theta) =  \-  \int_{-\i\infty}^{+\i\infty}
{\d \tiy \over 2\pi\i} \  f(\tiy, x\theta) \ e^{x\tiy} 
\ , 
\eqno(E.4)
\end{equation}
with,
\begin{equation}
f(\tiy,\theta) = \int_0^{t(\tiy)} \d t\ \zeta'(t)\ e^{\theta \over \zeta'(t)}\ . 
\eqno(E.5)
\end{equation}

This form is well suited for an expansion in powers of $\theta$ and 
allows one to get directly an expression for the vertices $\nu_{\Q}(x)$.
With the definition
\begin{equation} 
f_{\Q}(\tiy) = \int_0^{t(\tiy)} \d t \ \left[-\zeta'(t)\right]^{1-\Q} \ , 
\eqno(E.6) 
\end{equation}
we get $\nu_{\Q}$ in terms of $\zeta$,
\begin{equation} 
x h(x) \nu_{\Q}(x) =   x^{\Q} \int_{-\i\infty}^{+\i\infty}
{\d \tiy \over 2\pi\i}   \ f_{\Q}(\tiy) \ e^{x\tiy} \ .
\eqno(E.7)
\end{equation}
From the latter expression, 
it is readily seen that the functions $\varphi^{(\Q)}(y) $ can be formally
expressed as
\begin{equation} 
\varphi^{(\Q)}(y)=-(-1)^{\Q} \int_0^y {\d^{\Q} \over \d y^{\Q}} f_{\Q}(y)\ . 
\eqno(E.8) 
\end{equation}

\subsubsection{Small $x$ limit}

In the limit $x=0$, $y$ gets of order $1/x$ and $t$ as well gets very large. 
This immediately shows that in the latter limit, only the {\it 
model-independent } asymptotic form \eqref(2.15) of,
\begin{equation}
 \zeta(t) = c t^{-\ka},
\eqno(E.9)
\end{equation}
will contribute. Our calculation, thus, will be valid for any minimal 
tree hierarchical model satisfying the ``model-independent'' conditions.
The solution of \eqref(E.4) then is
\begin{equation}
t = (\ka c y)^{\alpha},\eqno(E.10)   
\end{equation}
with
\begin{equation}
\alpha = 1/(\ka + 2) \ . 
\eqno(E.11) 
\end{equation}
This leads to the expression of  $f_\Q(y)$ and we get,
\begin{equation} 
\varphi^{(\Q)}(y) = 
{\Gamma \left[\alpha(\Q-2)\right] \over 
\Gamma \left[-(1-\alpha)(\Q-2)\right] }(\ka c y)^{-\alpha(\Q-2)}\ .  
\eqno(E.12)  
\end{equation}
With the help of the relation
\begin{equation}
 \int_{-\i\infty}^{+\i\infty} {\d u \over 2\pi\i} 
\ u^{-\delta}\ e^u= 1/\Gamma(\delta) \ , 
\eqno(E.13) 
\end{equation}
the dressed vertices in the small $x$ limit can then be expressed as
\begin{equation} 
\nu_{\Q}(x) = {\Gamma \left[2(1-\alpha)\right] \over
\Gamma \left[-(1-\alpha)(\Q-2)\right] }  x^{\alpha \Q}\ . 
\eqno(E.14) 
\end{equation}
We see that
\begin{equation} 
\nu_1(x) = b(x) =  {\Gamma \left[2(1-\alpha)\right] \over
\Gamma \left[1-\alpha\right] }   x^{\alpha}\ ,  
\eqno(E.15) 
\end{equation}
which is the bias calculated in BeS92. Also 
\begin{equation} 
\nu_2(x) = b^2(x) Q_3(x) = 0 
\ ,  
\eqno(E.16) 
\end{equation}
that vanishes to leading order in $x$ and is given by higher terms 
than the one arising from the model-independent form \eqref(E.9), as 
discussed by the latter authors. This violates the lower 
bound  \eqref(2.22.4.5) on 
$Q_3$ (Peebles 1980), 
now applied to the moments of \eqref(2.53.9) that are directly related
to $\nu_{\Q}(x)$, 
and implies that some of the values of the associated  probabilities 
are negative in the small $x$ limit. This problem cannot be cured by
including the terms, Eq. (\ref{nu2compl}), that arise for $x \sim x_v $
since it is always possible to choose the volume $v$ small enough so
as to have $x_v$ arbitrarily small. 

Note however that the generating function of,
\begin{equation} 
h(>x,\theta) = \int_x^{\infty} h(x,\theta),
\eqno(E.18) 
\end{equation}
is $\varphi(y,\theta)/y$, so the corresponding vertices are
\begin{equation} 
\nu_{\Q}(>x) = {2 \Gamma \left[2(1-\alpha)\right] \over
(2-\Q)\Gamma \left[(1-\alpha)(2-\Q)\right] }  x^{\alpha \Q}\ . 
\eqno(E.19) 
\end{equation}
The corresponding $S_P$ parameters, the first
few are plotted on Figs \ref{S3}-\ref{S6}, are all finite
at small $x$ and do not violate the positivity
constraints \eqref(2.22.4.4).

In this case it is actually possible to get the generating
function for the vertices,
\begin{eqnarray}
&&\hmove\tzetax(x>0,\theta)=\nonumber\\
&&\hmove\ \ \ \disp{
\int_{-\i\infty}^{+\i\infty}
\d y\,y^{-\kappa/(\kappa+2)}e^{y+\beta\theta\,y^{(\kappa+1)/(\kappa+2)}}
\over
\int_{-\i\infty}^{+\i\infty}
\d y\,y^{-\kappa/(\kappa+2)} e^y}
\end{eqnarray}
with
\begin{equation}
\beta=-{\Gamma\left[-{1/(2+\kappa)}\right]\over
\Gamma\left[\kappa/2+\kappa\right]}
\end{equation}
so that $\tzetax(x>0,\theta)\sim 1-\theta$.
When $\theta\to\infty$ it has an asymptotic behavior given
by,
\begin{eqnarray}
&&\hmove\tzetax(x>0,\theta)\approx
{\kappa+2\over\kappa+1}{\Gamma\left[\kappa/(2+\kappa)\right]
\over\Gamma\left[(\kappa-1)/(\kappa+1)\right]}\,\beta^{-2/(\kappa+1)}
\times\nonumber\\
&&\hmove\ \ \times\theta^{-2/(\kappa+1)}.
\end{eqnarray}
It also corresponds to moments $S_{\P}(x)$ that satisfy the
constraints \eqref(2.22.4.4). 

We show in appendix that for all ``model-independent'' formulations
(that is all models satisfying a series of needed constraints
related to the count-in-cells, 
see BaS89 and discussion in Sect. \ref{ModIndProp}), 
some of the higher moments of the
cell correlations become unduly negative at small $x$ (see Figs. 
\ref{S4}-\ref{S6}).
This  indicates clearly that such a model for the
matter distribution is mathematically unsafe. Since there is no
problem for the larger values of $x$, we prefer to say that the
extension of the minimal model to calculate the cell correlations is
insufficient\footnote{It is due to the vanishing of the leading 
order of the expansion at small $x$. 
But the calculation of the higher orders of perturbation theory may
actually  not be necessary:  small deviations from the minimal model 
also may be sufficient to cure the problem. 
 Such studies, however, are left for future work.}
to describe the distribution of the less dense among the
over-dense cells. The form found
for $b(x)$ and $S_3(x)$ for $x\simgt x_*$ are attractive enough to explore more
completely the consequences of such a model on the whole galaxy clustering
properties. 

\subsection{The Hamilton model and extensions}

Hamilton (1988b) proposed another form for the vertex generating function
$\zeta$ of the minimal tree-hierarchical model, that does not enter into 
the class that we have called ``model-independent''. The latter obey 
conditions to be necessarily fulfilled so as to reproduce the known 
properties of the galaxy and matter count-in-cell distribution. We 
have however seen in the previous section that lead to difficulties 
for the minimal model. The question then may be raised whether by relaxing 
the model-independent conditions, probability laws that are acceptable for 
all values of $x$ can be found.

\subsubsection{Count-in-cells}

The vertex generating function of Hamilton (1988b), in its original form, is,
\begin{equation}
\zeta(\tau) = 1- \tau + {1 \over 2} k \tau^2, 
\eqno(E.21)
\end{equation}
with arbitrary $k>0$ .
The solution of the equation \eqref(2.13) relating the $\tau$ to the $y$
parameter is,
\begin{equation} \tau = {y \over 1 + k y},
\eqno(E.22) 
\end{equation}
and $\varphi(y)$ is, according to \eqref(2.12),
\begin{equation} \varphi(y) = y {1+(k-1/2)y \over 1+ky} \ . 
\eqno(E.23) 
\end{equation}
From the requirement for $\varphi(y)$ to be positive for all $y \ge 0$
(see discussion in BaS89), and specifically here for large $y$,
we get the constraint,
\begin{equation} k \ge 1/2 \ . 
\eqno(E.24) 
\end{equation}
The corresponding function $h(x)$ (for $x$ strictly positive) is then
(Eq. \ref{2.18}), 
\begin{equation}h(x) =  {1 \over 2 k^3} e^{-{x/k}} \ . 
\eqno(E.25) 
\end{equation} 
It is readily seen that, unless $k = 1/2$,  this form does not satisfy
the normalization condition  \eqref(2.21) and \eqref(2.22.4).
Nevertheless, the form \eqref(E.25) is perfectly suited to our discussion
for $k \ge 1/2$. Simply there is only a fraction $1/2k$ of the matter
in the dense cells, the remainder being in the underdense cells,
 whereas in the generic case the latter is negligible.
The form \eqref(E.25), because of its restriction to $x > 0$
amounts 
to write $\varphi(y)$, 
\begin{equation}
\varphi(y) = {1 \over 2k} {y \over 1+ky} + {k-1/2 \over k} y,
\eqno(E.26) 
\end{equation}
and to note that the last term does not contribute to $h(x)$.

If, however, we want to exclude scaling function with such peculiarities, 
we are restricted to,
\begin{equation}
k = 1/2\ , 
\eqno(E.27)
\end{equation}
a case that was specially discussed by Hamilton since it allows one to 
get many-body correlations with a structure that repeats itself under 
the grouping of a subset of points within an infinitesimal cell. In the 
case \eqref(E.26), the scaling function is,
\begin{equation}h(x) = 4 e^{-2x} \ . 
\eqno(E.28) 
\end{equation}

This is an important qualitative difference. 
In the actual counts, as discussed in BaS89 and confirmed by 
observations (Benoist et al. 1998 and references therein) as well 
as simulations (Valageas et al. 1999 and references therein), 
$\int_x^{\infty}h(x)\d x $
diverges for $x \to 0$, which means that the number of small 
objects increases without limit, so that their total number is 
not defined: this means that in realistic models
 the total number of objects depends on {\it other parameters 
than the ones introduced for the count-in-cells}.
For the models considered here on the other hand
$\int_0^{\infty}h(x)dx $ is finite and the total number of objects
is determined by the same physics that the dense cell. 
So, the models considered in this section are too different from 
these realistic models to be used as a modeling of the matter  distribution.

\subsubsection{The halo correlations}

The equation \eqref(2.47d) with the form \eqref(E.21) 
for $\zeta$ has the solution,
\begin{equation} 
\tau = {y (1-k\theta) \over 1 + k y}, 
\eqno(E.29) 
\end{equation}
and \eqref(2.47c) yields,
\begin{equation}
\varphi(y,\theta) = \left({1 \over 2k} -\theta+{1 \over 2} k \theta^2\right) 
{y \over 1+ky} + {k-1/2 \over k} \ . 
\eqno(E.30) 
\end{equation}
The last, constant, term does not contribute to the integral
\eqref(2.47b) that  
defines $\zetax(x,\theta)$ and we have,
\begin{equation}
\zetax(x,\theta) = 1-2k\,\theta+k^2\,\theta^2\ .
\eqno(E.31) 
\end{equation}
So, the model \eqref(E.21) corresponds to a constant bias,
\begin{equation}b(x) = 2k \ , 
\eqno(E.32) 
\end{equation}
normalized to unity since $xh(x)$ is normalized to $1/2k$.
Replacing $\theta$ by $\theta/2k$,
we get the vertex generating function for the dense cells,
\begin{equation} \tzetax(x,\theta) = 1 -\theta + {1\over4}\theta^2  \ . 
\eqno(E.33) 
\end{equation}
This is nothing but the $k=1/2$ model.
So, for an initial generating function with $k=1/2$, there is {\it no bias} 
for the distribution of the halos. Hamilton's condition for the correlation 
functions to repeat themselves when changing scale is equivalent to have 
an unbiased distribution. For $k>1/2$, the distribution of cells of size $v$
at the larger scale $V$ goes over to a $k=1/2$ distribution. 

\subsubsection{Other forms of $\zeta$ in  the minimal tree-hierarchical model}

Let us consider here the general
case where $\zeta'(t)$ has a zero for finite values of $t$ 
\begin{equation} 
\zeta'(t) \sim - c\ (1-kt)^s, \ \ \ \ s>0  \ . 
\eqno(E.34) 
\end{equation}
Around $t=1/k$, $\zeta$ has the expansion
\begin{equation} 
\zeta(t) 
\sim \zeta(1/k) + { c \over (s+1)k } (1-kt)^{(s+1)}\ . 
\eqno(E.35) 
\end{equation}
If we transform these relations into equalities,
for $s=1$, $c = 1$, this is the Hamilton model.

The equation \eqref(2.13) for $\varphi(y)$ has, 
in the vicinity of $t = 1/k$, the solution,
\begin{equation}1-kt \sim (yck)^{-1/s}  \ . 
\eqno(E.36) 
\end{equation}
So, the behavior of $\zeta$
at $t=1/k$ is related to the behavior at large $y$ of $\varphi$.
that, according to \eqref(2.12), has the expansion,  
\begin{equation} 
\varphi(y) \sim y\zeta({1 \over k}) + {1 \over 2k^2} - {s \over s+1}
{1 \over k^2}(yck)^{-1/s}  \ . 
\eqno(E.37) 
\end{equation}
This yields the behavior of $h(x)$ at small $x$,
\begin{equation} 
h(x) \sim {c^{2-1/s} \over s k^{2+1/s} \Gamma(2+1/s) }x^{1/s-1}\ . 
\eqno(E.38) 
\end{equation}

Then with the usual definition 
(\ref{2.47a}-\ref{2.47d}) of $y$ for the correlated cells
and the definition \eqref(E.1) for $t$ the equation \eqref(2.47c) reads,
\begin{equation}
t- \theta = -y \zeta'(t) 
\eqno(E.39) 
\end{equation}
and has the solution,
\begin{equation}
1-kt \sim ({1-k\theta \over yck} )^{1/s}\  . 
\eqno(E.40) 
\end{equation}
This yields the expansion, at large $y$,
\begin{eqnarray} 
&&\hmove
\varphi(y, \theta) \sim y\zeta({1 \over k}) + {(1-k\theta)^2 \over 2k^2}
\nonumber\\
&&\hmove\ \ \  - {s \over s+1}{(1-k\theta)^{1+1/s} \over k^2}(yck)^{-1/s}  \ . 
\eqno(E.41) 
\end{eqnarray}
We then get for $h(x,\theta)$,
\begin{equation} 
h(x, \theta) \sim {c^{2-1/s} \over s k^{2+1/s} \Gamma(2+1/s) }x^{-1+1/s}
(1-k\theta)^{1+1/s}\ . 
\eqno(E.42) 
\end{equation}
The vertex generating function for the 
cell correlations, in the limit  $x=0$ is then
\begin{equation}
\zetax(x=0,\theta) = (1-k\theta)^{1+1/s} \ . 
\eqno(E.43) 
\end{equation}
Its derivative has a zero of order $1/s$ at $\theta = 1/k$
\begin{equation}{\partial \zeta \over \partial \theta}(x=0,\theta) = (1-k\theta)^{1/s}
\ . 
\eqno(E.44)  
\end{equation}

Clearly
none of these forms enter in the ``model-independent'' category 
defined  in Sect. \ref{ModIndProp}
since $h(x)$ has not the required behavior at the origin.

\subsection{The factorized tree-hierarchical model}

In this paragraph we explore the consequences of the hypothesis
\eqref(2.55.1) on the generating function $\zeta(\tau,\theta)$.
We will see that this ``scale-factoraziblity'' hypothesis
will provide us with a phenomenological
model that has all the desired properties.

With the assumption \eqref(2.55.1), from 
(\ref{2.47a}-\ref{2.47d}), it is readily seen that,
\begin{equation} 
\varphi(y, \theta) = y \zeta(\theta) \left[\zeta(\tau) -
{1 \over 2} \tau \zeta'(\tau) \right]\ , 
\eqno(E.45) 
\end{equation}
with, 
\begin{equation}
\tau = -y \zeta(\theta) \zeta(\tau)   \  .  
\eqno(E.46) 
\end{equation}
This implies the simple result,
\begin{equation}
\varphi(y, \theta) = \varphi(y\,\zeta[\theta])  \ , 
\eqno(E.47) 
\end{equation}
We thus get,
\begin{equation}
h(x,\theta) = {1\over\zeta(\theta)} h\left[x\over\zeta(\theta)\right]
\ , 
\eqno(E.48) 
\end{equation}
and 
\begin{equation}
h(>x,\theta) = h\left[>{x\over\zeta(\theta)}\right]\ . 
\end{equation}
It leads to,
\begin{eqnarray}
\zetax(x,\theta) &=& {1\over h(x)\zeta(\theta)} 
h\left[x\over\zeta(\theta)\right]
\ ,\\ 
\zetax(>x,\theta) &=& {1\over h(>x)} 
h\left[>{x\over\zeta(\theta)}\right]
\ . 
\eqno(E.49) 
\end{eqnarray}

A systematic expansion in powers of $\theta$ gives
\begin{eqnarray} 
\varphi^{(1)}(y)  &=&y\varphi'(y); \\
\varphi^{(2)}(y)  &=&\nu_2y\varphi'(y)  + y^2\varphi''(y); \\
\varphi^{(3)}(y)  &=& \nu_3y\varphi'(y)  + 3\nu_2y^2\varphi''(y) 
+ y^3 \varphi'''(y);\\
\varphi^{(4)}(y)  &=& \nu_4y\varphi'(y)  +
(3\nu_2^2+4\nu_3)y^2\varphi''(y) \nonumber\\
&&\hmove\ \ \ + 6\nu_2y^3 \varphi'''(y) +y^4 \varphi''''(y)\ ,
\eqno(E.50) 
\end{eqnarray}
and
\begin{eqnarray}
\nu_1(x) h(x)&=& -{\d \over \d x} xh(x); \\
\nu_2(x) h(x)&=& -\nu_2{\d \over \d x} xh(x) + {\d^2 \over \d x^2}x^2h(x); \\
\nu_3(x) h(x)&=& -\nu_3{\d \over \d x} xh(x) + 3\nu_2 {\d^2 \over
\d x^2}x^2h(x) \nonumber\\
&&\hmove - {\d^3 \over \d x^3}x^3h(x);\\
\nu_4(x) h(x)&=& -\nu_4{\d \over \d x} xh(x) + 
(3\nu_2^2+4\nu_3){\d^2 \over \d x^2}x^2h(x)  \nonumber\\
&&\hmove - 6\nu_2{\d^3 \over \d x^3}x^3h(x) + {\d^4 \over \d x^4}x^4h(x)  \  . 
\eqno(E.51) 
\end{eqnarray}
The first of these equations implies an $x$-dependent bias
that is expressed in terms of the scaling function $h(x)$ as
\begin{equation}
b(x) = - {1\over h(x)}  {\d \over \d x} x h(x)  \ . 
\eqno(E.52) 
\end{equation}
Note also that,
\begin{equation}
b(>x)=-{x\over h(>x)}\ {\d\over \d x}h(>x)={x\,h(x)\over h(>x)}.
\end{equation}

\subsubsection{Large $x$ behavior}

For large $x$, from \eqref(2.19), the bias is
\begin{equation}
b(x)={x\over x_*}-\omega_s\ \ {\rm and}\ \ b(>x)={x\over x_*}-\omega_s+1,
\eqno(E.53)
\end{equation}
with generically $\omega_s = -3/2$.
With the form \eqref(E.53), it is readily seen from \eqref(E.49) that
\begin{equation} 
\tzetax(x,\theta) = \zetax\left[x, {\theta\over b(x)}\right] 
\sim e^{-\theta} \ .
\eqno(E.54)  
\end{equation}
We have the same limiting form for $\tzeta(>x,\theta)$.

\subsubsection{Small $x$ limit}

At $x=0$, we have,
\begin{equation}
\zeta(x\to 0,\theta)=\zeta^{1-\omega}(\theta)
\end{equation} which, with $h(x) \sim x^{\omega-2}$,  gives
\begin{eqnarray}
b(x\to 0)&=&1-\omega,\\
\tzetax(x=0,\theta)&=&\zeta^{1-\omega}\left({\theta \over 1-\omega}\right)\ . 
\eqno(E.55) 
\end{eqnarray}
The same results are obtained for $b(x > 0)$ and $\tzetax(x>0,\theta)$.

Finally for 
\begin{equation}
\zeta(\tau)=\left(1+{\tau \over \ka }\right) ^{-\ka}\ , 
\eqno(E.56) 
\end{equation}
we have
\begin{equation}
\tzeta(x=0,\theta)=\tzeta(x>0,\theta)=
\left(1+{\theta\over\tka}\right)^{-\tka}\ , 
\eqno(E.57) 
\end{equation}
with 
\begin{equation}
\tka =(1-\omega)\ka.
\end{equation}

\subsubsection{A very special case}

Interestingly, for $x\to 0$ or for $x \to \infty$,
\begin{equation}
\zeta(\tau,\theta) = e^{-\tau-\theta} \ , 
\eqno(E.58) 
\end{equation}
is invariant under the transformation of the correlations 
of elementary points into the cell correlations. In this case we have
indeed, 
\begin{equation}
\tzeta(x=0,\theta) = \tzeta(x\to\infty,\theta) = e^{-\theta}  \ .  
\eqno(E.59)
\end{equation}
Note however that this form is not preserved for any intermediate
value of $x$.

\subsection{Hierarchical models and tree-hierarchical models, comments}

The results presented throughout this paper give the qualitative
and quantitative behavior of the halo correlations in the evolved
nonlinear density field.
They are based on assumptions on the nonlinear matter correlation
functions that have been now widely checked. Nevertheless in this section
we aim to discuss the various results that we obtain with regards to the 
hypothesis that have been made at several stages.

The shape of the nonlinear mass distribution function is derived from
the assumption that the correlation functions follow a hierarchical pattern
\eqref(2.8). 
This form implies that the mass distribution function takes the form
\eqref(2.16).

The joint mass distribution function for two cells is determined by the 
behavior of the correlation functions when the points separate into two
different subsets that lie in two volumes $v_1$ and $v_2$ at the positions
$\vr_1$ and $\vr_2$. For any hierarchical clustering we obtain
\begin{eqnarray}
&&\hmove \xib_{p_1,p_2}(v_1,v_2)=\nonumber\\
&&\hmove \ \ \ C_{p_1,p_2}\left[\xib_2(v_1)\right]^{p_1-1}\ \xi_2(\vr_1,\vr_2)
\ \left[\xib_2(v_2)\right]^{p_2-1},
\eqno(2.57)
\end{eqnarray}
which implies that
\begin{eqnarray}
&&\hmove p(m_1,\vr_1;m_2\vr_2)\ \d m_1\ \d m_2=p(m_1)\ \d m_1\ p(m_2)\ \d m_2
\nonumber\\
&&\hmove \ \ \ \times \left[
1+\xi_2(\vr_1,\vr_2)C({m_1\over\Mcu},{m_2\over\Mcd})\right].
\eqno(2.58)
\end{eqnarray}
The resulting correlation between the halos at the position
$\vr_1$ and $\vr_2$ exhibits some of the general properties previously
mentioned:
\begin{itemize}
\item the distance dependence is the same as the one of the matter
correlation function;
\item the bias parameter is a function of $x$ only.
\end{itemize}

But at this level of hypothesis there is no property of
factorizability. This latter property is obtained with the tree hypothesis
which implies that $C_{p_1,p_2}=C_{p_1} C_{p_2}$. As a result the bias function
can be factorized since,
\begin{eqnarray}
&&\hmove h'(x_1,x_2)=\disp{
\int {\d y_1\over 2\pi \i}\ {\d y_2\over 2\pi \i}} e^{x_1y_1+x_2y_2}
\ \nonumber\\
&&\hmove \ \ \ \times\disp{
\left(\sum_{p_1,p_2} (-1)^{p_1+p_2} C_{p_1,p_2}\ {y_1^{p_1}\over p_1!}\ 
{y_2^{p_2}\over p_2!}\right)},
\eqno(2.59)
\end{eqnarray}
which under the tree-hierarchical hypothesis reads
\begin{eqnarray}
&&\hmove h'(x_1,x_2)=
\int {\d y_1\over 2\pi \i}\ 
\left(\sum_{p} (-1)^p C_{p}\ {y_1^p\over p!}\right)\ 
e^{x_1y_1}\nonumber\\
&&\hmove \ \ \ \times\ \int {\d y_2\over 2\pi \i}\ 
\left(\sum_{p} (-1)^p C_{p}\ {y_2^p\over p!}\right)\ 
e^{x_2y_2}.
\eqno(2.60)
\end{eqnarray}
Such remarks can be generalized to any order of the halo correlations. As a
result we get the following properties:

The hierarchical hypothesis implies that
\begin{itemize}
\item the halos are correlated in a hierarchical way (Eq. [\ref{2.39b}])
with the same scale dependence as the matter correlation functions;
\item the strength of the correlation functions, whatever the order,
is a function of the scaling parameter $x=m/\Mc$ only.
\end{itemize}

The {\it tree-hierarchical} hypothesis implies that
\begin{itemize}
\item the halo correlations follow a tree-hierarchical pattern too;
\item they take a specific form (Eq. [\ref{2.53}]) in the very massive halo
limit.
\end{itemize}

Note that in case of the minimal tree model, the halo correlations also follow
a minimal tree-hierarchical model.

\section{The observational consequences}
\label{ObsCons}

\subsection{ The two- and three-point correlation functions}

\begin{figure}
\centerline{
\psfig{figure=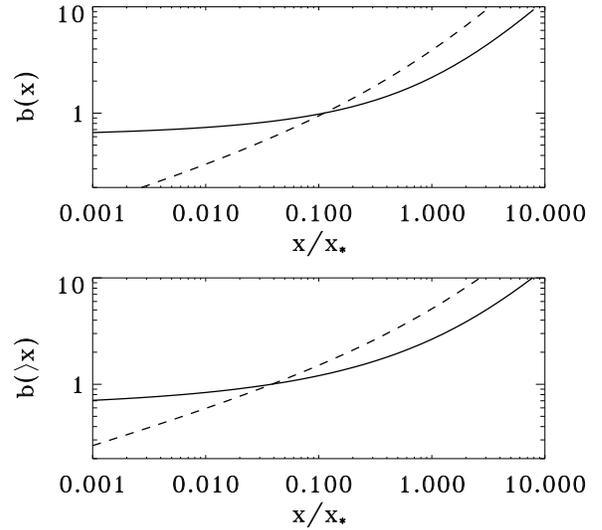,width=8cm}}
\caption{Bias factor as a function of the scaling parameter $x$. The solid 
lines correspond to the model given by the equation (\ref{2.56}) and the dashed
lines corresponds to a minimal tree-hierarchical model (Eq. [\ref{2.55}]).
In both cases we have used $\kappa=1.3$ for describing the matter field.
In the top panel we represent the bias as a function of $x$
(Eq. [\ref{2.32}]) and 
in the bottom one for a threshold in the density field.}
\label{bx}
\end{figure}

\begin{figure}
\centerline{
\psfig{figure=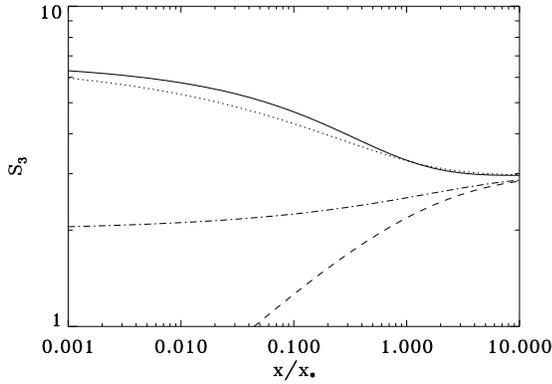,width=8cm}}
\caption{The three-point parameter as a function of $x$. 
The parameter $S_3\equiv3\,Q_3$
describes the strength of the three-point halo correlation function
(Eq. [\ref{2.42}]). The solid line (respectively dashed line)
gives $S_3(x)$ for the minimal tree (respectively factorized)
model, the dotted line (respectively dotted-dashed line) gives
$S_3(>x)$.}
\label{S3}
\end{figure}

\begin{figure}
\centerline{
\psfig{figure=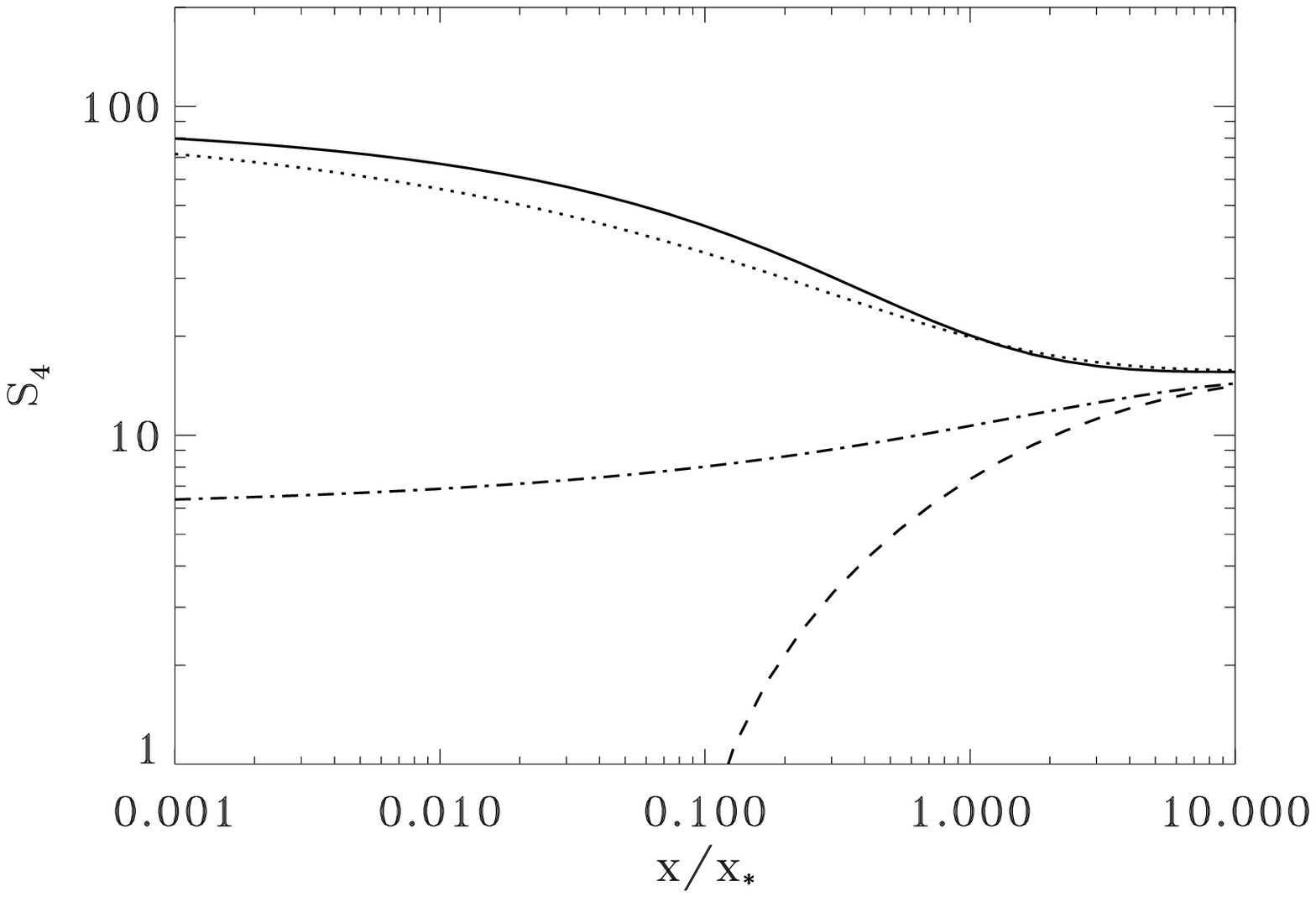,width=8cm}}
\caption{The same as in Fig. \ref{S3} for $S_4$.}
\label{S4}
\end{figure}

\begin{figure}
\centerline{
\psfig{figure=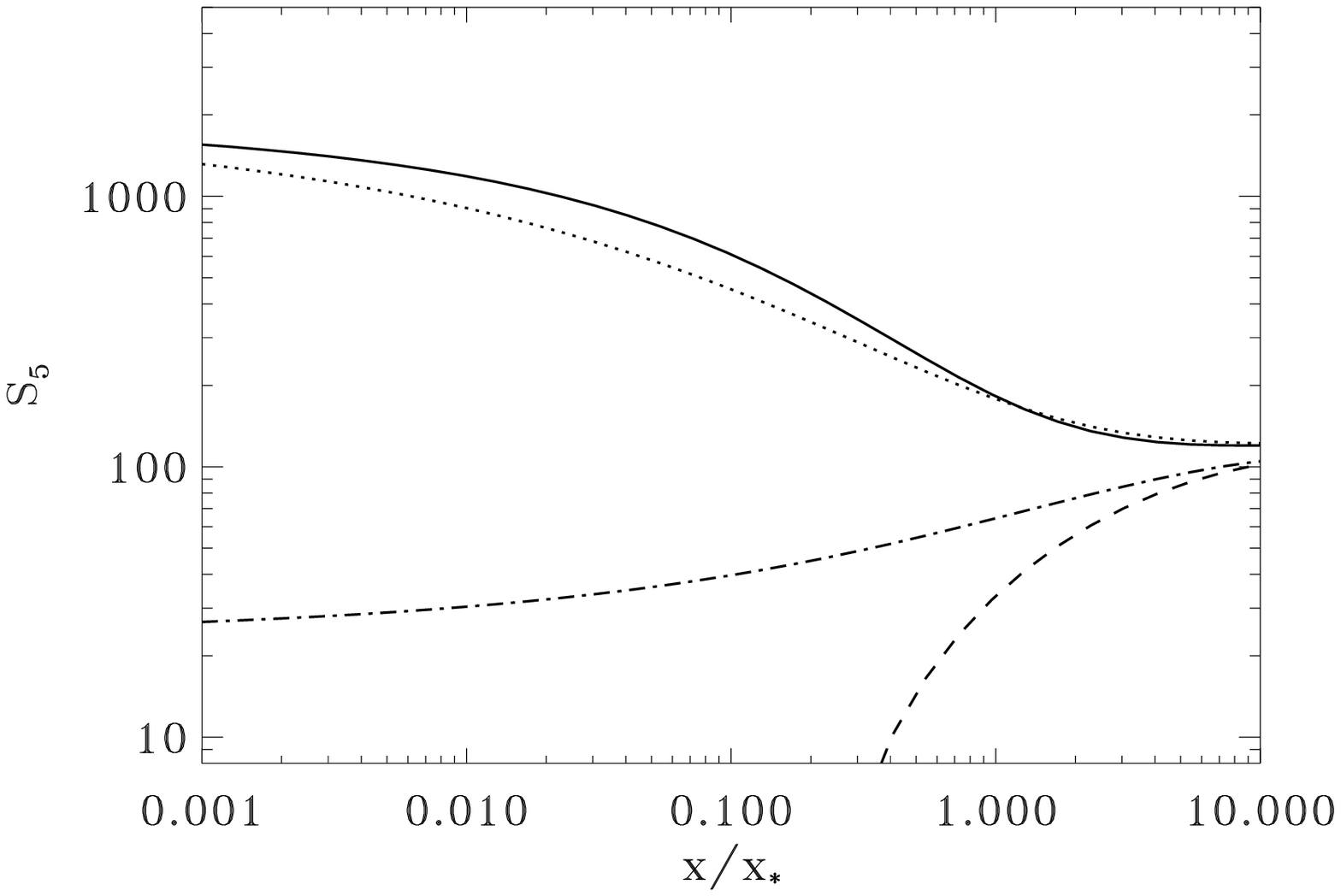,width=8cm}}
\caption{The same as in Fig. \ref{S3} for $S_5$.}
\label{S5}
\end{figure}

\begin{figure}
\centerline{
\psfig{figure=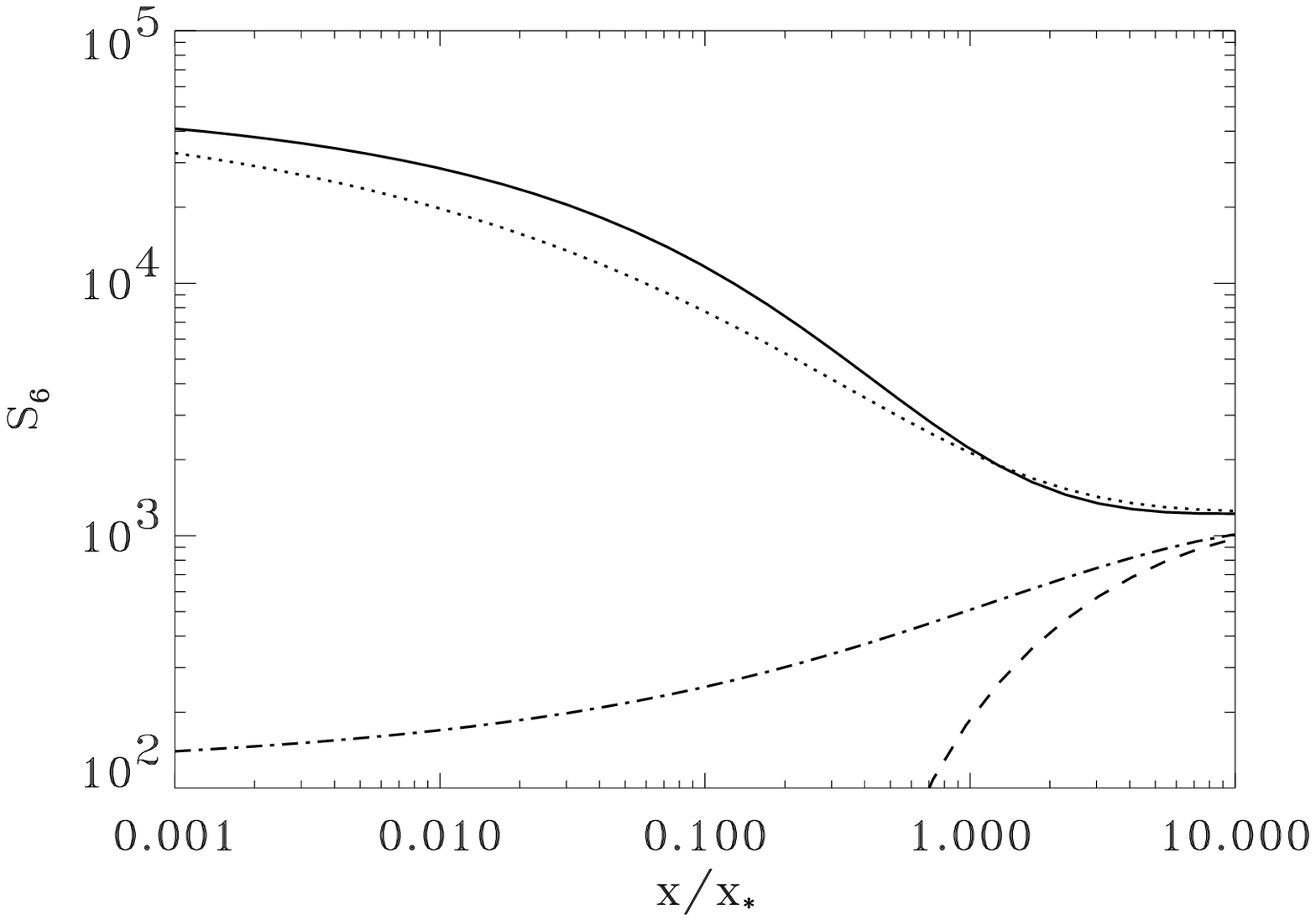,width=8cm}}
\caption{The same as in Fig. \ref{S3} for $S_6$.}
\label{S6}
\end{figure}

The results obtained for the two-point correlation
functions of the halos have already been discussed on an observational
point of view in a previous paper (BeS92). In that paper we only 
gave results for the minimal tree-hierarchical model.
In this paper we introduce
an other specific model (Eq. [\ref{2.56}]), assuming the small and large
scale-dependence of the generating function factorize,
for which we calculate the shape of the bias parameter and of the higher order
statistical properties.

In Fig. \ref{bx} we present the function $b(x)$
for the two models. We also present the results for $b(>x)$ 
that give the statistical properties of the objects
defined by a threshold. The difference between the two cases
is an indication of the magnitude of the variations that may 
occur within the hypothesis of hierarchical clustering. In both cases
the bias is proportional to $x$ for large values of $x$, and
is weakly varying or nearly constant for small values of $x$.
These results are natural consequences of the hierarchical clustering and
by no mean have been imposed a priori. The general properties of these
quantities have been given in (BeS92). Two properties of observational
interest (Eqs. [\ref{2.41a}-\ref{2.41b}]) are given in Sect. \ref{ModIndProp}.

Although we have not made any attempt here to transform the scaling parameter
$x$ which caracterises the halos into luminosity, it is very obvious 
from both the Fig. 5a of BeS92 and Fig. 5 of Benoist et al. (1996) that,
 within the uncertainty in defining the luminosity,
we can bring
 our predictions of the bias is in agreement with the data
(but it is also obvious from these figures that 
the process looks somewhat easier for the factorized model, e.g.
Fig. 4.7 of Bernardeau 1992, PhD thesis).
Numerical simulations (Munshi et al. 1999) seem to favor the
minimal model.

\subsection{High-order correlation functions} 

The new results that have been derived in this paper concern the
high order correlation functions. One can obviously 
calculate order by order the shape of the 4, 5-point correlation 
functions of the halos.

It requires the expansion of the generating function $\zetax(x,\theta)$
with $\theta$ (that can be done at the level of the equations 
[\ref{2.47a}-\ref{2.47d}]).
The coefficients of this expansion, $\nu_1(x)$, $\nu_2(x)$, $\nu_3(x)\dots$
give the values of $S_P(x)$ (Eq. [\ref{2.39b}]) for growing values
of $P$: 
\begin{eqnarray}
S_3(x)&=&3{\nu_2(x)\over\nu_1^2(x)};\nonumber\\
S_4(x)&=&4{\nu_3(x)\over\nu_1^3(x)}+12{\nu_2^2(x)\over\nu_1^4(x)};\nonumber\\
S_5(x)&=&5{\nu_4(x)\over\nu_1^4(x)}+60{\nu_3(x)\nu_2(x)\over\nu_1^5(x)}
+60{\nu_2^3(x)\over\nu_1^6(x)};\nonumber\\
&&\dots\nonumber
\end{eqnarray}
following a tree shape construction (e.g. Bernardeau 1992). 

The resulting values of $S_3(x)$-$S_6(x)$
are presented on Figs. \ref{S3}-\ref{S6}. They have been calculated
for the minimal and the factorized models using $\kappa=1.3$.
One can see that the values of $S_P$ for the factorized model
are always positive (solid lines) and above the positivity constraints.

This is not the case for the minimal-tree models (dashed lines).

\subsection{The void probability function}

An alternative
way to directly constrain the high order correlation functions
of the halos is to observe their void probability function.
This is the probability $\VP(0)$
that a 
spherical volume $V$ does not contain any object of a given kind, and it
is closely related to the behavior of the high-order correlation functions
of the objects (White 1979). 
In case of the tree-hierarchical models it is possible to relate
$\VP(0)$ 
to the generating function of the vertices $\zeta(\theta)$ (BeS92
-see also Schaeffer, 1984- and appendix B). For
the objects of kind $x$ we then obtain
\begin{eqnarray}
-{\ln \VP(0)\over n(x)V}&=&\tzetax(x,\theta)-{1\over 2}\theta\ 
{\partial \tzetax\over \partial \theta}\nonumber\\
\theta&=&-\Nc\ 
{\partial \tzetax\over \partial \theta}
\eqno(3.2)
\end{eqnarray}
where $N_c$ is the value taken by $n(x)Vb^2(x)\xib(V)$ depending
on the values of $x$ and $V$.

\begin{figure}
\centerline{
\psfig{figure=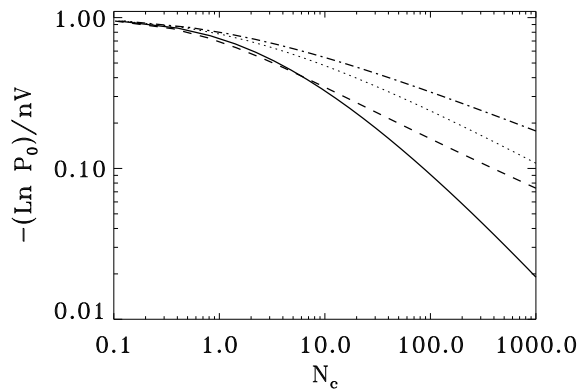,width=8cm}}
\caption{The void probability function $\VP(0)$ as a function of the
reduced variable $\Nc=n\,V\,\xib(V)$, where $n$ is the density of the
sample, $V$ the volume of a randomly placed cell and $\xib(V)$ is the mean
value of the two-point correlation function as measured in the sample and
within the volume $V$. The dotted line is the form used to
describe the matter 
distribution. The solid line is the limit we obtained for the
rare halo limit (Eq. [\ref{3.2}]). The dashed lines corresponds to the
minimal model (Eq. [\ref{2.55}]) for $x> 0$.
The dotted-dashed line corresponds to the factorized-tree model
(Eq. [\ref{2.56}]) for $x=0$ (or $x>0$ since it gives the same result).
}
\label{Sigma}
\end{figure}

The calculable properties of $\VP(0)$ are then direct consequences
of the properties of $\tzetax(x,\theta)$. In Fig. \ref{Sigma} we present
the shape of the function $- \ln\VP(0) / nV $  as a function
of $N_c$ for various values
of $x$ and for the two explicit models we considered. 
A change between the void probability function in the matter field
(assumed to be represented by a fair sample of points), and the
void probability function of halos is seen. This is due to the existence
of biases in the high order correlation functions, that lead to changes
in the values of $S_P$ from the matter field to the halo field, and are 
independent of the fact that $b=1$ or not. For the two models there
is a common limit for $x\to\infty$, corresponding to $S_P=P^{P-2}$, which
is independent of $x$.
In such a case the void probability function reads,
\begin{eqnarray}
-{\ln \VP(0)\over n(x)V}&=&
(1+{1\over 2}\theta) e^{-\theta}\nonumber\\
\theta e^\theta&=&\Nc.
\eqno(3.3)
\end{eqnarray}
The resulting form for $- \ln\VP(0) / n(x)V$ corresponds to the
model of Schaeffer (1984) with $\nu=0$. 

The dependence of $- \ln\VP(0) / n(x)V$  with $x$ is quite weak at this
level of description and most of the bias effects are contain in 
$b(x)$ (especially for the bright end). The weak dependence of $S_P(x)$
with $x$ is an illustration, at the level of the 
multi-point correlation functions,
of this feature. The kind of variation that is expected is,
however, dependent of the form assumed for $\zeta(\tau,\theta)$. For
the form \eqref(2.55) this is a growing function of $x$, for the form
\eqref(2.56) a decaying. We have thus no universal  answer for a set of
typical galaxies ($x\approx x_*$). However, for rich clusters or bright
galaxies we know what should be
the shape of the void probability function.

The main advance brought by our models is that it can reconcile
the measurements in catalogues and in numerical simulations in a unique
description. 
The properties of the field halos are mainly due to a threshold effect
in the nonlinear density field (specially for the rare halo limit),
at contrast with the matter field properties 
which are solely due to the dynamics.
It is however quite remarkable to note that the ``model-independent''
requirements \ref{ModIndProp} for the count-in-cells  
precisely call for the properties required (\ref{RaHalo}) 
for the above universal
asymptotic behavior to  hold. The Hamilton model, for instance, that does not
satisfy these requirements, also does not lead to the above behavior.
In this sense, the properties of the field halos definitely reflect
the underlying matter dynamics. Simply, the conditions we imposed
already for  the count-in-cells to reflect that dynamics also induce
the above behavior.

\subsection{ The mass function in clusters}

\begin{figure}
\centerline{
\psfig{figure=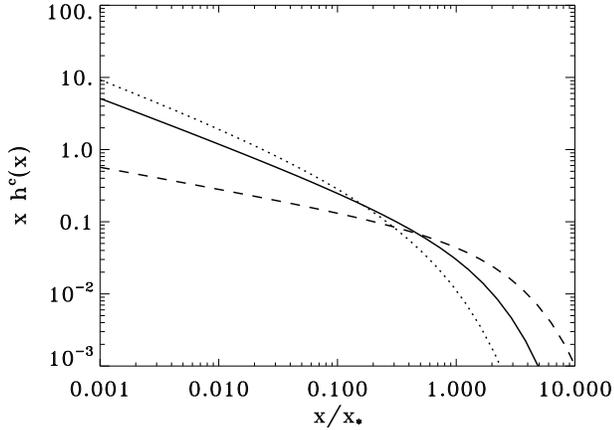,width=9cm}}
\caption{Change of the  mass functions of ``galaxies'' from
the ones in the field (dotted lines) to the ones in rich clusters 
for the factorized-tree model (solid line) and the minimal-tree model
(dashed line). The mass associated to a ``galaxy'' 
is assumed to be independent of the environment. The effect is
solely due to hierarchical clustering, the fluctuations 
of the non-linear density field being larger within clusters.
It
is seen
to be less important with the factorized-tree model (\ref{2.56}) in
agreement with the results (\ref{3.9b}, \ref{3.9a}).
}
\label{hc}
\end{figure}

\subsubsection{The general formalism}

In the previous section, correlation properties of the matter field,
and correlation properties of the halos were considered separately.
But in fact 
the general results we get hold to describe the auto-correlation
functions for a given kind of objects, as well as for the cross-correlations
between objects of different kinds.
We present the results that follow as an illustration of what can be
derived from this theory. 

The calculation of the mass multiplicity function of the halos,
representing galaxies, being in a cluster involves the statistics of a
two component field. One component 
is the matter field characterized by a density $n$, a two-point
correlation function $\xi_2(\vr,\vr')$ and $p$-body 
correlation functions following a tree-hierarchical form given
by the vertex generating function $\zeta(\theta)$; the second component 
is a halo field characterized by a parameter $x$ which leads to a
density $n(x)$, and correlation functions given by $\zetax(x,\theta)$.
As the latter function describes as well the auto-correlations
as the cross-correlations between the density field and the halo field,
one can compute the joint density distribution of the two fields,
$P(N(x),N)$, which is the probability that the volume $V$ contains $N$ points
of matter and $N(x)$ objects of scaling parameter $x$. The generating
function of such joint probability distribution is defined by
\begin{equation}
\sum_{N(x),N}\la_x^{N(x)}\,\la^N\,P(N(x),N)= e^{\chi(\la_x,\la)}
\eqno(F.1)
\end{equation}
and its expression reads,
\begin{eqnarray}
\chi(\la_x,\la)&=&(\la_x-1)\,n(x)\,V\,\zetax(x,\theta)\nonumber\\
&&\hmove -{1\over 2}(\la_x-1)\,
n(x)\,V\,{\partial\zetax\over\partial\theta}(x,\theta)\nonumber\\
&&\hmove +(\la-1)\,n\,V\,\zeta(\theta)-{1\over 2}(\la-1)\,
n\,V\,{\partial\zeta\over\partial\theta}(\theta)\nonumber\\
\theta&=&(\la_x-1)\,n(x)\,V\,\xib(V)\,{\partial\zeta\over\partial\theta}
(x,\theta)\nonumber\\
&&\hmove +(\la-1)\,n\,V\,\xib(V)\,{\partial\zeta\over\partial\theta}
(\theta).
\eqno(F.2)
\end{eqnarray}
We are interested in the expectation number of objects of kind $x$
being in a volume $V$ that contains a certain mass $M=N\,\rho$,
\begin{equation}
\mg N(x)\md_M={\sum_{N(x)}\,N(x)\,P(N(x),N)\over\sum_{N(x)}\,P(N(x),N)}.
\eqno(F.3)
\end{equation}

\subsubsection{The rich cluster limit}

The mass $M$ is chosen so that the volume $V$ is rich enough to
represent the inner part of a cluster ($N\simgt \Nc(V)$). 
The calculation
of $\mg N(x)\md_M$ from \eqref(F.2) is quite straightforward. 
The generating 
function of $\sum_{N(x)}\allowbreak P(N(x),N)$ is simply
$\exp(\chi(\la_x=0,\la))$ whereas the one of $\sum_{N(x)}\,\allowbreak
N(x)\,P(N(x),N)$ is 
\begin{equation}
\mP^c(\la)={\partial\chi\over\partial\la_x}(0,\la)\,e^{\chi(0,\la)}.
\eqno(F.4)
\end{equation} 
The latter expression can be deduced from \eqref(F.2),
\begin{eqnarray}
\mP^c(\la)&=&n(x)\,V\,\zetax(x,\theta)\,e^{\chi(0,\la)}\nonumber\\
\theta&=&(\la-1)\,n\,V\,\xib(V)\,{\partial\zeta\over\partial\theta}(0,\theta).
\eqno(F.5)
\end{eqnarray}
When we are the regime $N\simgt \Nc(V)$, we can use the continuous
variable $X=N/\Nc(V)$ to describe the content of the volume $V$ instead of
the variable $N$. We are thus led to calculate,
\begin{eqnarray}
\mg N(x)\md_X&=&n(x)\,V\,\rho_{\rm cluster}
{\int_{-\i\infty}^{+\i\infty}\d Y\,\zetax(x,\theta)\,e^{XY} \over
\int_{-\i\infty}^{+\i\infty}\d Y\,\zeta(\theta)\,e^{XY}},\nonumber\\
\theta(Y)&=&-Y{\partial\zeta\over\partial\theta}(0,\theta).
\eqno(F.6)
\end{eqnarray}
When $X$ is large one can compute more precisely the previous relation. Its
result is dominated by the singularity that lie in $y=y_s<0$ 
in the complex plane. 
This singularity is characterized by the equations,
\begin{eqnarray}
\theta_s&=&{\partial\zeta\over\partial\theta}(0,\theta_s)/
{\partial^2\zeta\over\partial\theta^2}(0,\theta_s)\nonumber\\
y_s&=&-1/{\partial^2\zeta\over\partial\theta^2}(0,\theta_s).
\eqno(F.7)
\end{eqnarray}
The integrations in \eqref(F.6) are then performed around the singularity
$Y=y_s$ which requires the expansion of $\zeta(\theta)$ and of 
$\zetax(x,\theta)$ around $\theta_s$. We eventually get,
\begin{equation}
\mg N(x)\md_X=n(x)\,V\,\rho_{\rm cluster}
\,{{\partial\zetax\over\partial\theta}(x,\theta_s)}/
{{\partial\zeta\over\partial\theta}(\theta_s)}.
\eqno(F.8)
\end{equation}

The mass distribution function in rich clusters $h^c(x)$ normalized so that
$\int x\,h^c(x)\,\d x=1$ is then given by,
\begin{equation}
h^c(x)={\mg N(x)\md_X\over n(x)\,V\,\rho_{\rm cluster}}\,h(x),
\eqno(F.9)
\end{equation}
that reduces to,
\begin{equation}
h^c(x)=h(x)\,{{\partial\zetax\over\partial\theta}(x,\theta_s)}/{
{\partial\zeta\over\partial\theta}(\theta_s)}.
\eqno(F.10)
\end{equation}
We can notice that the result is independent of $X=N/N_c$, say 
the richness of the cluster.

For any tree-hierarchical model the behavior of $\zetax(x,\theta)$ has been
shown to be close to $\exp[-b(x)\theta]$ with $b(x)$ proportional to $x$
(Eq. [\ref{2.53}]), so that $h^c(x)$ behaves roughly as $b(x)h(x)$ at least
when $x$ is large,
\begin{equation}h^c(x)\sim b(x)h(x).
\eqno(3.8)
\end{equation}
As a result the upper cut-off of $h(x)$, characterized
by $x_*$ (Eq. [\ref{2.19}]), is shifted towards greater values of $x$.
This change can be calculated as soon as a particular form for 
$\zeta(\tau,\theta)$ has been chosen. 

For the factorized-tree model it is possible to give an explicit
result. In this case, from Eq. (\ref{E.48}) 
we have in the rich cluster limit,
\begin{equation}
h^c(x)=-{1\over\zeta^2(\theta_s)}\left[
h\left(x\over\zeta(\theta_s)\right)+{x\over\zeta(\theta_s)}\,
h'\left(x\over\zeta(\theta_s)\right)
\right],
\end{equation}
which, from Eq. (\ref{E.51}),  can be written,
\begin{equation}
h^c(x)={1\over\zeta^2(\theta_s)}
b\left(x\over\zeta(\theta_s)\right)\,
h\left(x\over\zeta(\theta_s)\right).
\end{equation}
This is the function which is plotted on Fig. \ref{hc} with
a solid line.

Thus, for the factorized-tree model we get
the position of the cut-off for $h^c(x)$, $x_*^c$,
\begin{equation}
x_*^c=\left({\ka+2\over\ka+1}\right)^{\ka}\ x_*\approx 1.6\,x_*\ \ 
{\rm for}\ \ \ka=1.3.
\eqno(3.9b)
\end{equation}

In case of the minimal tree-hierarchical
model (Eq. [\ref{2.55}]),  the cut-off is given by,
\begin{equation}
x_*^c=\left({\ka+2\over\ka+1}\right)^{\ka+2}\ x_*\approx 3.3\,x_*\ \ 
{\rm for}\ \ \ka=1.3.
\eqno(3.9a)
\end{equation}
One can notice that the effect is similar, although somewhat more important.

This has important consequences. For a given cluster mass, the galaxy
multiplicity function within the cluster is expected to be biased towards
higher masses, irrespective of other phenomena that can change
the observed luminosity function. 
{\it  The $M/L$ ratio within clusters  is thus expected to be 
biased and smaller  than the average in the Universe}. 

\section{Conclusion}

In this paper we address the issue of the halo correlation properties
in a
strongly non-linear cosmic density field. This is a dramatic
extension of the results obtained by Bardeen et al. (1986) and
Mo et al. (1997) for Gaussian fields.

The difficulty is to have a reliable description of the matter density
fields and its high order correlation properties. Our analyses are all
based on the hierarchical properties (in the sense of Eq. [\ref{1.1}])
which seem well established in numerical simulations or in the
observable Universe.

We further assume a tree decomposition of the correlation
functions\footnote{Note that all these properties are verified at
leading order in 
perturbation theory (Bernardeau 1996).}(see Szapudi \&
Szalay 1997, 1998 for another approach). 
These assumptions lead to two crucial properties :
\begin{itemize}
\item factorazibility of the mass dependence of the correlation
properties;
\item the bias factors and vertices all depend only 
on the internal properties of the correlated halos, through a unique
scaling variable $x$. 
\end{itemize}
The central formula we derived
is given by Eqs. (\ref{2.47a}-\ref{2.47d}) which 
explicit the generating function of the vertices that
describe the halo auto-correlation function as well as halo-matter
cross-correlations. It provides adapted mathematical
techniques and tools to explore the relation between the intrinsic 
and  joint  properties
of the halos in a nonlinear density field. An illustration of the
strength of these methods is proposed in the last Section.

Generically, 
we find that in all relevant cases $S_P=P^{P-2}$ (which corresponds
to vertices that are all equal to unity) in the rare halo limit.
Also, the mass weighted statistics of the halo distribution reduces, when all dense halos are included, to 
the statistics of the matter field, as a generalisation of the properties
known (BeS92, Schaeffer 1987) for the two-body correlation function.
This is an indication that the halo correlation properties,
especially for the higher masses, reflect the
 selection effects more than  the intrinsic properties of the
matter field (this behavior is similar to what is expected in Gaussian
field, e.g. Mo et al. 1997).

In the small mass limit the behavior of the correlation
functions are more dependent on the underlying matter field.
Robust results in all regime therefore rely on precise assumptions
on the geometrical dependence of the correlation functions.
The simplest assumption is to suppose that the $p$-point functions are
exact trees in the sense that the vertices are pure numbers. This
corresponds to what we call the minimal tree-hierarchical model. In
this case it is possible to relate the correlation properties of
halos, and in particular to the $x$ dependence of $b(x)$,
to their mass function. It leads however to mathematical
inconsistencies (violation of positivity constraints) that we were not
able to cure within the minimal model. 

As an alternative, we develop
a purely phenomenological description, the factorized
tree-hierarchical model. We use this model for illustration purposes,
and because in this case no mathematical inconsistencies are
encountered. 

Our results for the bias are consistent with the observed 
luminosity dependence of the bias (see discussion in Bes92 and in
Benoist et al 1996) although we made no attempt here to determine the
luminosity of the halos we have considered.

Our results
can also be checked against
numerical simulations. Early results by
Bernardeau (1996) were focused on the quasilinear regime but more
recently Munshi et al. (1999) checked in some details the properties
of $b(>x)$ in the light of the predictions that were made in our
previous paper (BeS92) for the minimal model.
They found that the qualitative and quantitative behavior 
of $b(>x)$ were in perfect agreement to what we expected for the
minimal tree model, which make the mathematical inconsistencies we
found all the more puzzling. 

Obviously, further numerical investigations
would be precious for
constraining the models.

\appendix
\section{The generating function of the counts in cell}
\label{Aappend}

Let us divide the universe into cells of volume $v$. The 
probability to find $N$ particles of matter in a given cell, $p(N)$,
can be evaluated by means of the generating function,
\begin{equation}
\mP(\la)=\sum_{N=0}^{\infty}\la^N\,p(N),
\eqno(A.1)
\end{equation}
with,
\begin{equation}
p(N)={1\over 2\pi\i}\oint{\d \la\over \la^{N+1}}\,\mP(\la),
\eqno(A.2)
\end{equation}
where the integral is calculated in the complex plane around $\la=0$.
The generating function $\mP(\la)$ is related (BaS89, White 1979) to
the integrals of the generating functions
\begin{eqnarray}
\mP(\la)&=&e^{\chi(\la)},\eqno(A.3)\\
\chi(\la)&=&\sum_{p=1}^{\infty}{n^p(\la-1)^p\over p!}\int_v
\d^3\vr_1\dots\int_v\d^3\vr_p\,\times\nonumber\\
&&\ \ \ \times\xi_p(\vr_1,\dots,\vr_p).
\end{eqnarray}

\subsection{The minimal tree-hierarchical model}

For the moment, we assume that the $p$-body correlation functions take
the form \eqref(2.5) of the minimal tree-hierarchical model,
\begin{eqnarray} 
&&\hmove \xi_p(\vr_1,\dots,\vr_p)=\nonumber\\
&&\hmove \ \ \ \sum_{{\rm trees}\ (\alpha)} Q_p^{(\alpha)}
\sum_{{\rm labels}\ t_{\alpha}}
\prod_{\rm links} \xi_2(\vr_i,\vr_j),
\eqno(A.4)
\end{eqnarray}
where $(\alpha)$ is a particular tree topology connecting $p-1$ points
without making any closed
loop, $t_{\alpha}$ is a particular labeling of the point coordinates in a
given topology $(\alpha)$, the last product is made over the $p-1$ links
between the $p$ points (labeled by their coordinates $\vr_k$), and 
$Q_p^{(\alpha)}$ is a normalization parameter associated
with the order of the correlation and the topology $(\alpha)$ involved
that takes the form,
\begin{equation}
Q_{p}^{(\alpha)}=\prod_{{\rm vertices\ of\ }(\alpha)}\nu_q.
\eqno(A.5)
\end{equation}
The product is made over all the vertices of the tree, and is defined once
the topology $(\alpha)$ is specified, $\nu_q$ is the weight for each
vertex of the tree and depends on the number $q$ of outgoing lines
(see an example of graph in Fig. 1).

As shown in detail in a previous paper (Appendix A of BeS92), the function
$\chi(\la)$ can be constructed by the classical ``tree-graph'' summation Jannink and des Cloiseaux 1987)(\eg,
Jannink and des Cloiseaux 1987). Let us define,
\begin{equation}
\zeta(\tau)=\sum_{q=0}^{\infty} (-1)^q \nu_q {\tau^q\over q!},
\eqno(A.6)
\end{equation}
then $\chi(\la)$ is given by,
\begin{eqnarray}
\chi(\la)&=&\int_v\zeta[\tau(\vr)](\la-1)\,n\,\d^3\vr\nonumber\\
&&\hmove -{1\over2}\int_v(\la-1)\,n\,\d^3\vr
\,\tau(\vr){\d\zeta\over\d\tau}[\tau(\vr)]\eqno(A.7)\\
\xib_p(v)&=&S_p\left[\xib_2(v)\right]^{p-1},
\end{eqnarray}
with,
\begin{equation}
\tau(\vr)=\int_v(\la-1)\,n\,\d^3\vr'\xi(\vr,\vr')
{\d\zeta\over\d\tau}[\tau(\vr')].
\eqno(A.8)
\end{equation}

\begin{figure}
\centerline{
\psfig{figure=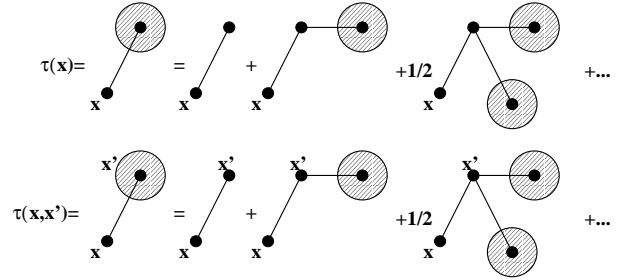,width=8cm}}
\caption{Schematic representation of the equations (\ref{A.8}, \ref{A.16}). 
Each point has the
weight $(\la-1)\,n$ and the lines the value of the two-point
matter correlation function. The shaded area represents the sum of all trees
connected to its inner point. 
The function $\tau(\vx)$ (in [a]) is the
sum of all the trees connected to $\vx$, and can be decomposed as a line
connected to a vertex, connected to as many similar sums. It leads
to the implicit equation in $\tau(\vx)$ given in \eqref(A.7).
The function $\tau(\vx,\vx')$ (in [b]) is the sum of all trees, the first line
of which is $\vx-\vx'$. It can be decomposed in a similar way and it leads
to the equation \eqref(A.16).}
\label{figA1}
\end{figure}

The function $\zeta[\tau(\vr)]$ represents the vertices at the point $\vr$ 
with an $\vr$-dependent weight $\tau(\vr)$ for each outgoing line. The latter
represents the sum of all tree graphs starting at the position $\vr$. Its
value is given by the implicit equation (\ref{A.8}) 
which states that $\tau(\vr)$
equals the specific contribution of the first line $\xi(\vr,\vr')$
multiplied by any kind of vertex $\nu_q$, the $q-1$ remaining lines being
dressed by as many $\tau(\vr')$ factors. The sum over all possible vertices 
introduces the function ${\d\zeta\over\d\tau}[\tau(\vr')]$ (and not
the function $\zeta[\tau(\vr')]$ since one line is treated apart) and an
integration over $\vr'$ is required to finally
construct  $\tau(\vr)$. The Fig. 
\ref{figA1}a presents a diagrammatic representation of the implicit equation
\eqref(A.8).

An important remark that can already be made at this stage is that, although
the matter particles outside the cell are strongly
correlated with the ones inside, {\it the generating function for the counts
in a cell depends only on the correlations within this cell.}

The next step is to simplify considerably the equations (\ref{A.7}-\ref{A.8})
by means of a ``mean field'' approximation that turned out to be
extremely accurate (about $1$ to $2\%$, see discussion in BeS92). We 
integrate \eqref(A.8) over $\vr$ and then approximate
 $\tau(\vr)$ by a constant $\tau$. The equation \eqref(A.8) is then changed
in,
\begin{equation}
\tau=(\la-1)\,n\,v\,\xib\,{\d\zeta\over\d\tau}(\tau),
\eqno(A.9)
\end{equation}
with
\begin{equation}
\xib=\int_v {\d^3\vr\over v}\,{\d^3\vr'\over v}\,\xi(\vr,\vr').
\eqno(A.10)
\end{equation}
The equation \eqref(A.7) then reads,
\begin{equation}
\chi(\la)=(\la-1)\,n\,v\,\zeta(\tau)-{1\over2}(\la-1)\,n\,v\,\tau
{\d\zeta\over\d\tau}(\tau).
\eqno(A.11)
\end{equation}
This is the form that is used for practical calculations.

In the limit of a continuous distribution, when $n$ goes to infinity for
a given mass density $\rho$, the product $(1-\la)\,n$ remains finite.
It will turn out 
to be convenient to define in this case
the variable 
\begin{equation}
y= (1-\la)\,n\,v\,\xib,
\eqno(A.12)
\end{equation}
which remains finite.
We are then led to use the function $\varphi(y)$ ,
\begin{equation}
\chi(\la)=-\varphi(y)/\xib,
\end{equation}
and the relations (\ref{A.9}-\ref{A.11}) then read,
\begin{eqnarray}
\varphi(y)&=&y\zeta(\tau)-{1\over2}y\,\tau{\d\zeta\over\d\tau}(\tau)
\eqno(A.13)\\
\tau&=&-y{\d\zeta\over\d\tau}(\tau).
\eqno(A.14)
\end{eqnarray}

\subsection{The general hierarchical model}

So far we  considered the minimal tree-hierarchical model, for which the
coefficient $Q_p^{(\alpha)}$ are given by \eqref(A.5). In case of a general
tree-hierarchical model, the equations (\ref{A.6}-\ref{A.8}) have to be 
modified to take into account the dependences of the coefficients
$Q_p^{(\alpha)}$ with the geometry of the vertices,
\begin{equation}
Q_p^{(\alpha)}(\vr_1,\dots,\vr_p)=\prod_{{\rm vertices}} \nu_q(\vr_{s_1}-\vr_i,\dots,\vr_{s_q}-\vr_i),
\eqno(A.15)
\end{equation}
where the product is still over all the vertices associated with the tree
topology $(\alpha)$, $\vr_i$ is the coordinate of the considered vertex and
$\vr_{s_1},\dots,\vr_{s_q}$ are the coordinates  of the end points of the
$q$ outgoing lines of the vertex $n$. The difference with the previous
calculation is that the vertex now depends on the geometry of the 
outgoing lines.

In this case we are led to define the function $\tau(\vr,\vr')$ 
representing the sum of all tree graphs starting with the line 
that joins $\vr$ and $\vr'$.
The function $\tau(\vr)$ defined previously is simply the integral over $\vr'$
of $\tau(\vr,\vr')$ within the volume $v$.

The analogous of the equation \eqref(A.8) now reads (see also
 Fig. \ref{figA1}b)
\begin{eqnarray}
&&\hmove \tau(\vr,\vr')=(\la-1)\,n\,\xi(\vr,\vr')\eqno(A.16)\\
&&\hmove \ \times \left[
1+\sum_{q=2}^{\infty}\int_v\d^3\vr_1\dots\d^3\vr_{q-1}
{(-1)^{q-1}\over(q-1)!}\right.\nonumber\\
&&\hmove \ \times\left.
\,\nu_q(\vr'\!-\!\vr,\vr'\!-\!\vr_1,\dots,\vr'\!-\!\vr_{q-1})\,
\tau(\vr',\vr_1)\dots\tau(\vr',\vr_{q-1})\right].\nonumber
\end{eqnarray}

It can be easily checked that this equation gives \eqref(A.8) when the vertices
are pure numbers by an integration over $\vr'$. The function $\chi(\la)$
is now given by ($\nu_1\equiv 1$),
\begin{eqnarray}
&&\hmove \chi(\la)=\int_v(\la-1)\,n\,\d^3\vr\nonumber\\
&&\hmove \ \ \ +\sum_{q=1}^{\infty}(1-{q\over 2})
\int_v(\la-1)\,n\,\d^3\vr\,\d^3\vr_1\dots\d^3\vr_q
{(-1)^q\over q!}\times\nonumber\\
&&\hmove \ \ \,\times\nu_q(\vr-\vr_1,\dots,\vr-\vr_p)
\,\tau(\vr,\vr_1)\dots\tau(\vr,\vr_q),
\eqno(A.17)
\end{eqnarray}
which also reduces to \eqref(A.7) in case of the minimal model.
In the general case there is still a possible ``mean field'' approximation
which just assumes that $\tau(\vr,\vr')$ is proportional to $\xi(\vr,\vr')$,
\begin{equation}
\tau(\vr,\vr')\approx {\tau\over\xib}\,\xi(\vr,\vr').
\eqno(A.18)
\end{equation}
We are then led to exactly the same equation (\ref{A.9}-\ref{A.14}) for
$\chi(\la)$ and $\varphi(y)$ with $\zeta(\tau)$ given by \eqref(A.6) and
the vertices $\nu_q$ defined by the expectation values
\begin{equation}
\nu_q=\int_v{\d^3\vr\over v}\,\nu_q(\vr-\vr_1,\dots,\vr-\vr_q)
\prod_{i=1}^{q}{\xi(\vr,\vr_i)\over\xib}\,{\d^3\vr_i\over v}.
\eqno(A.19)
\end{equation}

Obviously, the validity of such an approximation depends on the strength of the
geometrical dependences of $\nu_q(\vr-\vr_1,\dots,\vr-\vr_q)$. We
assume here that they are weak enough so that the mean field approximation
is valid.

\subsection{The calculation of $p(N)$}
\label{PN}

In the limit of
a continuous distribution of matter
we have the relations
\begin{eqnarray}
e^{-{\varphi(y)\over\xib}}&=&\int_0^{\infty}
\d m\ e^{-{m\over \Mc}y}\,p(m)\, \eqno(A.20)\\
p(m)\,\d m&=&\disp{{\d m\over \Mc}\,\int_{-\i\infty}^{+\i\infty}
{\d y\over2\pi\i\,}\, e^{-{\varphi(y)\over\xib}+{m\over\Mc} y}},
\eqno(A.21)
\end{eqnarray}
with,
\begin{equation}
\Mc=\Mb\,\xib,\ \ \ \Mb=\rho\,v,\label{A.23},
\end{equation}
where $\Mb$ is the mean mass contained in a cell of volume $v$. 
The counts in cells are then obtained by a simple Poisson
convolution of Eq. (\ref{A.21}).

Realistic models for $p(N)$ and hence $\varphi(y)$ or
$\zeta(\tau)$ have been
widely discussed  (Schaeffer 1985, BaS89, BeS92). Generically
$\varphi(y)$ must have singularities at  
negative $y$ and behave as $y^{(1-\omega)}$ at large $y$, with $0 \le
\omega \le 1$. In the strongly non-linear regime $\xia \gg 1$,
(\ref{A.21}) can be simplified into, 
\begin{equation}
p(m)\,\d m=\disp{-{\d m\over \Mc \xib}\,\int_{-\i\infty}^{+\i\infty}
{\d y \over 2\pi\i} \, \varphi(y) e^{{m\over\Mc}} y}.
\eqno(A.30)
\end{equation}
This form holds provided ${\varphi(y)/\xib}$ is small. For low values of $m$,
 however,  $y \propto {\Mc/ m} $ gets large and,
\begin{equation}
{\varphi(y)\over\xib} \propto  ({\Mc \over m})^{(1-\omega)}/\xia,
\eqno(A.31.0)
\end{equation}
may not be small enough. The form (\ref{A.30}) thus holds only provided,
\begin{equation}
m \gg \Mv = \Mc / \xia^{1 \over 1-\omega}.
\eqno(A.31)
\end{equation}
The halos referred to in this paper are those for which (\ref{A.31})
is satisfied.

\section{Joint counts in cells}
\label{Bappend}

We use the same partition of the universe into cells of volume $v_i$ (which
are not necessarily the same) as in the Appendix \ref{Aappend} 
and consider a set of
$\mN$ cells denoted by the index $i=1\dots\mN$. We aim to calculate
the probability
that $N_1$ elementary objects are in cell 1 at position $\vr_1$, $N_2$
in cell 2 at position $\vr_2$, ... The corresponding generating function is,
\begin{eqnarray}
&&\hmove e^{\chi(\la_1,\dots,\la_{\mN})}=\nonumber\\
&&\hmove \ \ \ \sum_{N_1}\dots\sum_{N_{\mN}}\ 
\left(\prod_{i=1}^{\mN}\la_i^{N_i}\right)\ p(N_1,\dots,N_{\mN}).
\eqno(B.1)
\end{eqnarray}
In case $\chi(\la_1,\dots,\la_{\mN})$ is known, $p(N_1,\dots,N_{\mN})$ 
can be obtained through,
\begin{eqnarray}
&&\hmove p(N_1,\dots,N_{\mN})={1\over(2\pi\i)^{\mN}}\ \oint
\prod_{i=1}^{\mN}\ {\d\la_i\over\la_i^{N_i+1}}\times\nonumber\\
&&\hmove \ \ \ \times e^{\chi(\la_1,\dots,\la_{\mN})}.
\eqno(B.2)
\end{eqnarray}

We may ask for the probability that cells $1,\dots,\P$, which are assumed
to all lie with a larger finite volume $V$, contain respectively 
$N_1,\dots,N_{\P}$ elementary particles, whatever the contents of the cells
$\P+1$ to $\mN$ are. This amounts to sum over $N_{\P+1},\dots,N_{\mN}$
in \eqref(B.1) with $\la_{\P+1},\dots,\la_{\mN}$ all equal to unity. It is
important to note that in this case all integrations over the variables
within cells $\P+1$ to $\mN$ drop out: the sums simply run from $i=1$ to $\P$.
In case $\P=1$ we simply recover the expressions (\ref{A.1},
\ref{A.2}) for the counts in one cell, the solution of which is given
by the equations (\ref{A.9}-\ref{A.14}). 

From now on we will always implicitly assume that the cells $1,\dots,\P$
all lie within a volume $V$, much larger than the volume $v_i$ (
$v_i\ll V$) of the elementary cells.

\subsection{General properties of the generating function $\chi$}

\subsubsection{The minimal tree-hierarchical model}
Once again the minimal tree-hierarchical model is simpler and
deserves to be treated separately. The function $\chi$ can be constructed in 
a similar way than in (\ref{A.7}, \ref{A.8}). 
The only change is that the value of $\la$ 
depends on the volume in which the corresponding point lies.

The equation for $\tau(\vr)$ now reads,
\begin{equation}
\tau(\vr)=\sum_{i=1}^{\P}\ \int_{v_i}(\la_i-1)\ n\ \d^3\vr'\xi(\vr,\vr')
{\d\zeta\over\d\tau}[\tau(\vr')].
\eqno(B.3)
\end{equation}
and the expression of $\chi$ is,
\begin{eqnarray}
&&\hmove \chi(\la_1,\dots,\la_{\P})=\sum_{i=1}^{\P}\,\int_{v_i}
(\la_i-1)\,n\,\d^3\vr\,\zeta[\tau(\vr)]\,\nonumber\\
&&\hmove \ \ \ \disp{-{1\over2}\sum_{i=1}^{\P}\,
\int_{v_i}(\la_i-1)\,n\,\d^3\vr\,
\tau(\vr){\d\zeta\over\d\tau}[\tau(\vr)]}.
\eqno(B.4)
\end{eqnarray}

Once again we can introduce a mean field approximation. We assume that
$\tau(\vr)$ has a constant value $t_i$ within each volume $i$. The equation 
for $t_i$ can be obtained from \eqref(B.3),
\begin{equation}
t_i=(\la_i-1)\,n\,v_i\,\xib_i\,{\d\zeta\over\d\tau}(t_i)+
\sum_{j\ne i}(\la_j\!-\!1)\,n\,v_j\,\xib_{ij}\,{\d\zeta\over\d\tau}(t_j)
\eqno(B.5)
\end{equation}
where $\xib_i$ is the mean value of $\xi(\vr,\vr')$ within the volume $i$ and
$\xib_{ij}$ is its means value when $\vr$ is in $v_i$ and $\vr'$ in $v_j$.

For convenience, we split $t_i$ into two terms $\tau_i$ and $\theta_i$,
\begin{eqnarray}
\tau_i&=&(\la_i-1)\,n\,v_i\,\xib_i\,{\d\zeta\over\d\tau}(\tau_i+\theta_i),
\eqno(B.6)\\
\theta_i&=&\sum_{j\ne i}(\la_j-1)\,n\,v_j\,\xib_{ij}\,{\d\zeta\over\d\tau}
(\tau_j+\theta_j).
\eqno(B.7)
\end{eqnarray}
The notation $\tau_i$ corresponds to graphs the first line of which starts
and ends within the same box $i$, whereas $\theta_i$ corresponds to graphs
that start in the box $i$ but immediately join another box $j\ne i$.
The function $\chi(\la_1,\dots,\la_{\P})$ is now given by, 
\begin{eqnarray}
&&\hmove \chi(\la_1,\dots,\la_{\P})=\sum_{i=1}^{\P}\,(\la_i-1)\,n\,v_i\,
\zeta(\tau_i+\theta_i)\nonumber\\
&&\hmove \ \ \ -{1\over2}\sum_{i=1}^{\P}(\la_i-1)\,n\,v_i\,
(\tau_i+\theta_i){\d\zeta\over\d\tau}(\tau_i+\theta_i).
\eqno(B.8)
\end{eqnarray}

The same summation can be written in a different way that will turn out to be
useful in the following. We can first focus on one cell. Let us define
$\chi(\la,\theta)$ by,
\begin{eqnarray}
&&\hmove \chi(\la,\theta)=(\la-1)\,n\,v\,\zeta[\tau(\la,\theta)+\theta]\nonumber\\
&&\hmove \ \ \ -{1\over2}(\la-1)\,n\,v\,\tau(\la,\theta){\d\zeta\over\d\tau}
[\tau(\la,\theta)+\theta],
\eqno(B.9)
\end{eqnarray}
with,
\begin{equation}
\tau(\la,\theta)=(\la-1)\,n\,v\,\xib(v)\,
{\d\zeta\over\d\tau}(\tau(\la,\theta)+\theta).
\eqno(B.10)
\end{equation}
These relations define $\chi(\la,\theta)$ as a function of two variables $\la$ 
and $\theta$ (and the volume of the cell $v$).
This function is the sum off all connected tree graphs within the cell 
(with vertices only in the cell, all being connected by lines)  having
an arbitrary number of outgoing lines, each weighted by a factor $\theta$.
The generating function $\chi(\la_1,\dots,\la_{\P})$ can be
built up with the functions $\chi(\la_i,\theta), i=1,\dots,\P$ 
once $\theta$ is specified for each values of $i$.
The basic property required for such a construction is that one
can relate the partial derivative of $\chi(\la,\theta)$ relatively
to $\theta$ (for a fixed $\la$) to the derivative of $\zeta$,
\begin{equation}
\left.{\partial \chi\over \partial\theta}\right.
(\la,\theta)=(\la-1)\,n\,v\,{\d\zeta\over\d\tau}(\tau(\la,\theta)+\theta)
\eqno(B.11)
\end{equation}
with $\tau(\la,\theta)$ given by \eqref(B.10).

As a result the equation \eqref(B.7) can be written using the function
$\chi(\la,\theta)$ instead of the function $\zeta$,
\begin{equation}
\theta_i=\sum_{j\ne i}\xib_{ij}\,\left.{\partial \chi\over \partial\theta}
\right.(\la_j,\theta_j),
\eqno(B.12)
\end{equation}
and the function $\chi(\la_1,\dots,\la_{\P})$ now reads
\begin{equation}
\chi(\la_1,\dots,\la_{\P})=\sum_{i=1}^{\P}\chi(\la_i,\theta_i)-
{1\over 2}\sum_{i=1}^{\P}\theta_i
{\partial \chi\over \partial\theta}(\la_i,\theta_i).
\eqno(B.13)
\end{equation}

The equations \eqref(B.12) and \eqref(B.13) are similar to the
equations (\ref{A.9}-\ref{A.14}),
where $\zeta(\tau)$ has been replaced by $\chi(\la,\theta)$ as a function of
$\theta$ and where the matter have been replaced by cells of weight
$(\la_i-1)\,n\,v_i$. The function $\chi(\la,\theta)$ is the sum of all
diagrams within a given cell when the environment of the cell has a
weight $\theta$. The function $\chi(\la)$ in the appendix A corresponds to 
the case when no constraints have been put in the vicinity of the cell $i$,
that is given by $\theta=0$, so that $\chi(\la)=\chi(\la,\theta=0)$.

The form $\chi(\la,\theta)$ is the central relation for the construction
of the statistics of the content 
of a large cell $V$ in terms of the elementary finite cells $v_i$, instead
of the elementary points of matter as in the Appendix \ref{Aappend}.
That will be the ground for joint count-in-cell calculations.

\subsubsection{The general tree-hierarchical model}

The case of the general tree-hierarchical models is basically the
same although technically more complicated.

The equation for $\tau(\vr,\vr')$ now reads when $\vr'$ is in the cell $j$,
\begin{eqnarray}
&&\hmove \tau(\vr,\vr')=(\la_j-1)\,n\,\xi(\vr,\vr')\left[1+\sum_{q=2}^{\infty}
\sum_{i_1=1}^{\P}\dots\sum_{i_{q-1}=1}^{\P}\right.\nonumber\\
&&\hmove \ \ \ \left.\times\int_{v_{i_1}}\d^3\vr_1\dots
\int_{v_{i_{q-1}}}\d^3\vr_{q-1} 
\nu_q(\vr'\!-\!\vr,\vr'\!-\!\vr_1,\dots,\vr'\!-\!\vr_{q-1})\right.\nonumber\\
&&\hmove \ \ \ \left.\times{(-1)^{q-1}\over(q-1)!}
\,\tau(\vr',\vr_1)\dots\tau(\vr',\vr_{q-1})\right].
\eqno(B.14)
\end{eqnarray}

The expression of $\chi$ has to be changed in a similar way,
\begin{eqnarray}
&&\hmove
\chi(\la_1,\dots,\la_{\P})=\sum_{i=1}^{\P}\int_{v_i}(\la_i-1)\,n\d^3\vr
+\sum_{q=1}^{\infty}(1-{q\over2})\nonumber\\
&&\hmove \ \ \ \times\sum_{i=1}^{\P}\sum_{i_1=1}^{\P}\dots\!\sum_{i_q=1}^{\P}
\int_{v_{i}}(\la_i\!-\!1)n\d^3\vr\int_{v_{i_1}}\d^3\vr_1\dots
\int_{v_{i_{q}}}\d^3\vr_{q\!-\!1}\nonumber\\
&&\hmove \ \ \ \times{(-1)^q\over q!}\nu_q(\vr-\vr_1,\dots,\vr-\vr_q)\,
\tau(\vr,\vr_1)\dots\tau(\vr,\vr_q).
\eqno(B.15)
\end{eqnarray}
The extension of the mean field approximation leads to assume that,
\begin{equation}
\tau(\vr,\vr')\approx \tau_{ij}{\xi(\vr,\vr')\over \xib_{ij}}
\eqno(B.16)
\end{equation}
where $\vr$ is in the volume $i$ and $\vr'$ in the volume $j$. The value
of $\tau_i$ corresponding to the definition \eqref(B.6) is,
\begin{equation}
\tau_i=\int_{v_i}{\d^3\vr\over v_i}\,\int_{v_i} {\d^3\vr'\over v_i}
\tau(\vr,\vr')=\tau_{ii},
\eqno(B.17)
\end{equation}
whereas,
\begin{equation}
\theta_i=\int_{v_i}{\d^3\vr\over v_i}\,\sum_{j\ne i}
\int_{v_j} {\d^3\vr'\over v_j}\tau(\vr,\vr')=\sum_{j\ne i}\tau_{ij}.
\eqno(B.18)
\end{equation}

Our aim is to calculate \eqref(B.15) with the mean field approximation. The 
quantities $\tau_{ij}$ in fact depend not only on $\la_1,\dots,\la_{\P}$
but also on the positions of the cells in the volume $V$. In case of the
minimal tree-hierarchical model this dependence is completely determined
by the values of $\xib_{ij}$. For the general case the vertices introduce some
extra geometrical dependences. We then are led to introduce a second mean field
approximation which is to assume that $\tau_{ij}$ is just proportional
to $\xib_{ij}$ as far as the positions of the cells are considered. It just
leads to change the equation \eqref(B.16) in,
\begin{equation}
\tau(\vr,\vr')=\tau_i {\xi(\vr,\vr')\over \xib_i},
\eqno(B.19)
\end{equation}
when $\vr$ and $\vr'$ are in $v_i$ and,
\begin{equation}
\tau(\vr,\vr')=\tau_{ij} {\xi(\vr,\vr')\over \xib(V)},
\eqno(B.20)
\end{equation}
when $\vr$ and $\vr'$ are in two different cells $v_i$ and $v_j$.
$\xib(V)$ is the mean value of the two-point matter correlation function
in the large volume $V$. The transformation of the relation \eqref(B.14) and
\eqref(B.15) introduces the geometrical averages,
\begin{eqnarray}
&&\hmove \nu_{q,\Q}= \nonumber\\
&&\hmove \ \ \ \int_V{\d^3\vr\over V}{\d^3\vr_1\over V}\dots{\d^3\vr_{\Q}\over V}
\int_v{\d^3\vr_{\Q+1}\over v}\dots{\d^3\vr_q\over v}      \nonumber\\
&&\hmove \ \ \ 
\times\nu_q(\vr_1-\vr,\dots,\vr_{\Q}-\vr,\vr_{\Q+1}-\vr,\dots,\vr_q-\vr)
\nonumber\\
&&\hmove \ \ \ \times{\xi(\vr_1,\vr)\over\xib(V)} \dots
{\xi(\vr_{\Q},\vr)\over\xib(V)}{\xi(\vr_{\Q+1},\vr)\over\xib(v)}\dots
{\xi(\vr_{q},\vr)\over\xib(v)}.
\eqno(B.21)
\end{eqnarray}
The index $q$ is the total number of lines and $\Q$ is the number of long
lines (going from one cell to another). As soon as $v\ll V$,
the averages $\nu_{q,\Q}$ are independent of the partition of $V$
in small cells. These generalized vertices are thus specific of the matter 
correlation properties.

We then define the generating function $\zeta(\tau,\theta)$ of this
vertices 
\begin{equation}\zeta(\tau,\theta)=\sum_{q,\Q}\,\nu_{q,\Q}
{(-1)^{q-\Q}\tau^{q-\Q}\over (q-\Q)!}{(-1)^{\Q}\theta^{\Q}\over \Q!}.
\eqno(B.22)
\end{equation}
For the minimal tree-hierarchical model we simply have
\begin{equation}\zeta(\tau,\theta)=\zeta(\tau+\theta).
\end{equation}
The equation for $\tau_i$ and $\theta_i$ now reads
\begin{eqnarray}
\tau_i&=&(\la_i-1)\,n\,v_i\,\xib_i\,{\partial\zeta\over\partial\tau}
(\tau_i,\theta_i);\eqno(B.23)\\
\theta_i&=&\sum_{j\ne i}(\la_j-1)\,n\,v_j\,\xib(V)\,
{\partial\zeta\over\partial\theta}(\tau_j,\theta_j)\eqno(B.24)
\end{eqnarray}
and 
\begin{eqnarray}
&&\hmove \chi(\la_1,\dots,\la_{\P})=\sum_{i=1}^{\P}\,(\la_i-1)\,n\,v_i\,
\zeta(\tau_i,\theta_i)\nonumber\\
&&\hmove \ \ \ -{1\over2}\sum_{i=1}^{\P}(\la_i-1)\,n\,v_i\,
\tau_i{\partial\zeta\over\partial\tau}(\tau_i,\theta_i)\nonumber\\
&&\hmove \ \ \ -{1\over2}\sum_{i=1}^{\P}(\la_i-1)\,n\,v_i\,
\theta_i{\partial\zeta\over\partial\theta}(\tau_i,\theta_i).
\eqno(B.25)
\end{eqnarray}
The expression of $\chi(\la_1,\dots,\la_{\P})$ can be transformed as previously
in a tree-graph summation. We define $\chi(\la,\theta)$ by
\begin{eqnarray}
&&\hmove 
\chi(\la,\theta)=(\la-1)\,n\,v\,\zeta[\tau(\la,\theta),\theta]\nonumber\\
&&\hmove \ \ \ -{1\over2}(\la-1)\,n\,v\,\tau(\la,\theta)
{\partial\zeta\over\partial\tau}[\tau(\la,\theta),\theta]
\eqno(B.26)
\end{eqnarray}
with
\begin{equation}
\tau(\la,\theta)=(\la-1)\,n\,v\,\xib(v)\,
{\partial\zeta\over\partial\tau}[\tau(\la,\theta),\theta].
\eqno(B.27)
\end{equation}
We get an equation similar to \eqref(B.11),
\begin{equation}{\partial \chi\over \partial\theta}
(\la,\theta)=(\la-1)\,n\,v\,{\partial\zeta\over\partial\theta}
(\tau,\theta) 
\eqno(B.28)
\end{equation}
and the function $\chi(\la_1,\dots,\la_{\P})$ is the same than in
(\ref{B.12}, \ref{B.13}): 
\begin{eqnarray}
\theta_i&=&\sum_{j\ne i}\xib(V)\,{\partial \chi\over \partial\theta}
(\la_j,\theta_j)\eqno(B.29)\\
\chi(\la_1,\dots,\la_{\P})&=&\sum_{i=1}^{\P}\chi(\la_i,\theta_i)-
{1\over 2}\sum_{i=1}^{\P}\theta_i
{\partial \chi\over \partial\theta}(\la_i,\theta_i).\eqno(B.30)
\end{eqnarray}

The only difference between the minimal tree model and the general case
is that in the latter we are led,
in order to get tractable calculations,
 to integrate out
the position dependence of the $\P$ cells, and thus replace $\xib_{ij}$
in \eqref(B.12) by $\xib(V)$. The information we get from the spatial
dependences of the halo correlation functions is then poorer in this case. 
Nevertheless, 
for a very large variety of applications (see Schaeffer 1984, 1985, 1987, 
and the extended discussion in BeS92)
we can accurately
use the averages of these correlation functions
in the volume $V$,
 as it is shown in the next appendix.

\subsubsection{Diagrammatic representations}

\begin{figure}
\centerline{
\psfig{figure=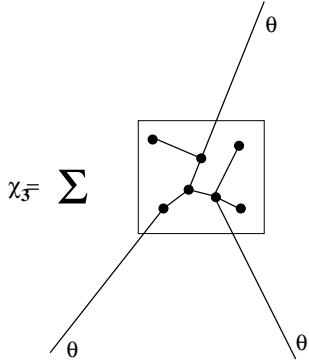,width=4cm}}
\caption{Example of a contribution to the function $\chi_3(\la)$. 
This is the sum
of all the trees within a given box with three external lines. These
lines each have the weight $\theta$.
}
\label{figB1}
\end{figure}

\begin{figure}
\centerline{
\psfig{figure=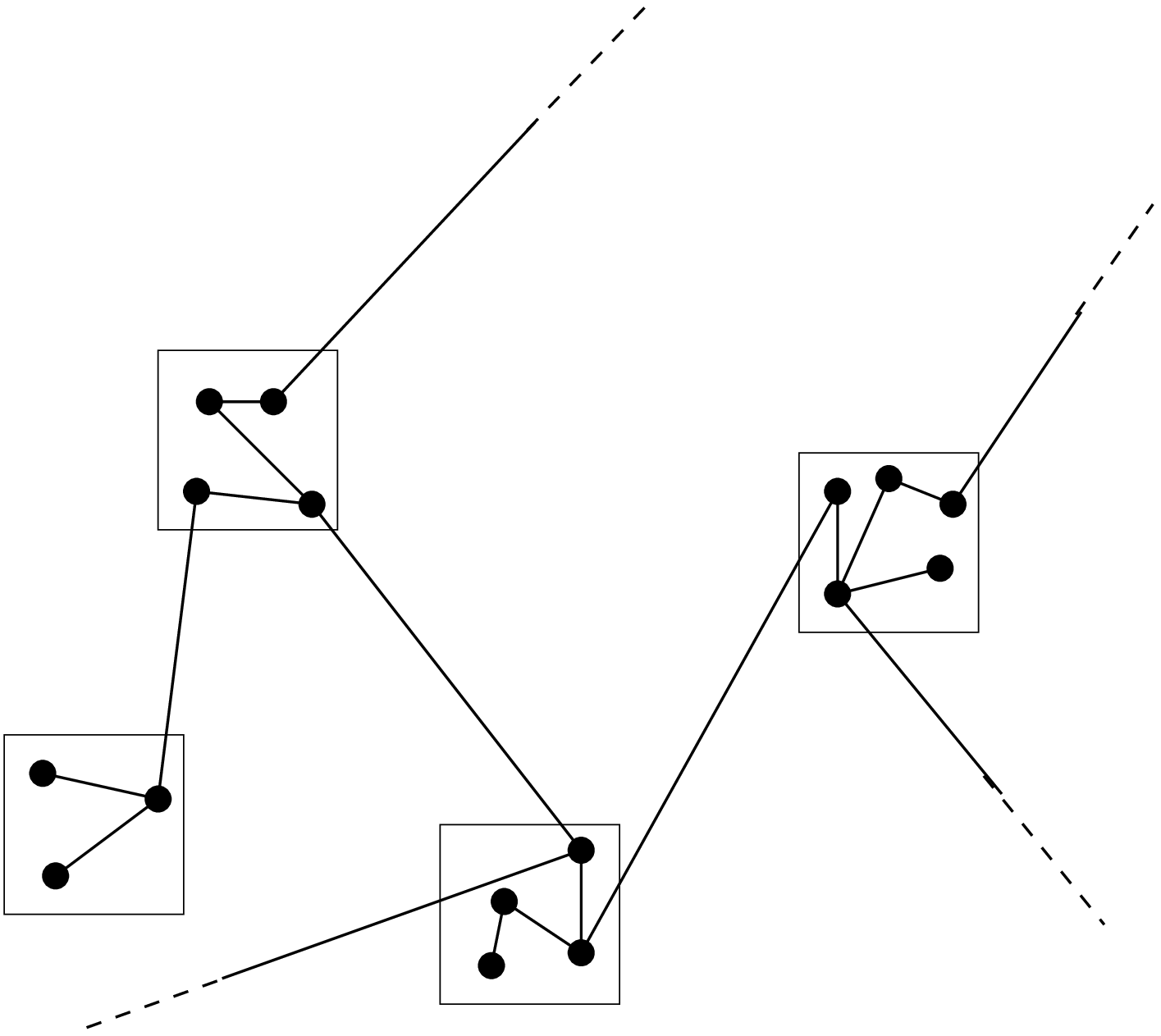,width=7cm}}
\caption{Decomposition of the sum $\chi(\la_1,\dots,\la_I)$ 
in product of functions
$\chi_{\Q}(\la_i)$. The term that appears in the figure involves 4 cells
with the product 
$\chi_1(\la_1)\allowbreak\xi_{12}\allowbreak
\chi_3(\la_2)\allowbreak\xi_{2j_1}\allowbreak\xi_{23}
\allowbreak\chi_3(\la_3)\allowbreak\xi_{3j_2}\allowbreak
\xi_{34}\allowbreak\chi_3(\la_4)\allowbreak\xi_{4j_3}\allowbreak
\xi_{4j_4}\dots.$
The external lines are connected to other cells of the volume $V$ 
(with the indices $j_1$,$j_2$, $j_3$ and $j_4$)
}
\label{figB2}
\end{figure}

The equations \eqref(B.13) and \eqref(B.30) are decisive results 
for joint counts in cells calculations. They deserve attention. 
The function $\chi(\la_1,\dots,\la_{\P})$ is the sum of all connected tree
graphs with their proper  weights, as explained in the
beginning of this section. These relations mean that such
a summation can be performed in two steps. Indeed one can
first compute the sum of all graphs having their vertices in a given
cell and a given number, say $\Q$, of outgoing lines (Fig \ref{figB1}). Let us
denote $\chi_{\Q}(\la)$ such a sum.
The generating function for these graphs
is precisely the function $\chi(\la,\theta)$ (Eqs. [\ref{B.9}] and 
[\ref{B.26}]) 
calculated in the
given cell, $\theta$ being the weight attached to each free line. The 
expansion of $\chi(\la,\theta)$ in powers of $\theta$,
\begin{equation}
\chi(\la,\theta)=\sum_{\Q} (-1)^{\Q} \chi_{\Q}(\la){\theta^{\Q}\over \Q!}
\eqno(B.31)
\end{equation}
then leads to the corresponding sums $\chi_{\Q}(\la)$.
Once the sum within an elementary cell is known,
$\chi(\la_1,\dots,\la_{\P})$ is obtained by connecting (Fig. \ref{figB2}) 
the different
cells in all possible ways, each connection between, say cell $i$ and $j$,
having a weight $\xib_{ij}$ for the case of the minimal tree model, or
simply $\xib(V)$ for the general case. These connections build a new tree
whose elementary vertices are the graphs already summed within one cell.

\begin{figure}
\centerline{
\psfig{figure=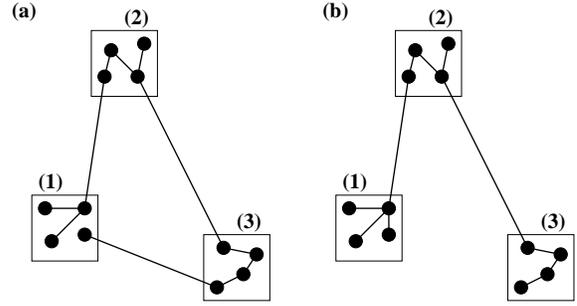,width=7.5cm}}
\caption{Two examples of contribution to the sum $\chi(\la_1,\la_2,\la_3)$
that make intervene the same number of points in each of the cells.
The first one (a) creates a loop between the cells although it is a
tree at the matter point level. The second one (b) is a tree even at the
cells level. The ratio between these two contributions is of the
order of $\xi_{13}/\xib(v)$ (where $\xib(v)$ is the mean value of the
two-point matter correlation function in the volume $v$ of a
cell). In most of the relevant cases the contributions with loops will then
be negligible compare to the trees.
}
\label{figB3}
\end{figure}

At this point however loops between cells
 are not excluded. They arise when a cell  is reached
again
after having been connected to various neighboring cells.
It can be seen in Fig. \ref{figB3}a that the original graph,
at the matter point level, has no loops according
to the rules of the general hierarchical model
although the graph at the cell level contains one loop.

\subsection{Correlations among cells}

\subsubsection{General expression}

From the expression of the generating function $\chi(\la_1,\dots,\la_{\P})$ 
of the connected graph we 
now seek for the probability $p(N_1,\vr_1;\dots;N_{\P},\vr_{\P})$
using the general formula \eqref(B.2). An explicit  calculation is 
possible when one assumes
that the distances between the cells are large compare to their
size so that we have,
\begin{equation}
\xib_{ij}\ll\xib(v).
\eqno(B.36)
\end{equation}

Let us first inspect the structure of $\chi(\la_1,\dots,\la_{\P})$
with the assumption  (\ref{B.36}), having in mind we seek  a perturbation
expansion in $\xi_{ij}/\xib(v)$.
The function $\chi(\la_1,\dots,\la_{\P})$ is a sum of tree graphs
(Fig.\ref{figB3}) connecting points within the various cell 
$i=1,\ldots,\P$, each connection being weighted by the average $\xib(v_i)$ of
the correlation function within the cell, and points belonging to two
different cells, say $i$ and $j$, with a weight $\xi_{ij}$. 
That this is the case in general
is readily seen from equation \eqref(B.31) where the weight of an 
external line appears indeed to be proportional to
$\xi_{ij}$ whereas an internal line has the weight $\xib$. 
The condition \eqref(B.36) insures that a graph that contains a loop
(that is coming back to a given cell after having visited other cells)
introduces a factor $\xi_{ij}/\xib\ll 1$
more than the same graph (same number of points in each of the cells)
(Fig. \ref{figB3}b) that does not contain the loop. 
In the
limit \eqref(B.36), at leadind order,
the remaining 
contribution to $\chi(\la_1,\dots,\la_{\P})$ is the sum with
the smallest number of external lines connecting the different
cells. Whence, in the limit \eqref(B.36), each cell appears only once in each 
remaining graphs. Each cell (as in Fig. \ref{figB2}) 
then appear as a dressed point (see Fig.  \ref{figB1})
with outgoing lines that connect this cell to other ones.

The sum of the remaining graphs is then a sum of product
of  factors $\chi_{\Q}(\lambda)$ corresponding to different cells:
\begin{eqnarray}
&&\hmove \chi(\la_1,\dots,\la_{\P})=\sum_{{\rm tree\ graphs}}\,
C_{{\Q}_1,\dots,{\Q}_{\P}}\nonumber\\
&&\hmove \ \ \ \times\chi_{{\Q}_1}(\la_1)\dots\chi_{{\Q}_{\P}}(\la_{\P}),
\eqno(B.37)
\end{eqnarray}
where $C_{{\Q}_1,\dots,{\Q}_{\P}}$ is independent of the parameters $\la$.

\begin{figure}
\centerline{
\psfig{figure=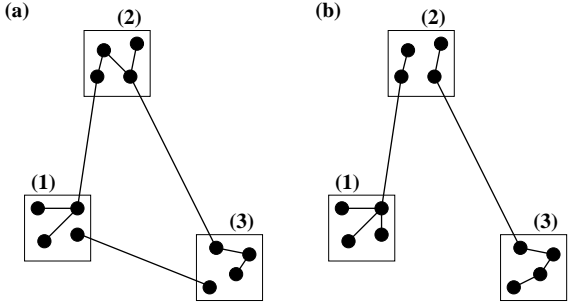,width=7.5cm}}
\caption{Two examples of contribution to 
$p(N_1,N_2,N_3)$ that appear when $\exp[\chi(\la_1,\la_2,\la_3)]$
is expanded. The first diagram is negligible compared to a diagram
where a $\xi_{ij}$ factor has been replaced by a $\xib$
factor that still connect the three boxes. The second
diagram cannot be a priori neglected : if a $\xi_{ij}$ factor is
replaced by a $\xib$ factor the three cells are not connected anymore.
}
\label{figB4}
\end{figure}

This, however is true for  $\chi(\la_1,\dots,\la_{\P})$
whereas \eqref(B.2) involves   
 $\exp[\chi(\la_1,\dots,\la_{\P})]$. We therefore should extend
the above examination to the latter. 
While the former is the sum of all connected tree graphs, the latter
is the sum of all, connected or disconnected, tree graphs where now
the contribution of graphs in which all points are not linked together
is also included (the same way that the generating function of
the cumulants is given by the logarithm of the generating function of
the moments).

From Fig. \ref{figB4}a it is clear the expansion of
$\exp[\chi(\la_1,\dots,\la_{\P})]$  into tree graphs, not only brings
in additional disconnected graphs but also, as previously, loops
between boxes. We now proceed by selecting the connected contributions
at the cell level, that is by identifying those graphs in which all
cells are connected together (but with possible disconnected parts within the
cells). These contributions may thus be  built from disconnected
diagrams at the matter level, that become connected  at the cell level. 
However for the same reason as for Fig. \ref{figB3}a, diagrams similar to
Fig. \ref{figB4}a that build up loops among the cells have a
negligible contribution : it is always 
possible to change a  $\xi_{ij}$ factor by a $\xib$ factor 
and to preserve the connection between the boxes. This is not the case
for diagrams like in Fig. \ref{figB4}b : in such a case changing a $\xi_{ij}$
factor by a $\xib$ factor breaks the connection between the cells.
It is therefore  not possible a priori to neglect the latter contributions.

Eventually we then are left with diagrams that form trees between cells
in which the ``dressed'' vertices of order $\Q$ corresponding to each
cell are obtained by the summation of all diagrams - connected or not
within the cell- having $\Q$ outgoing lines.

\begin{figure}
\centerline{
\psfig{figure=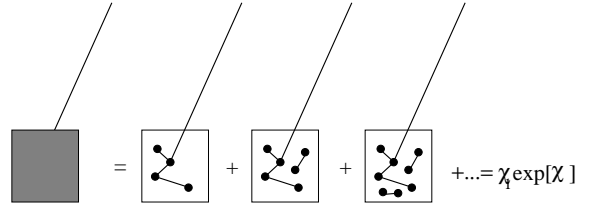,width=7.5cm}}
\caption{Contruction of the generating function of
$b(N)p(N)$. It involves all possible diagrams, connected or not,
that have one external line. It can be built from the generating
function of the connected diagrams with one external line, $\chi_1$, and the
generating function of the diagrams with no external lines, $\chi$.
}
\label{figB5}
\end{figure}

We know that the generating  function of the connected diagrams 
in a cell, with  outgoing lines weighted by $\theta$, is
$\chi(\lambda,\theta)$. The generating function of all  diagrams,
connected or unconnected,  is therefore $\exp[\chi(\la,\theta)]$ . 
This point can be illustrated for the diagrams with one
external line (e.g. on Fig. \ref{figB5}). The connected diagrams with
one external line are given by $\chi_1(\lambda)$. Adding
diagrams with no external lines amounts to multiply $\chi_1(\lambda)$ by the
generating function of those diagrams, $\exp[\chi(\lambda)]$ and the
fist derivative of $\exp[\chi(\lambda,\theta)]$ with respect to
$\theta$ is indeed $\chi_1(\lambda)\,\exp[\chi(\lambda)]$.

In a  similar way than at the end of the appendix \ref{Aappend},
we may introduce the function $\varphi(y,\theta)$ defined by,
\begin{equation}
\chi(\la,\theta)=-\varphi(y,\theta)/\xib(v),
\eqno(B.32)
\end{equation}
that changes the equation (\ref{B.26}, \ref{B.27}) in 
\begin{eqnarray}
\varphi(y,\theta)&=&y\zeta(\tau,\theta)-{1\over 2}y\,\tau
{\partial \zeta\over \partial \tau}(\tau,\theta)\eqno(B.33)\\
\tau&=&-y\,{\partial \zeta\over \partial \tau}(\tau,\theta)\eqno(B.34)
\end{eqnarray}
$y$ being defined in \eqref(A.12).
Then in case we are only interested in large values of $N$, in such a way
that,
\begin{equation}
N\gg\Nv=n\,v\,\xib^{-\omega/(1-\omega)},
\eqno(B.36.1)
\end{equation}
then $\chi(\la,\theta)$ is of the order of $1/\xib$ ($y$ is finite in
Eq. \ref{B.32}, see the discussion at the end of \ref{Aappend}
for the case where \eqref(B.36.1) is not fulfilled). So,
$\exp[\chi(\la,\theta)]\approx 1+\chi(\la,\theta)$ so that only connected
diagrams contribute to the dressed vertices. This is also the result
we could have obtained directly by writing 
 $\exp[\chi(\la_1,\ldots,\la_{\P})] \sim  1 + \chi(\la_1,\ldots,\la_{\P})$ ,
as can be inferred from the discussion at  the beginning of this paragraph.

In all cases, whether $N$ is larger than $\Nv$ or not, 
each cell appears only once in each term of the expansion
of $\exp[\chi(\la_1,\ldots,\la_{\P})]$ : at the cell
level all diagrams are connected trees.
The integration \eqref(B.2) over $\la_1,\dots,\la_{\P}$ is thus easily
performed, and  
amounts to replace each factor depending only on the variable $\lambda_i$
of one specific cell by its integral
${1/(2\pi\i)}\allowbreak\times
\int\d\la_i/\la_i^{N_i+1}$. This simply amounts to replace 
the sum in a given cell of all graphs with external lines 
$\exp[\chi(\la,\theta)]$ by, 
\begin{equation}
 p(N,\theta)={1\over 2\pi\i}\int{\d\la\over \la^{N+1}}e^{\chi(\la,\theta)}
\eqno(B.38)
\end{equation}
and to use these new generating functions as elementary vertices. 
For dense cells (with the meaning of 
Eq. [\ref{B.36.1}]) this  expression can be   simplified as
\begin{equation}
p(N,\theta)={1\over 2\pi\i}\int{\d\la\over \la^{N+1}}\chi(\la,\theta).
\eqno(B.38.b)
\end{equation}

As for the matter distribution (Eq. [\ref{A.6}]), we can define the
generating function of the vertices $\zeta(N,\theta)$ (App. \ref{Aappend}) by,
\begin{equation}
p(N,\theta)=p(N)\,\zeta(N,\theta)
\eqno(B.39)
\end{equation}
recalling that $p(N)\equiv p(N,0)$. The expansion
of $\zeta(N,\theta)$ in powers of $\theta$ defines new vertices $\nu_{\Q}(N)$,
\begin{equation}
\zeta(N,\theta)=\sum_{\Q} (-1)^{\Q} \nu_{\Q}(N)\,{\theta^{\Q}\over \Q!}
,
\eqno(B.40)
\end{equation}
that replace the parameter $\nu_q$ relevant for the matter distribution.
We have, for instance
\begin{eqnarray}
\nu_1(N)p(N)&\equiv& b(N)p(N)\nonumber\\
&=&-{1\over 2\pi\i}\int{\d\la\over
\la^{N+1}}\chi_1(\la)e^{\chi(\la)};\\
\nu_2(N)p(N)&=&{1\over 2\pi\i}\int{\d\la\over
\la^{N+1}}\left[\chi_2(\la)+\chi_1^2(\la)\right]\,e^{\chi(\la)}.
\end{eqnarray}
In the continuous limit we can further make the change of variable
$\lambda\to y$ (e.g. Eq. \ref{A.13}) and we have,
\begin{eqnarray}
b(m)p(m)&=&{1\over \Mc}\int_{-\i\infty}^{+\i\infty}{\d y\over
2\pi\i}{\varphi^{(1)}(y)\over\xib}\nonumber\\
&&\times\ e^{-{\varphi(y)/\xib}+y\,{m/\Mc}};\\
\label{nu1compl}
\nu_2(m)p(m)&=&-{1\over \Mc}\int_{-\i\infty}^{+\i\infty}{\d y\over
2\pi\i}\left[{\varphi^{(2)}(y)\over\xib}
-\left(\varphi^{(1)}(y)\over\xib\right)^2\right]\ 
\nonumber\\
&&\times\ e^{-{\varphi(y)/\xib}+y\,{m/\Mc}},
\label{nu2compl}
\end{eqnarray}
where $\varphi^{(1)}(y)$ and $\varphi^{(2)}(y)$ are successive derivatives of
$\varphi(y,\theta)$ with respect to $\theta$.

For the dense cells we have considered in this paper, 
$m \gg \Mv$, that is typically $m \sim \Mc$, these vertices, as well
as the ones of higher order depend on mass only through the ratio $x =
m/\Mc$. This limit will be examined in detail in Append. \ref{Dappend}.  
For the underdense cells, that cannot be called ``halos'',
that is for $m \ll \Mc$ and typically $m \sim \Mv$, in many cases,
there is another scaling law, similar to the one found for the
count-in-cells (BaS89) as well 
as for the bias parameter $b$ (BeS92). The vertices, in this case
depend on mass only through the  parameter $ z = m/\Mv$, with an
overlapping regime 
for $\Mv \ll m \ll \Mc$. This case, its conditions of existence and
its implications for the cell correlations, will be studied
elsewhere.

The sum over tree connections in all possible ways among the cells under
consideration is then performed by writing,
\begin{eqnarray}
&&\hmove p(N_1,\dots,N_{\P})=\nonumber\\
&&\hmove\ \ \exp\left[{-\sum_i p(N_i)\,\zeta(N_i,\theta_i)
+{1\over 2}\,p(N_i)\,\theta_i\,
{\partial \zeta\over \partial \theta}
(N_i,\theta_i)}\right]
\eqno(B.41)
\end{eqnarray}
with\footnote{Note that we use the same notation $\theta$ for the 
expression $\theta(N_1,\ldots,N_{\P})$
of this weight after integration over $\lambda$ 
(when the loops are not included) as for the 
weight $\theta(\lambda_1,\ldots,\lambda_{\P})$
in equation \eqref(B.30), although the two quantities are not the same.},
\begin{equation}
\theta_i=-\sum_{j\ne i}
\xib_{ij}\,p(N_j)\,{\partial \zeta\over \partial \theta}
(N_j,\theta_j).
\eqno(B.42)
\end{equation}
In case of the general tree-hierarchical model
$\xib_{ij}$ in \eqref(B.42) has to be replaced by $\xib(V)$.

We may similarly ask for the correlation among
cells which contain more than $N$ elementary particles, and explore the effect
of a threshold in the matter field. The same calculation shows that $P$,
and $\zeta$ are respectively replaced by
\begin{eqnarray}
p(>N,\theta)&=&{1\over 2\pi\i}
\int\,{\d\la\over \la^{N+1}(\la-1)}\,e^{\chi(\la,\theta)},\eqno(B.43)\\
\zeta(>N,\theta)&=&p(>N,\theta)/p(>N,0).\eqno(B.44)
\end{eqnarray}
The expansion of $\zeta(>N,\theta)$ in powers of $\theta$ defines the vertices
$\nu_q(>N)$ that are to be used to construct the many-body correlation
functions among cells containing  more than $N$ particles.

Similarly, we may also affect a multiplicity $N$ to each cell 
that contains $N$  particles. The related statistics can also
be described by tree correlations with the functions
\begin{eqnarray}
\bp(>N,\theta)&=&{1\over 2\pi\i}
\int\,{\d\la\over \la^{N+1}(\la-1)}\,{\la\d\over\d\la}\,
e^{\chi(\la,\theta)}\eqno(B.45)\\
\bzeta(>N,\theta)&=&\bp(>N,\theta)/\bp(>N,0).\eqno(B.46)
\end{eqnarray}
The vertex generating functions obey the normalization conditions,
\begin{eqnarray}
\sum_Np(N)\zeta(N,\theta)&=&1;\eqno(B.47a)\\
\sum_N{N\over \Nb}p(N)\zeta(N,\theta)&=&\zeta(\tau=0,\theta).
\eqno(B.47b)
\end{eqnarray}
In the first sum, the underdense cells are heavily weighted, with a
negligible contribution of order $1/\xib(v)$, for the same reasons as
the ones discussed in Sect. \ref{PN} for the count-in-cells.   
The last, on the other hand, is heavily weighted towards the dense cells, the 
underdense contribution vanishing as a power of $1/\xib(v)$.
These sum rules imply for the vertices $\nu_{\Q}(N)$, for $\Q >0$
($\nu_0(N)  = 1$ by convention)
\begin{eqnarray}
\sum_Np(N) \nu_{\Q}(N)  &=&0;\eqno(B.47b1)\\
\sum_N{N\over \Nb}p(N) \nu_{\Q}(N)  &=&\nu_{q,\Q=q}.
\eqno(B.47b2)
\end{eqnarray}

\subsection{Counts of halos}
Let us consider a volume $V$, made of finite elementary cells of finite
volume $v$. We assume that,
\begin{equation}V\gg v.
\eqno(B.48)
\end{equation}

In the spirit of the previous section where we calculated the correlation
among dense cells, we consider now the counts of dense cells in the 
volume $V$, that is the probability to find a given number of ``full''
sub-cells in the volume $V$. To be precise we may decide a cell
is ``full'' if it contains more than $N$ elementary particles (or
exactly $N$..) and ``empty'' when the converse is true. The cells in the
volume $V$ may be labeled from $i=1$ to $i=\P$. We assume the threshold
value $N$ to be the same for all cells although a generalization with
a mixture of thresholds is straightforward.

The generating function for the counts of cells in volume $V$,
\begin{equation}
e^{\chi_V(\la)}=\sum_K\,\la^K\,\VP (K),
\eqno(B.49)
\end{equation}
can be defined from the probability $\VP (K)$ of having $K$
full cells in $V$. From the probability $p(>N)$ for a cell
to be full, we may define a number density of cells
\begin{equation}
n(>N)={1\over v}p(>N).
\eqno(B.50)
\end{equation}
Now, since we know that the many-body correlation functions, in the
limit $V\gg v$, are given by a tree-graph series with a vertex generating 
function $\zeta(>N,\theta)$, we get the equation
\begin{eqnarray}
\chi_V(\la)&=&\sum_i
(\la-1)\,n(>N_i)\,v_i\,\zeta(>N_i,\theta_i)\nonumber\\
&&\hmove \ \ \ -{1\over2}\sum_{i}(\la-1)\,n(>N_i)\,v_i\,\theta_i\,
{\partial\zeta\over\partial\theta}(>N_i,\theta_i),\eqno(B.51)\\
\theta_i&=&\sum_j(\la-1)\,n(>N_j)\,v_j\,\xib_{ij}\,
{\partial\zeta\over\partial\theta}(>N_j,\theta_j),\eqno(B.52)
\end{eqnarray}
which are the same as \eqref(A.7), \eqref(A.8), but written for finite
cells and with 
a different vertex generating function $\zeta(>N,\theta)$ that replaces 
$\zeta(\tau)$ of equation \eqref(A.6). For the general tree-hierarchical models
$\xib_{ij}$ has to be replaced by $\xib(V)$.

In case all the volumes of the elementary cells $v_i$ 
are equal, and all thresholds
$N_i$ are the same, we get the analogue of equations (\ref{A.9}-\ref{A.14}):
\begin{eqnarray}
Y&=&(1-\la)\,n(>N)\,V\,\xib(V);\eqno(B.53)\\
\theta&=&-Y{\partial \zeta\over \partial \theta}(>N,\theta);\eqno(B.54)\\
\Phi(Y)&=&Y\zeta(>N,\theta)-{1\over2}Y\,\theta
{\partial \zeta\over \partial \theta}(>N,\theta);\eqno(B.55)\\
\chi_V(\la)&=&-\Phi(Y)/\xib(V).\eqno(B.56)
\end{eqnarray}
These equations together with the equations \eqref(B.26), \eqref(B.27)
and \eqref(B.43)
allows to derive the generating function of the counts of halos
in the volume $V$. Once $\chi_V(\la)$ is known, one can get, by the
inversion of the relation \eqref(B.49), the values of $\VP(K)$. In particular
the void probability function in $V$ (the absence of dense objects, not
the absence of matter) can be directly obtained by $\VP(0)=\exp[\chi_V(0)]$.

We can note that in case the threshold is given by $N=0$ we should
recover
the expression of the void probability function for the matter field
in the volume $V$ which reads,
\begin{eqnarray}
Y&=&n\,V\,\xib(V);\eqno(B.57)\\
\theta&=&-Y{\d \zeta\over \d \theta}(\theta);\eqno(B.58)\\
\varphi(Y)&=&
Y\zeta(\theta)-{1\over2}Y\,\theta{\d \zeta\over \d \theta}(\theta)
;\eqno(B.59)\\
\VP(0))&=&e^{\chi_V(\la=0)}=e{-\varphi(Y)/\xib(V)}.\eqno(B.60)
\end{eqnarray}

To check this we have to show that $n(>0)\zeta(>0,\theta)=n\zeta(\theta)$.
From \eqref(B.43) we get 
\begin{equation}
n(>0)\,v\,\zeta(>0,\theta)=-\chi(0,\theta).
\eqno(B.61)
\end{equation}
The latter, from \eqref(B.26), \eqref(B.27) is given by
\begin{eqnarray}
\chi(0,\theta)&=&-n\,v\,\zeta(\tau,\theta)+
{1\over2}n\,v\,\tau\,
{\partial\zeta\over\partial\tau}(\tau,\theta)
,\eqno(B.62)\\
\tau&=&-n\,v\,\xib(v)\,
{\partial\zeta\over\partial\tau}(\tau,\theta).\eqno(B.63)
\end{eqnarray}
If $\theta$ is finite, then $Y$ must be finite too, and this is only possible
when $y$ is small (since $v\ll V$)
and whence $\tau$ is small too. If $\theta$ is large, then according
to the equation \eqref(B.63), $\tau$ is necessarily small whatever the value
of $y$: its value is dominated by the environment. When $\tau$ is small
$\zeta(\tau,\theta)\approx\zeta(0,\theta)$ if the function
$\zeta(\tau,\theta)$ is regular enough, otherwise that would mean the
successive mean field approximations that we did are not valid (because
the dependence of the vertices is too strong).
Except in these cases we see that we 
recover the right form for $\zeta(>0,\theta)$.

\section{Scaling laws for dense cells}
\label{Dappend}

In this section we are interested in cases where dense
cells only are considered. It means that we assume that $N \gg \Nv$ so that
we can apply the formula (\ref{B.38.b}).
We further assume that the density of points is large enough so that
the continuous limit can be used. It amounts to make the change of
variable from $\lambda$ to $y$. One can then use the formulae
(\ref{B.32}-\ref{B.34}) for $\chi(\lambda,\theta)$.

\subsection{Continuous limit}

The continuous limit is obtained by letting the number density $n$ of
points go to infinity, while keeping the mass density $\rho$ finite.
When this limit applies, the cell content is described by its mass $m$
(see Sect. \ref{PN}).
Remaining in the dense cell regime,
the  results can be described with the
variable  $x=m/\Mc$, and we have,
\begin{eqnarray}
p(m,\theta)&=&-{1\over  \xib}\int_{-\i\infty}^{+\i\infty}
{d y \over 2\pi\i} \,\varphi(y,\theta)\,e^{xy}\equiv
{1\over \xib}\,h(x,\theta)\eqno(D.3)\\
\zeta(m,\theta)&=&{h(x,\theta)\over h(x)}\equiv \zeta(x,\theta).\eqno(D.4)
\end{eqnarray}
The function $\zeta(x,\theta)$ generates by a series expansion,
\begin{equation}
\zeta(x,\theta)=\sum_{\Q} \nu_{\Q}(x){(-\theta)^{\Q}\over {\Q}!},
\eqno(D.5)
\end{equation}
the vertices $\nu_p(x)$ that weight the vertices of the tree expansion
of the cells. The number density $n(x)$ of $x$-type cells is in this case
\begin{equation}
n(x)={\rho\over \Mc}\,h(x) .
\eqno(D.6)
\end{equation}

In case we define a ``full'' cell by a threshold $x$, $h(x,\theta)$ and $h(x)$
have to be replaced by,
\begin{eqnarray}
h(>x,\theta)&=&\int_x^{\infty}\,h(x,\theta)\,\d x\nonumber\\
&=&-\int_{-\i\infty}^{+\i\infty}
{\d y\over 2\pi\i}\,{\varphi(y,\theta)\over y}\,e^{xy},\eqno(D.7)\\
h(>x)&=&\int_x^{\infty}\,h(x)\,\d x= - \int_{-\i\infty}^{+\i\infty}
{\d y\over 2\pi\i}\,{\varphi(y)\over y}\,e^{xy},\eqno(D.8)
\end{eqnarray}
and in case each cell is weighted by its content they have to be replaced
by,
\begin{eqnarray}
\hb(>x,\theta)&=&\int_x^{\infty}\,x
\,h(x,\theta)\,\d x\nonumber\\
&=&-\int_{-\i\infty}^{+\i\infty}
{\d y\over 2\pi\i}\,{1\over y}\,
{\partial\varphi(y,\theta)\over \partial y}\,e^{xy},\eqno(D.9)\\
\hb(>x)&=&\int_x^{\infty}\,x\,h(x)\,\d x\nonumber\\
&=&-\int_{-\i\infty}^{+\i\infty}
{\d y\over 2\pi\i}\,{1\over y}\,{\d\varphi(y)\over\d y}\,e^{xy}.
\eqno(D.10)
\end{eqnarray}

The function $\zeta(x,\theta)$, Eq. \eqref(D.5), and the corresponding
generating 
functions deduced from $h(>x,\theta)$ or $\hb(>x,\theta)$ defines a new set 
of many-body correlation functions. For instance, the vertex,
\begin{equation}
\nu_1(x)=\left.{\partial
\zeta\over\partial\theta}(x,\theta)\right\vert_{\theta=0}, 
\eqno(D.11)
\end{equation}
or its analogues $\nu_1(>x)$, $\nub_1(>x)$ are simply the bias factors,
$b(x)$, $b(>x)$, $\bb(>x)$ respectively as presented in BeS92.

The normalization condition \eqref(B.47b) implies
\begin{equation}
\int xh(x)\zeta(x,\theta)\,\d x=\zeta(\theta)
\eqno(D.12)
\end{equation}
which is a generalization of the sum rule (BeS92),
\begin{equation}
\int\,xh(x)b(x)\,\d x=1,
\eqno(D.13)
\end{equation}
that can be deduced from \eqref(D.12) by the expansion of 
$\zeta(\theta)=1-\theta+\dots$.

There is no counterpart to \eqref(B.47a) for which the domain $N\simlt\Nv$
is predominant.

\subsection{The renormalized vertices}

In case of a single set of objects, the observable quantities, such as the
various moments of the density distribution, give information
only on $\nu_p(x)/b^p(x)$ since the absolute strength of the matter
correlation is not directly observable.

We are then led to define
\begin{eqnarray}
\tzetax(x,\theta)&=&1-\theta+{\nu_2(x)\over b^2(x)}\,{\theta^2\over 2}
-{\nu_3(x)\over b^3(x)}\,{\theta^3\over 3!}+\dots\nonumber\\
&=&\zeta(x,\theta/b[x]).
\eqno(D.14)
\end{eqnarray}
This function may be determined 
directly
from the observed galaxy distribution  through the void
probability function.

\subsection{Asymptotic form for $\tzetax(x,\theta)$}

The exact form of $\zetax(x,\theta)$ depends on the form taken by
$\zeta(\tau,\theta)$. There is however a limit which is independent
of the exact form of $\zeta(\tau,\theta)$ which corresponds to the
case $x\to \infty$. First of all we derive the expression
of the bias,
\begin{eqnarray}
b(x)\,h(x)&=&\int_{-\i\infty}^{+\i\infty}{\d y\over 2\pi\i}
\,y\,{\partial\zeta\over\partial\theta}(\tau,\theta)\,e^{xy},\nonumber\\
\tau(y)&=&-y{\partial\zeta\over\partial\tau}(\tau(y),0)
\eqno(D.15)
\end{eqnarray}
and $h(x)$ is given by the equation \eqref(D.4). The behavior of $b(x)$ 
for large values of $x$ can be calculated by the steepest descent method, 
taking advantage of the singularity in the implicit equation for $\tau(y)$.
We are then led to define $\tau_s$ and $y_s$ so that 
${\d y\over\d\tau}(\tau_s)=0$,
\begin{eqnarray}
\tau_s&=&{\partial\zeta\over\partial\tau}(\tau_s,0)/
{\partial^2\zeta\over\partial\tau^2}(\tau_s,0),\nonumber\\
y_s&=&-1/{\partial^2\zeta\over\partial\tau^2}(\tau_s,0).
\eqno(D.16)
\end{eqnarray}
Then we can derive the expression of the bias,
\begin{eqnarray}
b(x)h(x)&=&\int_{-\i\infty}^{+\i\infty}{\d y\over2\pi\i}\,y_s\,
{\partial^2\over\d\tau\d\theta}(\tau_s,0)(\tau-\tau_s)\,e^{xy},\nonumber\\
xh(x)&=&\int_{-\i\infty}^{+\i\infty}{\d y\over2\pi\i}\,
{\partial\zeta\over\partial\tau}(\tau_s,0)\,(\tau-\tau_s)\,e^{xy}.
\eqno(D.17)
\end{eqnarray}
As a result we get the behavior of $b(x)$ when $x$ is large,
\begin{equation}
b(x)=x\,y_s\,{\partial^2\over\d\tau\d\theta}(\tau_s,0)/
{\partial\zeta\over\partial\tau}(\tau_s,0).
\eqno(D.18)
\end{equation}
The form \eqref(D.4) 
can be calculated with a similar method. The position of the
singularity has to be determined as a function of $\theta$. So we consider
$y_s(\theta)$ and $\tau_s(\theta)$ so that 
${\d y\over\d\tau}(\tau_s(\theta),\theta)=0$:
\begin{eqnarray}
\tau_s(\theta)&=&{\partial\zeta\over\partial\tau}(\tau_s,\theta)/
{\partial^2\zeta\over\partial\tau^2}(\tau_s,\theta)\nonumber\\
y_s(\theta)
&=&-1/{\partial^2\zeta\over\partial\tau^2}(\tau_s,\theta).
\eqno(D.19)
\end{eqnarray}
We obviously  have $y_s(\theta=0)=y_s$ and $\tau_s(\theta=0)=\tau_s$.

We can then write the form of the generating function of the vertices,
\begin{equation}
\zetax(x,\theta)={{\partial\zeta/\partial\tau}(\tau_s,\theta)\over
{\partial\zeta/\partial\tau}(\tau_s,0)}e^{(y_s(\theta)-y_s)x}.
\eqno(D.20)
\end{equation}
The resulting expression for $\tzetax(x,\theta)$ is given by 
$\tzetax(x,\theta)=\zetax(x,\theta/b(x))$ and as $b(x)$ is proportional
to $x$, in the limit $x\to \infty$ we have $\theta/b(x)\ll 1$. It implies
that 
\begin{equation}
y_s(\theta)=y_s+\left.{\d y_s\over \d \theta}\right\vert_{\theta=0}
{\theta\over b(x)}
\eqno(D.21)
\end{equation}
which leads to
\begin{equation}
\tzetax(x,\theta)\approx e^{-\theta}
\eqno(D.22)
\end{equation}
in the limit $x\to\infty$. This is a generalization of previous results
obtained by BeS92 where it was shown that $Q_3(x)\to 1$ in such a limit.
Here we obtain that all the vertices $\nu_{\Q}(x)\to 1$. This result is valid
for the minimal tree-hierarchical model as well as for the general case.
In the latter case however we can only give the mean behavior of the
many-body correlation functions of the halos as a function of scale and
the geometrical dependence is not known. We thus rather have to define
the averages of the halo correlation function in a volume $V$,
\begin{equation}
\xib_{\P}^{{\rm halo}}=\int{\d^3\vr_1\over V}\dots{\d^3\vr_p\over V}\
\xi_{\P}^{{\rm halo}}(\vr_1,\dots,\vr_p)
\eqno(D.23)
\end{equation}
and the coefficient $S_{\P}(x)$ by
\begin{equation}
S_{\P}(x)={\xib_{\P}^{{\rm halo}}\over 
\left(\xib_2^{{\rm halo}}\right)^{\P-1}}.
\eqno(D.24)
\end{equation}
These coefficient are mere combinations of the vertices $\tnu_{\Q}(x)$, and
in the limit $x\to\infty$ we get that $S_{\P}(x)$ is simply the number of
trees that can be constructed with $p$ points, that is
\begin{equation}
S_{\P}(x)=\P^{\P-2}\ \ \ {\rm when}\ \ \ x\to\infty.
\eqno(D.25)
\end{equation}
The latter limit is valid for any tree-hierarchical models,
provided some regularity conditions hold for the generating function
$\zeta(\tau,\theta)$ so as the above expansions are justified.

\end{document}